\documentclass[aps,pra,reprint,twocolumn,superscriptaddress,floatfix,nofootinbib,longbibliography]{revtex4-1}
\usepackage{amsthm}
\usepackage{amsmath,amssymb,color,comment,physics}
\usepackage[makeroom]{cancel}
\usepackage[utf8]{inputenc}
\usepackage{newunicodechar}
\newunicodechar{́}{\'}
\usepackage[caption=false]{subfig}
\usepackage{mathrsfs}
\usepackage{graphicx}
\usepackage{subfig}
\usepackage[countmax]{subfloat}
\usepackage[english]{babel}
\usepackage{dsfont}
\usepackage[normalem]{ulem}
\usepackage[bookmarks=true,colorlinks,linkcolor=OrangeRed,urlcolor=NavyBlue,citecolor=RoyalBlue]{hyperref}
\usepackage[dvipsnames]{xcolor}
\usepackage{braket}
\usepackage{bm}
\usepackage{cancel}
\usepackage[utf8]{inputenc}
\definecolor{mypurple}{rgb}{0.49,0.18,0.56}

\DeclareMathOperator{\dg}{\dagger}
\DeclareMathOperator{\uarr}{\uparrow}
\DeclareMathOperator{\darr}{\downarrow}
\DeclareMathOperator{\br}{ {\bf r} } % Bold vector: r
\DeclareMathOperator{\cP}{ \hat{c} } % Projected (onto LLL) fermion operator

\begin{document}

\title{Quantum Hall Ferromagnetism in a Cavity Vacuum}

\author{Ceren~B.~Dag}
\email{cbdag@iu.edu}

\affiliation{Department of Physics, Indiana University, Bloomington, Indiana 47405, USA}
\affiliation{Quantum Science and Engineering Center, Indiana University, Bloomington, Indiana 47405, USA}
\affiliation{ Department of Physics, Harvard University, Cambridge, Massachusetts 02138, USA}

\author{Luis Brey}
\email{brey@icmm.csic.es}
\affiliation{Instituto de Ciencia de Materiales de Madrid (CSIC), Cantoblanco, 28049 Madrid, Spain}

\author{Ganpathy Murthy}
\email{murthy@g.uky.edu}
\affiliation{Department of Physics and Astronomy, University of Kentucky, Lexington, KY 40506, USA}

\author{H.A. Fertig}
\email{hfertig@iu.edu}
\affiliation{Department of Physics, Indiana University, Bloomington, Indiana 47405, USA}
\affiliation{Quantum Science and Engineering Center, Indiana University, Bloomington, Indiana 47405, USA}

%\date{\today}

\begin{abstract}

We uncover a continuous phase transition in a quantum Hall ferromagnet (QHF) at filling factor $\nu=1$, driven by vacuum fluctuations of a cavity. Our analysis starts with a Landau level projection in dipole gauge, where we find the states to be well-represented by a tensor product of the electronic and photonic degrees of freedom. 
Through analytic spin wave calculations, mean-field theory and density matrix renormalization group (DMRG) simulations, we show that for a spatially antisymmetric cavity field, the uniform QHF state is stable only for weak light-matter coupling and gives way to states of inhomogeneous electron density above a critical coupling. These states involve ``flanks'' of uniform QHF fluids, separated by a ``compact core'' of doubly occupied orbitals with the core size being the order parameter, which we dub as ``compact-core phases". 
While the fully spin polarized electronic states are product states, entanglement builds up between the uniform QHF flanks across the compact core in the $S_z=0$ magnetic sector, motivating an ansatz for the compact-core electronic states. The transition boundary is exactly derived for product electronic states in terms of matter and cavity parameters, and numerically confirmed by DMRG. We also study the thin-cylinder limit near the critical point where quantum fluctuations are enhanced, and the many-body excited states in both phases by focusing on the entanglement spectrum degeneracies. Remarkably, the photon number is found to probe the order parameter of the transition, providing a possible experimental signature of the electronic transition and the compact-core states. Our study offers a rare example of a phase of electrons stabilized solely by coupling to the enhanced vacuum fluctuations of a cavity mode.

\end{abstract}

\maketitle

\section{Introduction}

The emerging field of quantum cavity material seeks to modify electronic many-body systems by coupling them to the enhanced vacuum fluctuations of quantized electromagnetic modes in a cavity \cite{garcia2021manipulating,Latini2021,Bloch2022,Schlawin2022,LuEtAl2025AOPA}.
Among the most fascinating possibilities being explored in this area is the modification of phases and phase boundaries of electron systems by coupling them to \textit{undriven} vacuum fluctuations of photon fields \cite{PhysRevB.84.195413,bartolo2018vacuum,Sentef2018,CurtisEtAl2019PRL,
PhysRevB.99.235156,
Li_2020,Ashida_2020,
Chiocchetta_2021,ThomasEtAl2021NL,
PhysRevB.106.205114,
Bostrom_2023,weber2023cavity,Arwas_2023,nguyen2023electronphoton,PhysRevLett.132.166901,Dag2024,PhysRevB.109.195173,pxkp-trx5,Thomas_2025,KerenEtAl2026N,ghorashi2025tunabletopologicalphasesmultilayer,karle2025hybridlightmatterboundariesgraphene,mochida2026cavitycontrolquantumphase,yang_2026,montanaro2026cavityenhancedsuperconductingresponseunderdoped}. Increasingly,  quantum Hall systems are being explored as candidates for such phenomena. Quantum Hall effects occur in two-dimensional electron systems in a strong perpendicular magnetic field at low temperature, and encompass a broad variety of states \cite{DasSarma_1997,Jain_2007} that may be accessed by varying the filling factor $\nu$, defined as the ratio of electron density to the magnetic flux density passing through the system.  For many of these states, the Hall conductance in the low-temperature limit is quantized at integer or rational fractional fractions of $e^2/h$, indicating a deep connection between this quantity and a topological index -- the Chern number -- associated with the state \cite{Thouless_1982,Girvin_Yang_2019}.  The effect of coupling this system to one or more cavity modes on the quantization of $\sigma_{xy}$ remains a subject of theoretical debate \cite{CiutiHopping,Appugliese_2022,Rokaj_2022,rokaj2023topological,RokajPRX,enkner2024enhancedfractional,Bacciconi_2025,Winter_2025,cardosoCavityQuantumHall2026,Yang_2026_QHE,boriçi2026compositefermionstudycavitymodifiedfractional}.  In particular, it is known that if the electric field of the cavity mode is constant in the plane of a translationally uniform electron gas, the cavity only couples to the center-of-mass (Kohn) mode of the quantum Hall system, without affecting any of the interaction physics~\cite{rokaj2023topological}. By contrast, a cavity mode field that varies spatially {\it can} couple to the internal dynamics of the electrons in this type of system~\cite{Bacciconi_2025}. 

Experimental studies have realized such spatially non-uniform cavity fields, with  results that suggest the impact of a cavity mode coupled to a  quantized Hall state is not universal, and depends on the particular integer or fraction involved: in some cases the degree of quantization appears weakened at finite temperature \cite{Appugliese_2022}, while in others it is enhanced \cite{enkner2024enhancedfractional}.  Moreover, coupling to a cavity mode can impact the electronic state at fillings between quantum Hall states \cite{Graziotto_2026}.

\begin{figure*}[t]
 \centering
\includegraphics[width=1.9\columnwidth]{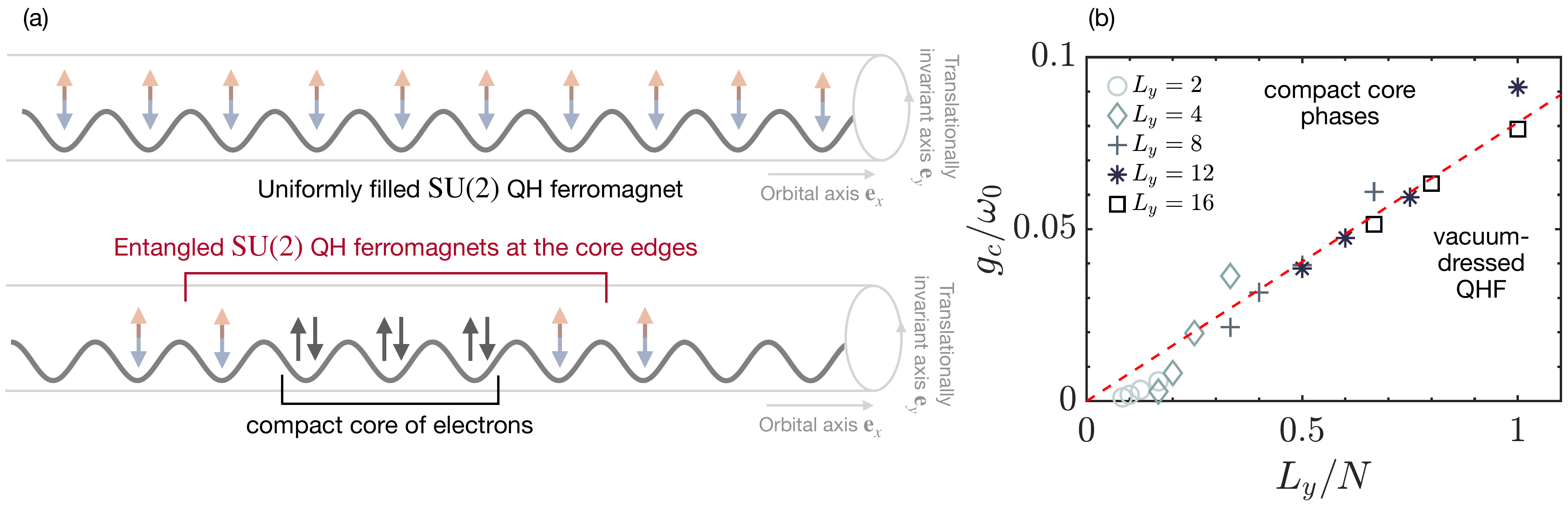}  
    \caption{(a) The schematics on the ground state phases of 2DEG on a cylinder embedded in a cavity with linear field profile. For smaller light-matter coupling than a critical coupling $g_c$, the matter is a vacuum-dressed Quantum Hall Ferromagnet (QHF), which is a uniform state with SU(2) symmetry. For larger light-matter couplings, a new phase of matter emerges with non-uniform states featuring a compact core of electrons occupying the core of 2DEG with entangled uniform QHF states at the core edges. (b) The phase diagram computed with density-matrix renormalization group with all data collapsing on a line in the plane of critical coupling normalized by cavity frequency $g_c/\omega_0$ and system size parameter $L_y/N$, where $L_y$ and $N$ are cylinder circumference and electron number, respectively.}   \label{fig:intro}
\end{figure*}

Quantum Hall ferromagnetism (QHF) occurs in systems in which the electrons have a discrete degree of freedom, the simplest example being the electron spin. In some cases, QHF may be central to the stability of a particular state.   In GaAs systems, for example, the Zeeman energy associated with spin is the smallest energy scale in the system. In fact, even when the  Zeeman energy vanishes, a system with odd integer $\nu$ -- for which a pair of degenerate Landau levels at the Fermi energy are only half-filled overall --
exhibits clear manifestations of spin polarization and a gap in the charged excitation spectrum \cite{Usher_1990,Zhitomirsky_2004,Lupatini_2020}.  This is the phenomenon of quantum Hall ferromagnetism, and may be understood as spontaneous symmetry breaking of the SU(2) spin-rotation symmetry due to interactions.  Interactions in a QHF also produce a local spin stiffness, leading to low-lying spin wave excitations, skyrmions, and other collective spin-related behavior \cite{Sondhi_1993,Fertig_1994,DasSarma_1997,Ezawa_2009}.  

In this work, we  investigate the following questions: Can enhanced vacuum fluctuations of a photon mode coupled to the electrons qualitatively change the state of a quantum Hall ferromagnet? What is the impact of a coupled cavity mode on low-lying excited states of the electronic system? Recent experiments \cite{Appugliese_2022} on such systems in fact exhibit differences in transport for quantum Hall systems at even and odd integer fillings which are impacted by coupling to a cavity, suggesting that spin indeed plays a non-trivial role. Hence, the focus of our study is on one of the simplest realizations of QHF, the $\nu=1$ quantum Hall system. In particular we evaluate the impact of coupling to a non-uniform cavity mode, focusing on the strong magnetic field limit and projecting the state to the lowest Landau level (LLL). The top panel in Fig.~\ref{fig:intro}(a) shows a schematic of a uniform QHF state --a fully filled Landau level, realizing an SU(2)-symmetric ferromagnet.  

Recalling that a non-trivial coupling of internal degrees of freedom of the quantum Hall system to the cavity mode becomes possible only when the electric field of the latter is spatially-varying, we consider the simplest realization of this: a field that varies linearly with position.  We find that, indeed, 
for sufficiently strong coupling, the ground state reorganizes dramatically from the uniform ferromagnetic state.  The change in behavior is signaled by an instability in the spin wave spectrum, which we show  may be computed exactly may be computed exactly within the LLL approximation.  The results are qualitatively the same for both short-range electron-electron interactions and for Coulomb interactions. To investigate the new ground states and the nature of the transition from the uniform QHF state, we undertake density-matrix renormalization group (DMRG) studies for the simplest non-trivial case of a spatially antisymmetric, linearly varying cavity potential. The results indicate that when an instability sets in,  charge is transferred away from regions of high cavity electric field to regions where its magnitude is low.  This results in a state with a ``compact core" of doubly-occupied orbitals, singly-filled orbitals on its flanks that are essentially uniform QHF states, and unoccupied electron orbitals at the physical system edges. 
We show that such
\textit{`flank-core-flank'} states may host entanglement across the compact core. This phase is schematically sketched in the bottom panel of Fig.~\ref{fig:intro}(a).

To implement the LLL approximation in a consistent way, we work in dipole gauge.  Importantly, the dipole gauge photon degree of freedom does not by itself correspond to the physical photon, which involves an admixture of electron degrees of freedom.
Remarkably, neither the uniform QHF nor the compact-core states develop entanglement between the electrons and photons in dipole gauge, as we show using DMRG.  The instability of the uniform QHF state is instead largely induced by an effective single particle potential that emerges when the Hamiltonian is properly projected into the LLL. Motivated by its relative simplicity, we undertake mean-field studies of the ground state, and demonstrate the compact core nucleates in a continuous phase transition for large system sizes, provided the thermodynamic limit is taken in a way that preserves the cavity field profile across the sample.

Extensive DMRG simulations for various system sizes then demonstrate that, when the electric field of the cavity mode is antisymmetric [$\vec{E}(x)=-\vec{E}(-x)$], the critical light-matter coupling  $g_c \propto L_y/N$, where $L_y$ is the electron gas dimension perpendicular to the electric field direction and $N$ is the electron number, provided the system is in the large circumference limit $L_y \gg \ell$, with $\ell$ being the magnetic length. This result is highlighted in Fig.~\ref{fig:intro}(b). We find that both a spin wave and a compact core Hartree-Fock mean-field theory (MFT) analyses lead to a simple expression for the phase boundary that is consistent with the linear relation in Fig.~\ref{fig:intro}(b), 
\begin{equation} 
g_c=\frac{1}{2\pi\ell }\sqrt{\frac{v^{\rm ee}_0 \omega_0}{\pi}} \frac{L_y}{N}, \qquad \hbar=1.
\label{eq:SW_instability_g}
\end{equation} 
Here $v^{\rm ee}_0$ is the electron-electron interaction strength, which is assumed to be a contact interaction, and $\omega_0$ is the cavity frequency.  
An order parameter characterizing the compact-core phase is simply found to be the core size normalized by the electron number. 

We extensively examine the properties of the various quantum phases not only in terms of their electronic density and magnetization profiles, but also their entanglement spectra with respect to an entanglement cut in the center of the system. In particular, we present an analytical expression which provides a signature of the entanglement spectrum for a uniform QHF state with zero total magnetization, regardless of the coupling strength to the cavity mode. Moreover, we demonstrate that the first neutral many-body excitation of such a QHF state is doubly-degenerate throughout its entanglement spectrum, and show that this must be the case due to its symmetry properties under inversion.
Finally, we show that in the thin cylinder limit, $L_y \sim \ell$,  the system exhibits enhanced quantum fluctuations, particularly at the breakdown of the uniform QHF state.

At larger coupling, the compact core states, while not entangled with the photon mode in dipole gauge, nevertheless impact the photon state. This is realized as a suppressed physical photon number compared to the one in the uniform QHF phase, and an accompanying change in cavity energy which emerges with the more complicated electron ordering.  The change in the physical photon state exemplifies one of the most interesting behaviors of a quantum cavity  material~\cite{kass2026polarizationresolvedphotonstatisticscavity}: a smoking gun signature in the photon state of the transition from the uniform QHF to a compact-core phase. We analytically show that physical photon number tracks the order parameter of the electronic transition in the thermodynamic limit.

Our study proposes a novel quantum cavity material, in which a uniform quantum Hall ferromagnet state fragments into multiple such states across a compact core, under sufficiently strong coupling to a single cavity mode with an appropriate field configuration. Our presentation is organized as follows.  We first set the notation by reviewing some relevant facts about quantum Hall ferromagnetism, and then derive the effective Hamiltonian for this cavity quantum material in Sec.~\ref{sec:QHFMain}. We then start the discussion with the system's phase diagram based on the DMRG results in Sec.~\ref{sec:PhaseDiagram}, which motivates a spin wave analysis of the uniform QHF state in Sec.~\ref{sec:QHFSec-SW}.  Sec. \ref{sec:CompactCoreSec} provides further details of the compact core states, including a mean-field analysis of the states, an analysis of their entanglement structure, and a discussion of the thin-cylinder limit, in which quantum fluctuations of the system are enhanced.  We conclude with a summary and discussion in Sec.~\ref{sec:Conclusion}.

\section{Quantum Hall Ferromagnet in a Cavity: Hamiltonian \label{sec:QHFMain}}

The system we focus upon is a two dimensional electron gas (2DEG) in a strong magnetic field, at filling factor $\nu=1$, with a spin degree of freedom for which the Zeeman coupling is negligibly small.  The system is immersed in a cavity hosting a single mode with frequency $\omega_0$ which may have a non-uniform electric field profile.  In first quantization we adopt the Hamiltonian ($\hbar=1$)
\begin{eqnarray}
H &=& 
\frac{1}{2m_e}\sum_{j=1}^{N}\left[\frac{1}{i} \bm{\nabla}_j+e {\mathbf{A}}_B\left({\mathbf{r}}_j\right)+ e{\mathbf{\hat A}}_c\left({\mathbf{r}}_j\right)\right]^2 \notag \\
&+&H_{\rm int}^{\rm ee}+\omega_0\left(\hat a^\dagger \hat a+\frac{1}{2}\right),
\label{eq:H_Gen}
\end{eqnarray}
where $m_e$ is the effective (band) mass of the electron, $H_{\rm int}^{\rm ee}$ represents the electron-electron interaction, $\mathbf{A}_B \equiv -B(0,x,0)$ is the vector potential for a uniform magnetic field $-B\hat{z}$ in Landau gauge, and $\omega_0$ is the cavity frequency.  Note that each electron carries a spin-$1/2$ degree of freedom that does not enter $H$ in an explicit way, as we assume the Zeeman coupling in the system is negligibly small.  In addition to the electrons, there is a cavity mode with photons annihilated by the operator $\hat a$. The vector potential of that mode, $\mathbf{\hat A}_c$, couples it to the electrons through the kinetic energy operator,   and has the explicit form
\begin{eqnarray}
   \hat{\bm A_c} &=&  A_0\  u(\textbf{r}_j)\left[ \textbf{e}^* {\hat a}^{\dagger}+\textbf{e} {\hat a}\right],
\end{eqnarray}
where the mode strength is given by 
$A_0 =\left(2\epsilon_0\mathcal{V}\omega_0\right)^{-1/2}$ with the cavity mode volume $\mathcal{V}$, $\epsilon_0$ is the vacuum electric permittivity, $\textbf{e}$ is the polarization vector of the cavity mode, and $u(\textbf{r}_j)$ is an electric field profile, normalized to an effective area $S$ of the cavity in the plane of the 2DEG \cite{Todorov_2012,Bacciconi_2025},
$$\int d^2r\ u^2({\bf r}) = S.$$
It is important to recognize that while the electric field is present in a three-dimensional volume defined by the cavity geometry, the 2DEG may be present only in some fraction of that volume, due to its two-dimensional nature and its finite lateral size.  For simplicity, in what follows we consider a field that varies no faster than linearly within the region of the 2DEG, and has a constant direction \cite{Bacciconi_2025}. Outside the 2DEG the field profile will be more complicated: in particular, it will necessarily have a component directed perpendicular to the 2DEG which does not impact the Hamiltonian of the system. 

Our focus is to understand how the low-energy collective modes of the $\nu=1$ quantum Hall system are modified by the cavity field, as well as whether the field impacts the ground state, and if so, how. In the absence of the cavity, and in the strong magnetic field limit, this state is expected to be a spin-polarized, uniformly occupied Landau level, which we refer to as a uniform quantum Hall ferromagnet state \cite{Girvin_MacDonald_1997}. We thus simplify the problem by taking the advantage of projection into the lowest Landau level. In the following subsection, we briefly review this projection in the absence of cavity coupling, and set  notation for the remainder of the paper. 

\subsection{Quantum Hall Ferromagnetism}

Our choice of Landau gauge preserves the translational symmetry in the $y-$direction, so that single particle states have a momentum $\hbar k_y$. Adopting periodic boundary conditions along $\hat{y}$ implies that the sample is topologically equivalent to a cylinder, with a cross-section of circumference $L_y$, with quantized wavevectors $k_y$.  The lowest energy single-particle eigenstates of the electronic kinetic energy operator (without coupling to the cavity), $\frac{1}{2m_e}\left[\frac{1}{i}{\bm \nabla} + e\bm{A}_B(\bm{r}) \right]^2$, 
can be written in terms of  uniformly spaced guiding centers, with values $X_n \equiv \ell^2 k_y=\frac{\ell^2}{L_y}2\pi n$,
in the form
\begin{equation}\label{Eqn:LLLwavefunction}
    \phi_{X_n}(\br) = \frac{1}{\pi^{1/4}\sqrt{L_y \ell}} \exp\Big( iyX_n/\ell^2 - \frac{1}{2\ell^2}(x - X_n)^2 \Big), 
\end{equation}
where $\ell \equiv \sqrt{1 /eB}$ is the magnetic length, and the static magnetic field is assumed to be in the $-z$ direction. The wavefunctions above form a complete basis for the lowest Landau level (LLL).   The total number $N$ of allowed values of $X_n$ sets the orbital degeneracy of the Landau level, and is determined by the size of the system along the $x$-direction, $L_x$.  Allowing only values of $X_n$ such that $|X_n| \le L_x/2$ implies a total orbital degeneracy of $N_{\phi} = L_xL_y/2\pi \ell^2$, in which $N_{\phi}$ is also the number of magnetic flux quanta penetrating the area of the electron system \cite{DasSarma_1997,Jain_2007,Girvin_Yang_2019}.  For $\nu=1$ states, $N=N_{\phi}$.
To preserve the $X \to -X$ symmetry of the electron system, we assume $n\in\mathbb{Z}$ for an odd number of orbitals $N_{\phi}$ and $n\in\mathbb{Z}+\frac{1}{2}$ for even $N_{\phi}$. 
Note that,
within the LLL, the electronic Hilbert space naturally has a quasi-1D character, in which the guiding centers $X_n$ may be regarded as ``sites'' in a chain
with nearest neighbor separation
$\Delta X = 2\pi \ell^2/L_y$.
We emphasize that, because of the spin degree of freedom, each such site can accommodate up to two electrons.

We next turn to the electron-electron interaction.  The energy spectrum of the cavity-free kinetic energy operator has a harmonic oscillator form, with characteristic frequency $\omega_c=eB/m_e$ (the cyclotron frequency), and Landau level energies $(m+{1 \over 2})\omega_c$, where the integer $m=0,1\dots$ labels the Landau level.  In the strong field limit (large $\omega_c$), one expects only the lowest energy states in kinetic energy spectrum ($m=0$) to be relevant.
Projecting into the LLL yields
\begin{eqnarray}\label{Eqn:Vpairpair}
    \bar{H}_{\rm int}^{\rm ee} &\equiv& P_{\rm LLL} H_{\rm int}^{\rm ee}P_{\rm LLL} \nonumber \\
    &=& \frac{1}{2}\sum_{\br\br'} \hat{c}_{\alpha}^{\dg}(\br)\hat{c}_{\beta}^{\dg}(\br') V(\br-\br') \hat{c}_{\beta}(\br')\hat{c}_{\alpha}(\br),
\end{eqnarray}
where $P_{\rm LLL}$ is the LLL projection operator,
$V(\br-\br')$ is a pair interaction, and $ \cP_{\alpha}(\br) = \sum_{n} \phi_{X_n}(\br) d_{\alpha,n}$ is the LLL-projected electron field operator, with $X_n = n\Delta X$ and $\alpha$ denoting spin. The LLL operator $d_{\alpha,n}$ annihilates an electron with spin $\alpha\in\{\uarr,\darr\}$ and momentum $k_y=-\frac{2\pi n}{L_y}$. When LLL-projected electron operators are substituted, $ \bar{H}_{\rm int}^{\rm ee}$ can be reorganized into
\begin{equation}\label{Eqn:H_halfQH}
    \bar{H}_{\rm int}^{\rm ee} = \sum_{n,k,m} \frac{V_{km}}{2} \sum_{\alpha,\beta}d^{\dg}_{\alpha,n+k}d^{\dg}_{\beta,n+m}d_{\beta,n+m+k}d_{\alpha,n} ,
\end{equation}
where $n \in \mathbb{Z}$ ($n \in \mathbb{Z}+\frac{1}{2}$) if the number of sites is odd (even), and $k,m\in\mathbb{Z}$. The matrix elements for a general projected two-body interaction are (Appendix \ref{appSec:contact}) 
\begin{eqnarray}\label{Eqn:Vkm}
    V_{km} &=&  \frac{1}{\sqrt{2\pi} L_y\ell} \int_{-\infty}^{\infty} d\tilde{x} \int_{-L_y/2}^{L_y/2} d\tilde{y}\; \\
    && \times V(\tilde{x},\tilde{y}) e^{-i\tilde{y}X_k/\ell^2} e^{-\frac{1}{2\ell^2}(\tilde{x}-X_m)^2} e^{-{X_k^2}/{2\ell^2}} \notag 
\end{eqnarray}
with $(\tilde{x},\tilde{y})\equiv(\br-\br')$.  Note in writing this we have assumed $L_x \gg \ell$ and ignored edge effects.

We begin by considering short-range interactions, which can be interpreted qualitatively as  screened Coulomb interactions.  This dictates the form of $V_{km}$. In the simplest case one may assume a screening length much smaller than the magnetic length $\ell$, leading to a contact interaction, 
\begin{equation}
V_{\delta}(\tilde{x},\tilde{y})=v_0^{\rm ee}\ell^2\delta(\tilde{x})\delta(\tilde{y}),
\label{eq:V_delta}
\end{equation} with the intrinsic interaction strength $v_0^{\rm ee}$ which we can adopt as our unit of energy.
In this case the interaction strength, Eq.~\eqref{Eqn:Vkm}, becomes \cite{Eugenio_2020}
\begin{eqnarray}\label{Eqn:Vkm_contact}
  {V}^{\delta}_{km} &=&  \frac{\ell}{\sqrt{2\pi} L_y } e^{ -\left(X_k^2+X_m^2\right)/2\ell^2 } .
\end{eqnarray}

In the absence of the cavity, the only active element of the Hamiltonian is $\bar{H}^{\rm ee}_{\rm int}$, 
Eq.~\eqref{Eqn:H_halfQH}. For Coulomb and short range interactions, the ground state of the Hamiltonian is ferromagnetic \cite{Sondhi_1993, Yang_1994,MacDonald_1996}, and, in the absence of Zeeman coupling, is $(N+1)$-fold degenerate.  Importantly, because of the SU(2) spin symmetry of the interaction, this collection of states represents an \textit{exact} ground state manifold \cite{MacDonald_1996}.  Moreover, low-energy excited states can also be constructed exactly in terms of single spin-flip excitations with non-vanishing total momentum.  In what follows, we will exploit these properties of the low-energy electron Hamiltonian to construct a low-energy Hamiltonian for the system coupled to the cavity mode, valid when the ground state is fully ferromagnetic.  The results of this analysis indicate that 
this ground state becomes unstable towards states of non-uniform electron density, when the light-matter coupling exceeds a critical value.  We then explore the new ground states and their low-lying excitations using finite size systems, with both analytic mean-field approaches and DMRG.

\subsection{Coupling to the Cavity}

We consider the coupling of the quantum Hall system to a cavity mode.  For concreteness, we assume the cavity field is polarized in the $x$-direction with a vector potential \cite{Todorov_2012} 
\begin{eqnarray}
   \hat{\bm A_c}(\bm r) &=&  A_0 \left(\hat a^\dag+\hat a\right)u(x)\ \textbf{e}_x,
\end{eqnarray}
that varies spatially only in the $x-$direction, with $u(x)$ a dimensionless function of $x$. When the field is uniform, $u(x)=1$; moreover, we will
consider linearly varying cavity fields of the form $$u(x) = \varphi^{(1)}+2\varphi^{(2)}x.$$ Note that $\varphi^{(1)}$ is dimensionless, while $\varphi^{(2)}$ has units of inverse length. We further characterize the strength of the vector potential as it enters the Hamiltonian by a dimensionless parameter $\zeta$, defined by the relation 
\begin{equation}
eA_0=\frac{\zeta}{\ell}.
\end{equation}
In our analysis below we will in some cases retain the parameter $\varphi^{(2)}$ to allow direct examination of how the cavity field gradient impacts the system. However, the parameter set characterizing the cavity field can be simplified if one assumes the cross-sectional dimensions of the 2DEG, $L_x$ and $L_y$, are the same as those of the cavity, and moreover that the cavity field is spatially antisymmetric over the plane of the 2DEG, so that $\varphi^{(1)}=0$.  Under these assumptions, one finds (Appendix~\ref{sec:AppBSSec2})
\begin{equation}
\varphi^{(2)}=\frac{\sqrt{3}}{2\pi \ell^2}\frac{L_y}{N}.
\label{eq:varphi2value}
\end{equation}
Finally, for this cavity configuration, we shall see below that the energy scale for the effective light-matter interaction is given by
\begin{equation}
g=\zeta\varphi^{(2)}\ell\omega_0.
\label{eq:gdef1}
\end{equation}

\subsection{Projection to Lowest Landau Level}

We work in the limit where $\omega_c$ is much larger than any other energy scale in the problem, which motivates the projection of the Hamiltonian into the LLL. Note that we have already projected $H^{\rm ee}_{\rm int}$ into the LLL in Eq.~\eqref{Eqn:Vpairpair}.  We now turn to projecting the terms  involving the cavity degree of freedom that appear in the kinetic energy term of Eq.~\eqref{eq:H_Gen} as well.  While LLL projection is associated with high magnetic fields, it is important to note that the $B \to \infty$ limit also entails $\ell \to 0$, which becomes problematic  when there is both a cavity coupling and energy scales involving $\ell$ (e.g., Eqs.~\eqref{eq:V_delta},  \eqref{Eqn:Vkm_contact} and \eqref{eq:gdef1}). To avoid  these issues we will implement $\omega_c \to \infty$ by taking $m_e \to 0$ while keeping $B$, and thus $\ell$, fixed. For the kinetic energy term in  Eq.~\eqref{eq:H_Gen}, the $m_e\to0$ limit leads to a divergent coupling between the electron and photon degrees of freedom. To circumvent this issue, we transform to the dipole gauge, in which the electron-photon coupling is transferred to a part of the Hamiltonian independent of $m_e$. The transformation is
\begin{eqnarray}
\widetilde{H}&=&UHU^{\dagger},
\label{eq:dipole_gauge1}\\
U&=&e^{i\frac{\zeta}{\ell}\sum_{j}\varphi\left(x_j\right)\left(\hat a+\hat a^\dagger\right)},
\label{eq:dipole_gauge2}
\end{eqnarray}
where $\varphi(x_j)$ is given by the solution to $\frac{\zeta}{\ell}\bm{\nabla}_j\varphi\left(x_j\right)\left(\hat a+\hat a^\dag\right)\textbf{e}_x=\hat{\bm A_c}\left({\mathbf{r}}_j\right)$. Specifically, for field profiles that vary at most linearly with $x$, one may generally take 
\begin{equation}
\varphi(x) = \varphi^{(0)}+ \varphi^{(1)} x+ \varphi^{(2)}x^2.
\end{equation}
Note that the function $\varphi(x)$ has units of length. We will see below for our spin wave analysis performed on the quantum Hall ferromagnet, 
it is convenient to choose $\varphi^{(0)}$ such that $\sum_n\langle X_n |\varphi(x) | X_n\rangle=0$, where $\langle{\bf r}|X_n\rangle \equiv \phi_{X_n}({\bf r})$ is the lowest Landau level state, while for the DMRG analysis it is simplest to take $\varphi^{(0)}=0$.  Results are independent of the value chosen for $\varphi^{(0)}$, since only the gradient of $\varphi(x)$ has physical meaning.

In the dipole gauge, the Hamiltonian becomes
\begin{eqnarray}
\widetilde{H}&=&\frac{1}{2m_e}\sum_{j=1}^{N}\left[\frac{1}{i}{\bm{\nabla}}_j+ e{\mathbf{A}}_B\left({\mathbf{r}}_j\right)\right]^2+H_{\rm int}^{\rm ee} \notag \\
&+&\omega_0\left[\left(\hat a^\dag+i\Pi\right)\left(\hat a-i\Pi\right)+\frac{1}{2}\right].
\label{eq:tildeH}
\end{eqnarray}
As promised, the electron-photon coupling is no longer associated with $m_e$, allowing us to safely take the $m_e\to0$ limit. The dimensionless operator $\Pi$ is defined as
\begin{equation}
    \Pi\ \equiv\frac{\zeta}{\ell}\sum_{j}\varphi\left(x_j\right).
\end{equation} 
For the special case of a spatially constant cavity $E$ field, one can identify $\Pi$ with the dipole operator, but such a simple identification fails for spatially nonuniform fields. The total Hamiltonian can be more explicitly written as 
\begin{widetext}
\begin{eqnarray}
\widetilde{H}&=&\frac{1}{2m_e}\sum_{j=1}^{N}\left[\frac{1}{i}{\bm{\nabla}}_j+ e{\mathbf{A}}_B\left({\mathbf{r}}_j\right)\right]^2+H_{\rm int}^{\rm ee} + \omega_0 \left[\hat a^\dagger\hat a + \frac{1}{2} \right] 
+ i \omega_0  \frac{\zeta}{\ell} \left(\hat a - \hat a^{\dagger} \right)\sum_j \varphi\left(x_j\right) + \omega_0 \left(\frac{\zeta}{\ell}\right)^2 \bigg(\sum_j \varphi\left(x_j\right) \bigg)^2.\notag \\
\end{eqnarray}
\end{widetext}
The transformed Hamiltonian exhibits a dipole-like coupling of the cavity field to electronic degrees of freedom (hence the name dipole gauge) and cavity-mediated electron-electron interactions. The LLL projection operator $P_{\rm LLL}$ may now be applied, 
\begin{widetext}
\begin{eqnarray}
\label{eq:Hbar}
\bar{H} &\equiv& P_{\rm LLL}
\left[
\widetilde{H} - {N \over 2}\omega_c \right] P_{\rm LLL}=P_{\rm LLL}\left\{H_{\rm int}^{\rm ee}+\omega_0\left[\left(\hat a^\dag +i\Pi\right)\left(\hat a-i\Pi\right)+\frac{1}{2}\right]\right\}P_{\rm LLL},\nonumber\\
&=&P_{\rm LLL} H_{\rm int}^{\rm ee} P_{\rm LLL} 
+ i\omega_0\left(\hat a - \hat a^{\dagger} \right) P_{\rm LLL} \Pi P_{\rm LLL} 
+  \omega_0 P_{\rm LLL} \Pi^2 P_{\rm LLL}+ \omega_0\left(\hat a^{\dag} \hat a+\frac{1}{2}\right) ,
\end{eqnarray}
\end{widetext}
where we have removed the trivial electron kinetic energy in the LLL, ${N \over 2} \omega_c$.
We next need to consider the operators $\bar{\Pi} \equiv P_{\rm LLL} \Pi P_{\rm LLL}$, $\overline{\Pi^2} \equiv P_{\rm LLL} \Pi^2 P_{\rm LLL}$, and $\bar{H}_{\rm int}^{\rm ee} \equiv P_{\rm LLL} {H}_{\rm int}^{\rm ee} P_{\rm LLL}$. The explicit form for the last of these was discussed in the previous subsection. In second quantization, the first of these operators take the form
\begin{equation}
\bar\Pi = \frac{\zeta}{\ell}\sum_{\alpha,n} \varphi_nd_{\alpha,n}^\dagger d_{\alpha,n}.
\label{eq:pibardef}
\end{equation}
where $\varphi_n=\langle X_n | \varphi(x) | X_n\rangle$ is the expectation value with respect to a single particle state of the form in Eq.~\eqref{Eqn:LLLwavefunction}, with $X_n=2\pi n \ell^2/L_y$. 
It is convenient to write the operator $\overline{\Pi^2}$ in terms of the operator $\bar\Pi$, and doing so requires some care.   One may show
\begin{eqnarray}
  \overline{\Pi^2} &=& \left(\bar\Pi\right)^2 +\left(\frac{\zeta}{\ell  }\right)^2\sum_{n}\sum_{\alpha}
\delta\varphi^2_nd^\dag_{\alpha,n}d_{\alpha,n},
\label{eq:Pisq_main}
\end{eqnarray}
where
$\delta\varphi^2_n \equiv \langle X_n |\varphi^2(x)| X_n \rangle - \varphi_n^2$; details leading to Eq.~\eqref{eq:Pisq_main} may be found in Appendix~\ref{sec:SW_Details}. The last term in this equation  is an effective static potential for the 2DEG, that emerges from the combined effect of transforming to dipole gauge and LLL projection. Up to an overall constant, the total projected Hamiltonian is 
\begin{equation}
\bar{H} = \bar{H}_{\rm int}^{\rm ee}
+ V_\varphi+ \omega_0 \left[  i\left(\hat a - \hat a^{\dagger} \right)\bar{\Pi} + \left( \bar{\Pi} \right)^2
+\hat a^{\dag}\hat a+\frac{1}{2}
\right],\label{eq:FullHam1}
\end{equation}
where 
\begin{equation}
V_\varphi=\omega_0\left(\frac{\zeta}{\ell}\right)^2\sum_{n}\sum_{\alpha}
\delta\varphi^2_n\ d^\dag_{\alpha,n}d_{\alpha,n},
\label{eq:V_varphi}
\end{equation}
is the effective static potential. 
It is important to note that, in this dipole gauge, the bosonic operator $\hat a^\dagger$ does not by itself create physical cavity photons: the appropriate operator for this is $U\hat a^\dagger U^\dagger=\hat a^\dagger + i\Pi$ (see Appendix~\ref{sec:appD}).  We will nevertheless see that there is great simplification in expressing the states of the full system in terms of the photon states upon which  $\hat a^\dagger$ acts, and states of the electrons in the LLL.
Having obtained the LLL projected Hamiltonian in dipole gauge, we next proceed to examine its eigenstates. 

\section{Cavity altered ground state phase diagram of 2DEG with fluctuating spins \label{sec:PhaseDiagram}}

\begin{figure*}[t]
 \centering
\includegraphics[width=1.9\columnwidth]{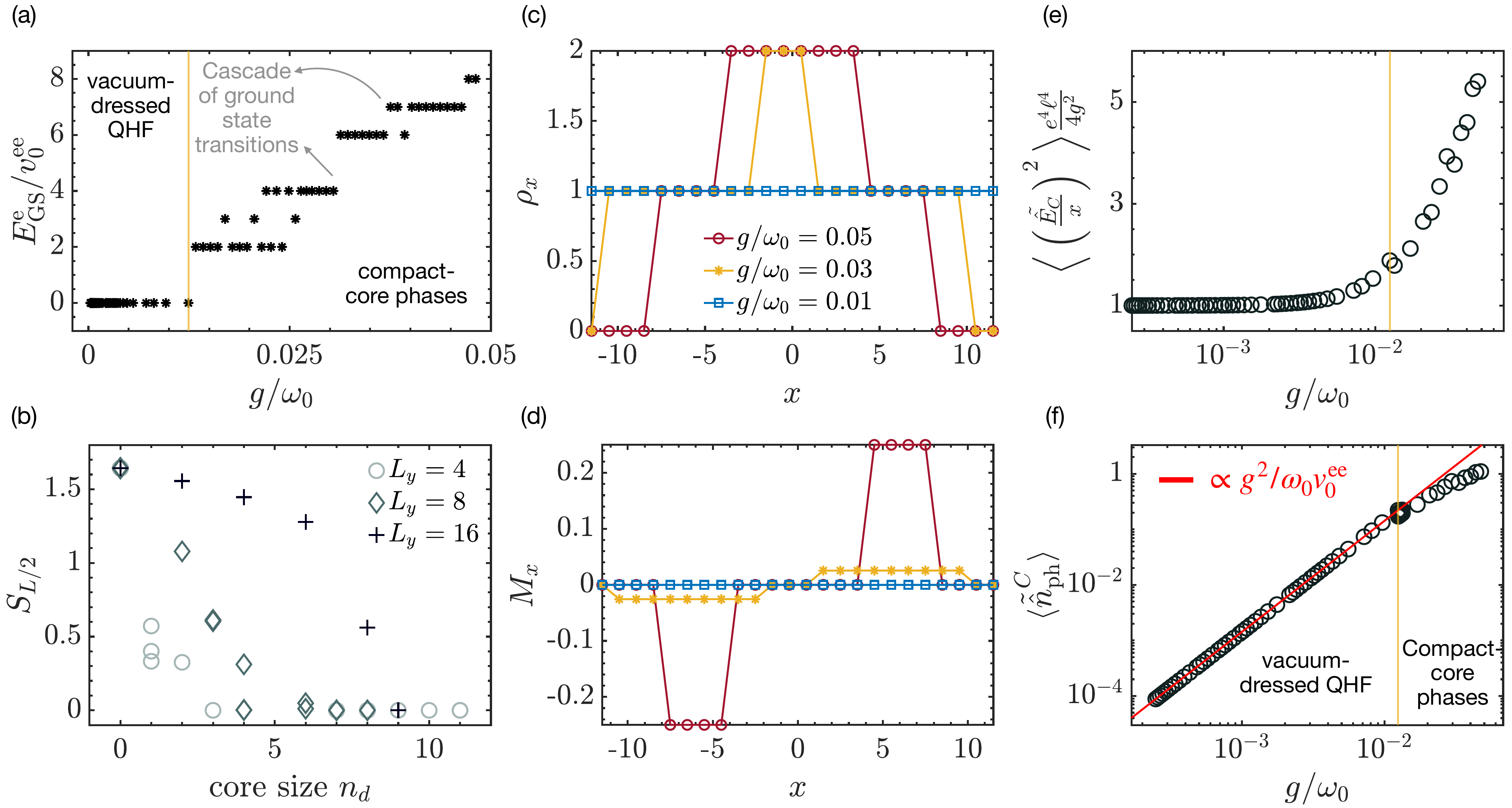}  
    \caption{(a) Electronic interaction energy $E^{\rm e}_{\rm GS}=\langle \Psi_{\rm gs} \vert \bar{H}_{\rm int}^{\rm ee} \vert \Psi_{\rm gs} \rangle$ with respect to light-matter coupling strength normalized by cavity frequency $g/\omega_0$. (b) Half-chain von Neumann entanglement entropy with respect to compact-core size $n_d$ for different cylinder circumferences at $N=24$. The core size $n_d=0$ corresponds to the vacuum-dressed QHF. As $L_y$ increases, more compact-core states with nonzero entanglement entropy occur. (c) Electronic density and (d) magnetization with respect to orbital position for different $g/\omega_0$, showing the vacuum-dressed QHF (uniform-density and magnetization) and two compact-core phases (non-uniform density and magnetization with a compact core). Subfigures (a,c,d) are for $N=24$ orbitals (electrons) and $L_y=8$. (e) Electric field fluctuations and (f) the physical photon number with respect to $g/\omega_0$. The red line corresponds to vacuum field fluctuations in (e), and $g^2/\omega_0^2$ in (f). The orange vertical line in (a,e,f) shows the transition line between the QHF and the compact-core phases.}   \label{fig:Eenergy-EE-density}
\end{figure*}

We begin by considering states of the system for 2DEGs of finite size.  In these calculations, we employ DMRG to investigate the many-body ground state of the Hamiltonian, Eq.~\eqref{eq:FullHam1}, and
focus on antisymmetric fields
($\varphi^{(1)}=0$), i.e.,~linearly varying cavity fields, in the geometry leading to Eq.~\eqref{eq:varphi2value} (see Appendix \ref{sec:AppBSSec2}). 
Assuming a contact interaction Eq.~\eqref{eq:V_delta},
the Hamiltonian takes the form (up to a constant) 
\begin{widetext}
    \begin{eqnarray}
    \bar H &=& 
\sum_{\substack{\alpha,\beta \\n,k,m}} \frac{V_{km}}{2}\ d^{\dg}_{\alpha,n+k}d^{\dg}_{\beta,n+m}d_{\beta,n+m+k}d_{\alpha,n} + \omega_0  \left(a^{\dagger} \hat a + \frac{1}{2}\right) + i\ g\ (\hat a-\hat a^{\dagger})  \sum_{\alpha,n}  \tilde{\varphi}_n \ d^{\dg}_{\alpha,n}d_{\alpha,n} \nonumber \\
&+ &\frac{g^2}{\omega_0} \bigg\lbrace \sum_{\alpha,\beta}\sum_{n,m} \tilde{\varphi}_n   \tilde{\varphi}_m \ d^{\dg}_{\alpha,n}d_{\alpha,n} d^{\dg}_{\beta,m}d_{\beta,m} + 2  \sum_{\alpha,n}\tilde{\varphi}_n \ d^{\dg}_{\alpha,n}d_{\alpha,n} 
\bigg\rbrace, 
\label{eq:DMRG_Ham}
\end{eqnarray}
\end{widetext}
where $\tilde\varphi_n=X_n^2/\ell^2+{1 \over 2}$. 
Eq.~\eqref{eq:gdef1} defines $g$ in this Hamiltonian. We set the effective cavity mode volume in $g$ with the 2DEG size $\mathcal{V} = \chi L_y L_x=\chi N 2\pi \ell^2$,
where $\chi$ is light confinement parameter on the plane~\cite{paravicini2019magneto}. For the DMRG, we proceed by fixing the magnetic length $\ell=1$ as our unit of length, and the cavity frequency to $\omega_0=10$ in units of $v_0^{\rm ee}/2\pi$. 
Our focus is on identifying the phases of this hybrid light-matter system with respect to $g$, as the total orbital number $N$ and cylinder circumference $L_y$ are varied. Although we present results for these specific parameters, we find that the qualitative nature of the phases, as well as the transitions between them, are independent of the parameter choice. 

The Hamiltonian is encoded as a matrix product operator on a mixed tensor of spinful fermions and a single bosonic mode coupled to every fermionic orbital site.  We note that due to the nature of the nonlocal photon-matter coupling, more DMRG sweeps than typical are needed for satisfactory convergence. 
More details on the simulations and how we achieve convergence can be found in Appendix~\ref{sec:appC}. 

Our first key finding is that the many-body ground state is always a tensor product of the 2DEG and the (dipole gauge) photon degrees of freedom, $\vert \Psi_{\rm gs} \rangle = \vert \psi_{\rm e} \rangle \otimes \vert \psi_{\rm p} \rangle $, regardless of the light-matter interaction strength. Hence, no entanglement builds up between the 2DEG and  photons, as quantified by the entanglement entropy $S_e = \textrm{Tr}\left[\rho_{e}\log \rho_{e}\right]=0$, where $\rho_{e}$ is the electronic state operator, $\rho_{e} = \textrm{Tr}_{p} \vert \Psi_{\rm gs}  \rangle \langle \Psi_{\rm gs}  \vert $, with $\textrm{Tr}_p$ denoting a trace over the photon states. However, their interaction can still alter the properties of the 2DEG and the cavity. For example, the electron system drives the photon mode into a coherent state 
(Appendix~\ref{sec:AppE}), and the uniform QHF ground state one finds in the absence of the cavity mode breaks down for sufficiently large light-matter coupling $g$, such that electronic system forms  ``compact-core phases" Fig.~\ref{fig:intro}(a), as we now explain. 

For $g < g_c$, where $g_c$ is a critical light-matter coupling, the uniform QHF phase is a stable phase of the 2DEG. In this state, for contact interactions, the electronic interaction energy of the ground state $\vert \Psi_{\rm gs} \rangle$, given by $E^{\rm e}_{\rm GS}=\langle \Psi_{\rm gs} \vert \bar{H}_{\rm int}^{\rm ee} \vert\Psi_{\rm gs}\rangle$, precisely vanishes, as illustrated in Fig.~\ref{fig:Eenergy-EE-density}(a). Because this is a state of a Hamiltonian with SU(2) spin symmetry, the $S_z=0$ state sector for even $N$ -- the focus of our DMRG studies -- may be arrived at starting from a fully spin-polarized state and acting upon it with the total spin lowering operator $N/2$ times \cite{MacDonald_1996}. 
The resulting ground state is an equal superposition of all states with a single electron in each orbital $X_n$, with half the electrons having each of the two spin values.  This results in a finite bi-partite entanglement entropy (Fig.~\ref{fig:Eenergy-EE-density}(b) with $n_d=0$), $S_{L/2}=\textrm{Tr}\left[\rho_{L/2}\log \rho_{L/2}\right]$, where $\rho_{L/2} = \textrm{Tr}_{\overline{L/2}} \vert \psi_{\rm e} \rangle \langle \psi_{\rm e}\vert $ is the reduced density matrix after tracing over the half the electron system, as well as the bosonic degree of freedom.  The electrons in this state have uniform density (Fig.~\ref{fig:Eenergy-EE-density}(c), blue-squares), and, moreover, a vanishing magnetization density [Fig.~\ref{fig:Eenergy-EE-density}(d)], since the DMRG is restricted to the $S_z=0$ sector. 
The connection between this state and the fully polarized uniform QHF state motivates our analytical spin wave analysis for the hybrid system in Sec.~\ref{sec:QHFSec-SW}. 

For $g > g_c$, the uniform QHF state breaks down, and transitions into a cascade of compact core phases as $g$ increases. An important property of these phases is that their electron density is non-uniform,  hosting a compact core of doubly occupied states,
as illustrated in Fig.~\ref{fig:Eenergy-EE-density}(c). We denote the core size by $n_d$, the number of doubly occupied orbitals $X_n$. 
At the breakdown of the uniform QHF state, i.e.,~the first transition, $n_d$ is small, and the compact core is flanked by singly occupied orbitals on either side, as illustrated qualitatively in Fig.~\ref{fig:intro}(a). The spinful regions in the two flanks can be entangled, as seen in Fig.~\ref{fig:Eenergy-EE-density}(b). Interestingly, the edges of the compact core are uniform [cf. Fig.~\ref{fig:Eenergy-EE-density}(c)] with a fixed magnetization density [Fig.~\ref{fig:Eenergy-EE-density}(d)]. It should be noted that the magnetization profile for a compact core state at a given $g$ such as those illustrated in Fig.~\ref{fig:Eenergy-EE-density}(d) is only one possibility among several at the same energy.  The degeneracies of such states is discussed below.
 Thus, the new phase of the electron system can be understood as spatially fragmented uniform QHF states, separated by a compact core, whose ramifications will be discussed in Sec.~\ref{sec:CompactCoreQHFFlanks}.

The electron-electron interaction energy of this state can be understood by finding the size of the compact core, Fig.~\ref{fig:Eenergy-EE-density}(a), because the uniform QHF flanks do not contribute to the electronic interaction energy within the uniform QHF flanks. 
As the light-matter coupling increases, the size of the compact core increases, as illustrated in Figs.~\ref{fig:Eenergy-EE-density}(a) and (c).  At large enough coupling this leads to a fully compact state, i.e.,~all orbitals $X_n$ doubly-occupied or empty. This final state is a product of doubly occupied $X_n$ orbitals, and hence has bipartite entanglement entropy $S_{L/2}=0$. We also observe that  this entanglement monotonically drops to zero with increasing core size, as illustrated in Fig.~\ref{fig:Eenergy-EE-density}(b). Note that for a given core size $n_d$, the state tends to host more entanglement across the core as the cylinder circumference $L_y$ increases. This may be understood as resulting from the decrease in system width $L_x=2\pi \ell^2 N/L_y$ and accompanying smaller geometric compact core width $n_d\ell$ when $N$ is fixed, so that the two flanks are closer together. 
These results motivate our analytical mean-field analysis on the compact core phases in Sec.~\ref{sec:MFTCompactCore}.

By sweeping through $L_y$ and $N$, we observe a relation for the critical light-matter coupling for the breakdown of the uniform QHF state, $g_c \propto L_y/N$, which is depicted in Fig.~\ref{fig:intro}. Below, we will see this relation may be understood analytically via both the spin wave analysis of the uniform QHF state, and the mean-field theory of the compact core phases. 

Just as the cavity has a profound effect on the QH system, inducing a phase transition at sufficiently large couplings, so also does the QH system have a back-action on the cavity. We derive in Appendix~\ref{sec:appD} a relation for the electric field fluctuations, of the form
\begin{eqnarray}
\bigg\langle  \left(\frac{\tilde{\hat E}_C(x)}{x}\right)^2\bigg\rangle &=& \frac{4g^2}{e^2\ell^4} \left[1 + \frac{8g}{\omega_0}\sqrt{\langle \hat n^D_{\rm ph}\rangle} - \frac{2g^2}{\omega_0^2} N \right].\label{eq:EfieldFluc}
\end{eqnarray}
Note for uncoupled electrons and photons, this quantity simply becomes
$4g^2/e^2\ell^4$.
 Therefore, the remaining terms in Eq.~\eqref{eq:EfieldFluc} are induced by the coupling to the 2DEG. Here $\langle \hat n^D_{\rm ph}\rangle$ is the mean photon number in the dipole gauge, which turns out to satisfy $\langle \hat n^D_{\rm ph}\rangle = \langle \bar \Pi \rangle^2$ (Appendix~\ref{sec:appD}). The electric field fluctuations normalized by the vacuum fluctuations, i.e.,~$4g^2/e^4\ell^4$, are plotted in Fig.~\ref{fig:Eenergy-EE-density}(e) with respect to light-matter coupling constant $g$. For weak light-matter coupling, the fluctuations are dominated by the vacuum fluctuations. As the light-matter coupling increases, we observe a monotonic divergence from the vacuum fluctuations as a signature of the 2DEG. We note that the fluctuations are most enhanced in the compact-core phases, however the enhancement is also present in the uniform QHF phase, since $\langle \hat n^D_{\rm ph}\rangle \neq 0$ as for any $g > 0$. In addition, we derive the physical photon occupation of the cavity to be
\begin{eqnarray}
\langle \tilde{\hat n}_{\rm ph}^C\rangle &=&\frac{2g}{\omega_0}\sqrt{\langle \hat n^D_{\rm ph}\rangle} - \frac{g^2}{2\omega_0^2} N. \label{eq:physPhotonNo}
\end{eqnarray}
When there is no coupling between matter and light, this expression reduces to $\langle \tilde{\hat n}_{\rm ph}^C\rangle=0$. Because $\langle \hat n^D_{\rm ph}\rangle$ is susceptible to the electronic phase transition (Appendix~\ref{sec:appD}), the physical photon number in the cavity carries a signature of the transition. Concretely, we observe $\langle \tilde{\hat n}_{\rm ph}^C\rangle \propto g^2/\omega_0 v_0^{\rm ee}$ when the matter is in the uniform QHF phase, whereas the photon number is suppressed once the 2DEG transitions into compact-core phases, as depicted in Fig.~\ref{fig:Eenergy-EE-density}(f). Because of this, the cavity photon number can be used to detect the breakdown of the uniform quantum Hall ferromagnet state and the onset of compact-core states. An approximate expression for $\langle \tilde{\hat n}_{\rm ph}^C\rangle$ in the thermodynamic limit, e.g.,~both $L_y,N\to \infty$ at fixed $L_x$, will be given in the MFT Sec.~\ref{sec:MFTCompactCore} below.

\section{Quantum Hall Ferromagnet in a Cavity: Spin Waves \label{sec:QHFSec-SW}
}

\subsection{Spin Flip Hamiltonian}

The numerical results described above indicate that the ground state of the cavity-coupled $\nu=1$ quantum Hall system is a uniform ferromagnet when the light-matter coupling is not too large.  This insight may be used to compute the spin wave spectrum of the system, and to understand how the instabilities found in the DMRG set in, as one considers increasingly large couplings.

Our approach involves exploiting the SU(2) spin-rotation  symmetry of the interaction Hamiltonian $\bar{H}^{\rm ee}_{\rm int}$ to simplify the problem of finding the low-energy spin wave excitations of the system in the presence of the cavity. Choosing the ground state $|\Omega\rangle$ such that all the electrons are polarized in the $\alpha = \uparrow$-direction, a low-energy subspace of single spin flips is spanned by states of the form
\begin{equation}
|X_n,Q_{y,m}\rangle=d^\dagger_{\downarrow,n+m}d_{\uparrow,n}|\Omega\rangle,    \label{eq:single_spin_flips}
\end{equation}
where $X_n = 2\pi\ell^2n/L_y$ and $Q_{y,m}=2\pi m/L_y$.  Note that the latter quantity is the $\hat{y}$ component of the momentum of the particle-hole state. 
Crucially, $\bar{H}$ [Eq.~\eqref{eq:Hbar}] acting on any such state produces a state of the same form, a result of the SU(2) symmetry encoded in $\bar{H}^{\rm ee}_{\rm int}$.  Moreover, states of this form are eigenstates of $\bar{\Pi}$, 
\begin{eqnarray}
\bar{\Pi}|X_n,Q_{y,m}\rangle &=& \frac{\zeta}{\ell}
\left[\varphi_{n+m}\mathrm{-\ } \varphi_n\right]|X_n,Q_{y,m}\rangle \nonumber \\
&\equiv&{\pi}_\varphi\left(X_n,Q_{y,m}\right)|X_n,Q_{y,m}\rangle,
\label{eq:pi_eval}
\end{eqnarray}
where in writing this, we have assumed $\sum_n \varphi_n=0$, which is always possible since $\varphi(x_j)$ is only defined up to an overall constant.  In addition, $\{|X_n,Q_{y,m}\rangle \} $ are also eigenstates of $V_{\varphi}$,
\begin{equation}
\label{Eq:v_varphi}
V_\varphi|X_n,Q_{y,m}\rangle = 
\hbar\omega_0\left[v_{\varphi} (X_n,Q_{y,m})+v_{\varphi}^{(0)} \right]|X_n,Q_{y,m}\rangle,
\end{equation}
where 
\begin{equation}
v_{\varphi}(X_n,Q_{y,m})=\left(\frac{\zeta}{\ell} \right)^2\left[\delta\varphi_{n+m}^2-\delta\varphi_{n}^2\right], 
\label{eq:vphi}
\end{equation}
and $v_{\varphi}^{(0)} \equiv \left(\frac{\zeta}{\ell} \right)^2\sum_n \delta\varphi_{n}^2$ obeys $V_{\varphi}|\Omega\rangle = \hbar\omega_0 v_\varphi^{(0)} |\Omega\rangle$; i.e., $\hbar\omega_0v_\varphi^{(0)}$ is the eigenvalue of the potential $V_{\varphi}$ when acting on the ground state.
Thus, in this basis,
\begin{widetext}
\begin{eqnarray}
    \bar{H}|X,Q_y\rangle
    &=&
    \bar{H}_{\rm int}^{\rm ee}|X,Q_y\rangle +
    \hbar\omega_0 
    \left\{\left[\hat a^\dagger+i\pi_{\varphi}(X,Q_y)\right]\left[\hat a-i\pi_{\varphi}(X,Q_y)\right]+\frac{1}{2}+v_{\varphi}(X,Q_y)+v_\varphi^{(0)}\right\}|X,Q_y\rangle.
\end{eqnarray}
\end{widetext}
If one adopts periodic boundary conditions $X_{n+N}=X_n$, exact eigenstates of $\bar{H}_{\rm int}^{\rm ee}$ within the low energy subspace can be written as well.  One finds \cite{Fertig_1989,Yang_1994}
\begin{equation}
    \bar{H}_{\rm int}^{\rm ee}|Q_x,Q_y\rangle = \left[\varepsilon_{SW} (Q_x,Q_y) + E_0^{\rm ee} \right] |Q_x,Q_y\rangle,
\end{equation}
where $|Q_x,Q_y\rangle = \frac{1}{\sqrt{N}}\sum_X e^{iQ_x X} |X,Q_y\rangle$, $E_0^{\rm ee}$ is the ground state energy in the absence of the cavity, and $\varepsilon_{SW} (Q_x,Q_y)$ is the spin wave spectrum. Because of this, we can approximately represent $\bar{H}_{\rm int}^{\rm ee}$ as
\begin{widetext}
\begin{equation}
P_{\rm SF}\bar{H}_{\rm int}^{\rm ee}P_{\rm SF} -E_0^{\rm ee}= \frac{1}{N}\sum_{Q_y}\sum_{X,X'} |X,Q_y\rangle\sum_{Q_x} e^{-iQ_x(X-X')} \varepsilon_{SW} (Q_x,Q_y)\langle X',Q_y|,
\end{equation}
\end{widetext}
where $P_{\rm SF}$ indicates we have projected into the space of single spin flips.

Several remarks are in order.  First, remarkably, the \textit{only} information about the electron-electron interaction that is actually needed in this expression is the spin wave spectrum in the absence of the cavity.  Second, the expression is approximate in that we are using a form appropriate for periodic boundary conditions, whereas the set of guiding center states $X$ retained has largest and smallest values.  Thus this form for $\bar{H}_{\rm int}^{\rm ee}$ should be quantitatively accurate for bulk spin wave excitations, but will only be qualitatively correct for matrix elements involving guiding centers near the system edges.  Our DMRG results for small system sizes produce results consistent with what emerges from this model, supporting this approximate treatment of the edges. Going further, since $|X,Q_y\rangle$ involves states localized both at $X$ and $X+Q_y\ell^2$, if either state is outside the system, terms involving that state should be dropped.  This turns out to be important when $Q_y\ell^2$ becomes comparable to the system size.  Finally, because our geometry preserves translational invariance in the $y$-direction, $Q_y$ is a good quantum number for a spin wave state.

The bosonic part of $\bar{H}$ can be simplified by one further gauge transformation, essentially bringing the system to a LLL version of the original Coulomb gauge. Writing $H_{\rm SF}=U_\pi P_{\rm SF}\tilde{H}P_{\rm SF}U_\pi^\dagger$
with
$$U_\pi=\sum_{X,Q_y}{{|X,Q_y \rangle}e^{-i
\pi_\varphi\left(X,Q_y\right)(\hat a+\hat a^{\dagger})}{\langle X,Q_y} |},$$
one finds
\begin{widetext}
\begin{eqnarray}
    H_{\rm SF}-E_0^{\rm ee}-\hbar\omega_0 v_\varphi^{(0)}= 
\sum_{X,X^\prime,Q_y}|X,Q_y\rangle&\Bigg\{&\frac{1}{N}\sum_{Q_x}e^{-i \left[   \pi_\varphi\left(X,Q_y\right)-\pi_\varphi\left(X',Q_y\right)\right]\left(\hat a + \hat a^\dagger\right) -iQ_x\left(X-X'\right)}  \varepsilon_{SW}\left(\mathbf{Q}\right) \nonumber \\
&+&\delta_{X,X'} \hbar\omega_0\left[ v_\varphi(X,Q_y)+\hat a^{\dagger}\hat a+\frac{1}{2}\right]
    \Bigg\}\langle X^\prime,Q_y|.
    \label{eq:HSF}
\end{eqnarray}
\end{widetext}
Eq.~\eqref{eq:HSF} is the central result of this section.  $H_{\rm SF}$ may be conveniently expanded in tensor product states $|X,Q_y\rangle \otimes|m\rangle$, 
with $|m\rangle \equiv \frac{1}{\sqrt{m!}}\left(a^{\dagger}\right)^m|0\rangle$, where $|0\rangle$ is the photon vacuum. The eigenvalues of the resulting matrix represent the spin wave spectrum above a $\nu=1$ quantum Hall ferromagnet coupled to a cavity mode in its uniform state, as a function of $Q_y$.  We next turn to the results of this analysis.

\begin{figure*}[t]
    \centering
\includegraphics[width=2\columnwidth]{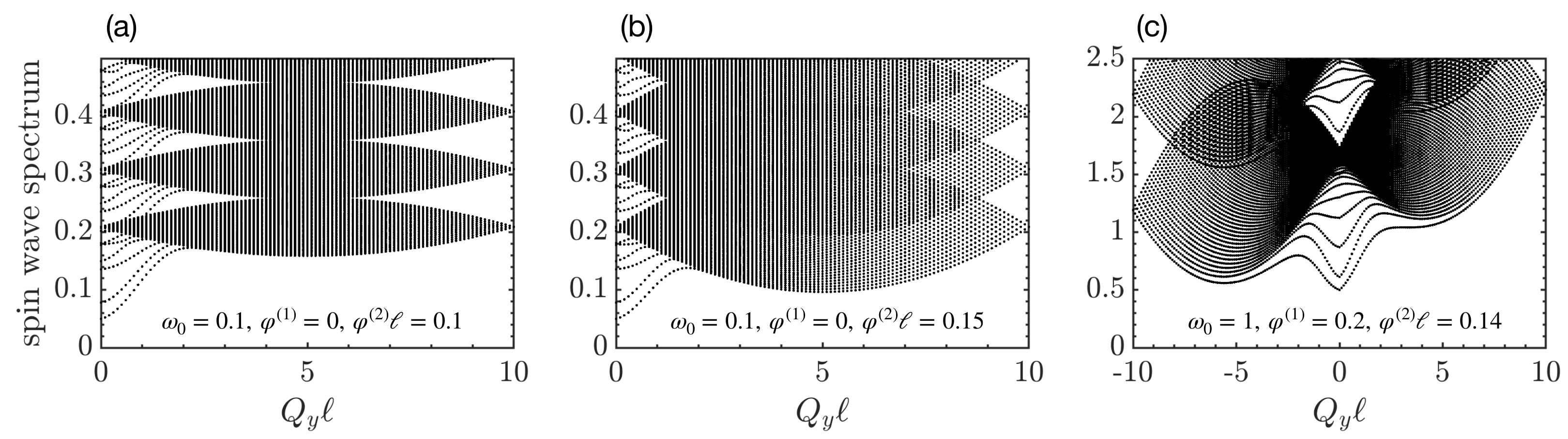} 
\caption{Spin wave spectra $ H_{\rm SF}-E_0^{\rm ee}-\hbar\omega_0 v_\varphi^{(0)}$ (Eq.~\eqref{eq:HSF}) for a linearly varying cavity field.  Calculations carried out for $N=101$ electrons and 2DEG dimensions $L_x=10.1e\ell$, $L_y=20\pi\ell$.   Highest photon occupation number retained was 20 and $\zeta=1$. Coupling strength between 2DEG and cavity photon field was varied using parameters $\omega_0$, $\varphi^{(1)}$ and $\varphi^{(2)}\ell$ as labeled.  In (a) and (b), electron-electron interaction is assumed to be a delta function contact potential, with scale $v^{\rm ee}_0$. Results are in these cases symmetric when $Q_y \to -Q_y.$ In (c), electron-electron interaction is taken to be Coulomb, with scale $e^2/\ell$. For stronger couplings $\varphi^{(2)}\ell$, the spectra indicate an instability.}   \label{fig:SW_spectra}
\end{figure*}
\subsection{Spin Wave Spectrum and Ground State Stability}

We begin with a general cavity field profile which varies at most linearly in $x$ over surface of the 2DEG,  for which
$\varphi(x)=\varphi^{(0)}+\varphi^{(1)}x + \varphi^{(2)}x^2$, as discussed above. Direct calculations yield the various parameters entering Eq.~\eqref{eq:HSF}; explicit forms for these are provided in Appendix \ref{sec:phis}.
For the important example of $\varphi^{(1)}=0$ (antisymmetric cavity field),
we find that the uniform ferromagnetic state becomes unstable to a spin wave excitation at a critical coupling $g \equiv g_c$ that is independent of $\varphi^{(2)}$ and satisfies
\begin{equation} g_c=\frac{\sqrt{2\hbar\omega_0\varepsilon_{SW}^\infty}}{L_x/\ell} = \frac{\sqrt{2\hbar\omega_0\varepsilon_{SW}^\infty}}{2\pi \ell}\frac{L_y}{N}
\label{eq:SW_instability_v1},
\end{equation} 
where $\varepsilon_{SW}^{\infty}$ is the spin wave energy of the quantum Hall ferromagnet at very large momentum $\hbar Q$, Eq.~\eqref{eq:ESW} in the $Q\rightarrow\infty$ limit. 
For contact interactions,  $\varepsilon_{SW}^{\infty}$ this energy scale acquires the simple form $\varepsilon_{SW}^{\infty}=v_0^{\rm ee}/2\pi$, which leads to one of our central results, quoted in Eq.~\eqref{eq:SW_instability_g}. Thus, the linear relation between $g_c$ and $L_y/N$, obtained numerically with DMRG, can be understood from the spin wave analysis. We note that Eq.~\eqref{eq:SW_instability_g} yields a critical coupling value $g_c=0.71\ L_y/N$ for contact interactions and parameters set in DMRG calculations, whereas we extract $\tilde{g}_c\sim 0.81\pm 0.03\ L_y/N$ from Fig.~\ref{fig:intro}(b). The difference between these values might arise from finite-size and finite meshing effects in the DMRG calculations, as well as the mean-field nature of the spin wave analysis. We explain the origin of Eq.~\eqref{eq:SW_instability_v1} below.  

\subsubsection{Symmetric and Antisymmetric Cavity Fields}

Consider first a uniform (i.e., spatially symmetric) cavity field, for which $\varphi^{(2)}=\varphi^{(0)}=0$.  As has been emphasized previously \cite{rokaj2023topological,Bacciconi_2025}, coupling a cavity degree of freedom with a uniform electric field does not modify the excitation spectrum of a quantum Hall system when when only a single Landau level has been retained in the Hamiltonian.  This result is equally true for the spin wave spectrum.  In this case one sees the eigenvalues of $\bar{\Pi}$ have the form $\pi_{\varphi}(X,Q_y)=\zeta\varphi^{(1)}Q_y\ell$, which has the special property of being independent of the guiding center $X$.  Moreover, the cavity-induced potential vanishes: $v_\varphi=0$.  It immediately follows that the  spectrum of Eq.~\eqref{eq:HSF} breaks up into commuting spin wave and bosonic mode contributions, just as one finds, when their coupling $g$ is set to zero. We have  confirmed the tensor product form of the many-body ground and excited states of 2DEG in a uniform cavity field using exact diagonalization and DMRG. 

We next consider an antisymmetric, linearly varying cavity field, for which $\varphi^{(1)}=0$, while $\varphi^{(0)},\varphi^{(2)} \ne 0$.  To compare with the DMRG results, 
we adopt the contact electron-electron interaction form, $v^{\rm ee}({\bf r})=v_0^{\rm ee}\ell^2 \delta({\bf r})$. The spin wave dispersion of the $\nu=1$ QHF may generally be written as \cite{MacDonald_1990}
\begin{equation}
    \label{eq:ESW}
\varepsilon_{SW}(Q)=\int \frac{d^2q}{\left(2\pi\right)^2}e^{-q^2\ell^2/2}\left[1-e^{i\hat{z}\cdot \left(\mathbf{Q} \times \mathbf{q}\right)\ell^2}\right]
\tilde{v}^{\rm ee}(q),
\end{equation}
where $\tilde{v}^{\rm ee}(q)$ is the Fourier transform of the electron-electron pair interaction.  For the contact interaction, 
this generates a spin wave dispersion (in the absence of the cavity) of the form 
\begin{equation}
    \varepsilon_{SW}(Q)=\frac{v_0^{\rm ee}}{2\pi} \left[1-e^{-\frac{Q^2\ell^2}{2}}\right].
\end{equation}
Note that $\varepsilon_{SW}
\sim Q^2$ at long wavelengths, consistent with the spontaneously broken SU(2) spin symmetry of the ground state \cite{Girvin_Yang_2019}.  For $Q\ell \gg 1$, one sees $\varepsilon_{SW}(Q)$ asymptotes to a constant, as must be the case for a state consisting of a particle-hole pair with well-defined momentum that is restricted to the LLL \cite{Fertig_2025}.

Representative results for the spin wave spectrum are illustrated in Fig. \ref{fig:SW_spectra}, in which we have taken a system size of $N=101$ electrons with dimensions $L_x=10.1\ell$ and $L_y=20\pi\ell$.  Panels (a) and (b) present specific results for the contact interaction.
For these examples we set have adopted $v_0^{\rm ee}$ as our unit of energy.  The cavity frequency is taken to obey $\hbar\omega_0/v_0^{\rm ee} = 0.1$, and the Hamiltonian diagonalization involved states containing up to $20$ photons. The important unitless quantity characterizing the coupling of the cavity to the spin waves turns out to be $\zeta\varphi^{(2)}\ell$, set by the scale of Eq.~\eqref{Eq:v_varphi}. We control this via $\varphi^{(2)}\ell$, while keeping $\zeta$ fixed at 1.  The ground state energy for the system (i.e., no spin flip) is $\frac{1}{2}\omega_0$, i.e.,~only the vacuum zero-point energy.  For excitation energies, this zero-point energy of the cavity mode should be subtracted away.

The behavior of the coupled cavity-spin wave spectrum may be understood from the general structure of Eqs.~\eqref{eq:HSF} and \eqref{Eq:v_varphi}.  The cavity photons impact the electronic degrees of freedom in Eq.~\eqref{eq:HSF} in two ways: dynamically, through the phase factors multiplying $\varepsilon_{SW}({Q})$, and statically, through the effective potential $\omega_0v_{\varphi}$. In practice, we find at that, at each $Q_y$, the lowest energy state obeys $\langle a^{\dagger} a \rangle \ll 1$ at couplings where the ground state is stable.  This is also true for the physical photon numbers in the QHF ground state determined by DMRG,~Fig.~\ref{fig:Eenergy-EE-density}(f).  Thus the phase factors in Eq.~\eqref{eq:HSF} have little impact compared to the effects of $v_{\varphi}$.  For $Q_y \to 0$, $v_{\varphi}=0$, and Eq.~\eqref{eq:HSF} breaks up into decoupled spin wave and cavity Hamiltonians.  The former yields discrete excitations because of the finite system size $L_x$, while the latter yields excited states precisely at $(n+\frac{1}{2})\omega_0$.  For small but non-vanishing $Q_y\ell$, the spin wave modes disperse from their quantized $Q_y=0$ values, while a continuum of modes opens up around each of the cavity states, broadening as $Q_y\ell$ increases.  These continuum states arise because, at high energies, $\varepsilon_{SW}(Q)$ becomes independent of $Q$, so that the first term on the right hand side of Eq.~\eqref{eq:HSF}
becomes nearly diagonal in $X$, yielding a spectrum that is broadened by the range of $\omega_0v_{\varphi}(X, Q_y)$.  Since $v_{\varphi}$ is an increasing function of $Q_y$, the broadening initially increases with increasing $Q_y$.

Balancing against this latter behavior is the limitation on $Q_y\ell$ that results from the finite system size $L_x$.  Recalling the spin flip states that are combined to form the spin waves of the system, Eq.~\eqref{eq:single_spin_flips}, one recognizes each of these has a particle at $X+Q_y\ell^2$ and a hole at $X$, both of which must fall inside the system.  This introduces a constraint on the basis states,
$$-\frac{L_x}{2} < X,X+Q_y\ell^2<\frac{L_x}{2}.
$$
Thus the number of particle-hole states shrinks to zero as the maximum value of $|Q_y|$ is approached, and the width of the particle-hole continuum vanishes.  Note that this change in size of the Hilbert space as $|Q_y|$ increases is also responsible for the apparent deviation of the gapless spin wave mode as $Q_y \to 0$ from the quadratic dispersion expected for an infinite system.

As is apparent in Fig. \ref{fig:SW_spectra}, with increasing $\varphi^{(2)}\ell$, the spin wave spectrum falls below the ground state energy, at $Q_y\ell^2 \approx L_x/2$.  This indicates an instability in which the uniform ground state gives way to a different state.  Recalling that the spin wave excitations involve single particle-hole pairs separated in real space by $Q_y\ell^2$, and, as explained below, the fact that $v_{\varphi}(X,Q_y)$ has a minimum when $|X|=L_x/2$, the vanishing energy excitation suggests the new ground state will involve electrons being displaced from the system edge to the system center. Again, this is consistent with the nature of the states to which the uniform ferromagnet becomes unstable in the DMRG results. An estimate of the parameters for which the instability sets in may be inferred from the form of the spin wave spectrum, a discussion which we defer to the next subsection.

\subsubsection{Asymmetric Linear Cavity Field}

In the most general case, for a cavity field profile $u(x)$ with spatial linear variation one may have $\varphi^{(0)},\varphi^{(1)},\varphi^{(2)} \ne 0$.  Typical results for Coulomb interactions are illustrated in Fig. \ref{fig:SW_spectra}(c).  As in the case of $\varphi^{(1)}=0$, for the lowest energy states, the photon occupation in dipole gauge (expectation value of $ a^{\dagger}a$ appearing in Eq.~\eqref{eq:HSF}) is negligibly small for all $Q_y$.   Several features distinguish the results from the antisymmetric cavity field case.  First, the spectrum is no long symmetric under $Q_y \to -Q_y$, with one particular sign of $Q_y$ (depending on the sign of $\varphi^{(1)}\varphi^{(2)})$ hosting the lowest energy excitations.  Local minima develop in the bulk of the spin wave spectrum, but are displaced from $Q_y\ell^2 = \pm L_x/2$.  Finally, it is notable that the two energy scales entering $\omega_0 v_\varphi$ for this case are 
$\omega_0\left(\zeta\varphi^{(2)}\ell\right)^2$ and 
$\omega_0\zeta^2\varphi^{(1)}\varphi^{(2)}\ell$.
Since the photon number in the low energy states is extremely small, the stability of the uniform ferromagnetic state is determined solely by the ratio of these parameters to the scale of the electron-electron interactions ($e^2/\epsilon_0\ell$ for the Coulomb case.)  

The observation that the $Q_y$ at which the instability occurs is large enough that $\varepsilon_{SW}({\bf Q})$ is essentially independent of $Q_y$ allows one to develop a criterion for the instability.  Setting $\varepsilon_{SW}({Q \to \infty}) \to \varepsilon_{SW}^{\infty}$ in Eq.~\eqref{eq:HSF}, the Hamiltonian becomes diagonal in the basis $|X,Q_y\rangle$, with a local minimum at $X=X^{\rm min}=-sgn(Q_y)L_x/2$.  Minimizing $v^{\varphi}(X^{\rm min},Q_y)$ with respect to $Q_y$ then yields $Q_y\ell^2=Q_y^{\rm min}\ell^2\equiv\pm\frac{L_x}{2}-\frac{\varphi^{(1)}}{2\varphi^{(2)}}$ \cite{com_phi1size}; the presence of the these two minima is apparent, for example, in Fig. \ref{fig:SW_spectra}(c).  The value of $v_{\varphi}$ at the lower of these two minima is $v_{\varphi}^{\rm min}=-2(\varphi^{(2)} \zeta)^2\left(\frac{L_x}{2}+\left|\frac{\varphi^{(1)}}{2\varphi^{(2)}}\right|\right)^2$ and the spin wave energy at the local minimum can be estimated as $E^{\rm min}=\varepsilon^{\infty}_{SW}+\omega_0v_\varphi^{\rm min}$.  Setting this to zero yields a criterion for instability of the uniform ferromagnetic state,
\begin{equation}
    \frac{1}{2}\omega_0\zeta^2\left(\varphi^{(2)}L_x+\left|\varphi^{(1)}\right|\right)^2=\varepsilon^{\infty}_{SW}.
    \label{eq:instability}
\end{equation}
As an example, comparison of this criterion with Fig. \ref{fig:SW_spectra}(b), for which the instability just sets in, shows that the relevant parameters obey Eq.~\eqref{eq:instability} to better than $2\%$. Note that rewriting $\zeta$ in terms of $g$ (Eq.~\eqref{eq:gdef1}) in Eq.~\eqref{eq:instability}, with $\varphi^{(1)}=0$, leads to Eq.~\eqref{eq:SW_instability_v1}.

\subsection{Scaling of many-body gap and entanglement spectrum}

\begin{figure}[t]
 \centering
\includegraphics[width=1\columnwidth]{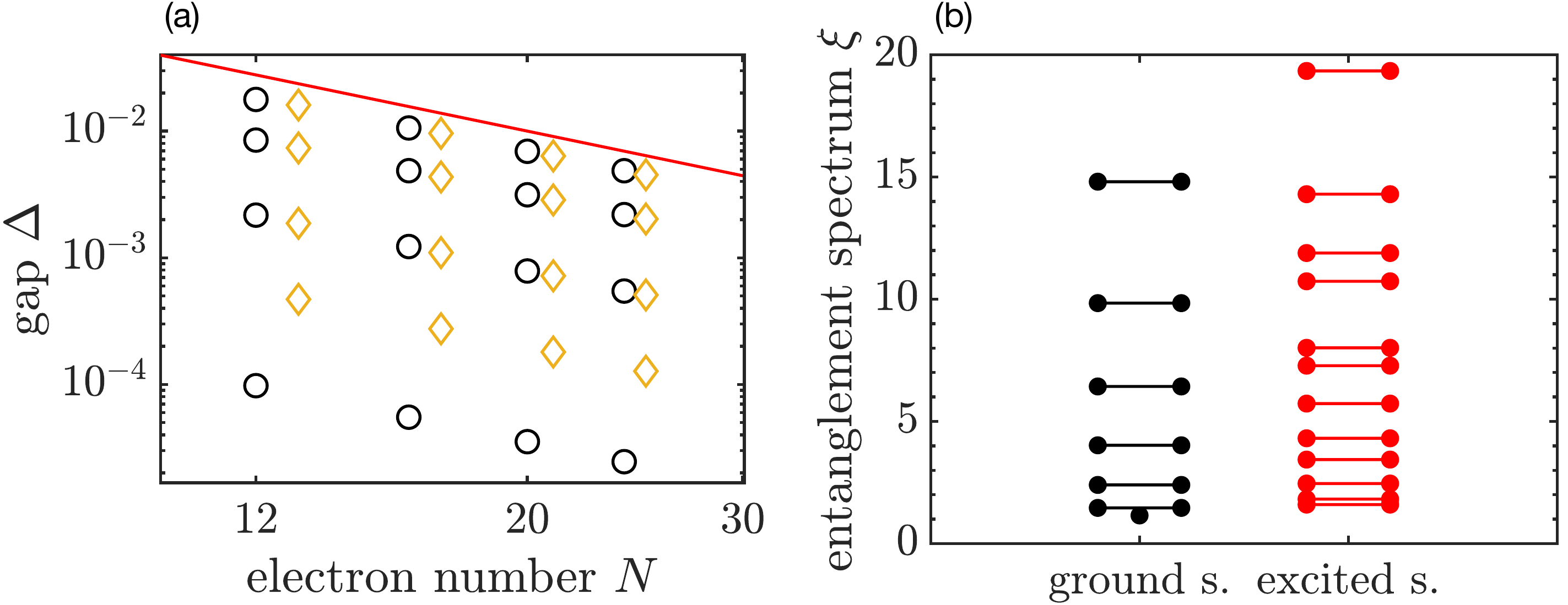}  
    \caption{(a) Many-body gap of vacuum-dressed QHF decreases with electron number $N$ and increases with cylinder circumference $L_y$ in a power-law fashion $\propto L_y^2 N^{-2}$ as found by both density matrix renormalization group (DMRG) (black-circles) and spin wave analysis (orange-diamond) with the same parameter set. The blue-solid line is $\propto N^{-2}$, and $L_y=2,4,8,12$ in an ascending order. The gaps match perfectly for all $L_y \gg \ell$. The gap is in the unit of $v_0^{\rm ee}/2\pi$.
    (b) The entanglement spectrum $\xi = -  \log \lambda_i$ for the QHF ground state (black-dots) and the first excited state (red-dots) at $N=24,\ L_y=16$ after a cut in the middle of the cylinder. The degenerate pairs are marked with a solid line.}   \label{fig:QHFgapEntSpec}
\end{figure}

Diagonalization of the spin flip Hamiltonian, Eq.~\eqref{eq:HSF}, allows for comparison with results from DMRG studies of the full Hamiltonian, Eq.~\eqref{eq:DMRG_Ham}.  Towards this end,
we extract the neutral many-body gap, $\Delta=E_1-E_0$ where $E_{0(1)}$ is the energy of the ground (first excited) state.   In the spin wave approach, this is computed for excitations with spin $\Delta S_z=1$ relative to the maximally spin-polarized ground state.  By contrast, for the DMRG studies, we focus on the $S_z=0$ symmetry sector of Eq.~\eqref{eq:DMRG_Ham}. Importantly, the DMRG excited state was found to have total spin quantum number $S=N/2-1$, so that the SU(2) symmetry of the Hamiltonian guarantees a state with the same $E_1$ with $S_z=N/2-1$.  Thus  
the minimal neutral many-body gap should be the same in both cases. This is indeed  observed for $L_y \gg \ell$, as illustrated in Fig.~\ref{fig:QHFgapEntSpec}(a), where the black circles and orange diamonds are the results from the DMRG and spin wave analysis, respectively. The lower neutral gap result of the DMRG for $L_y=2$, a thin cylinder, provides a signature of enhanced quantum fluctuations in this quasi-1D system. This significantly impacts results not only in the uniform QHF phase, but also in the compact core phases, as will be discussed in more detail in the next section.

Another observation is that, for the uniform QHF state, the many-body gap $\Delta$  does not change with light-matter coupling $g$, suggesting its scale is set solely by the electron-electron interaction.  Moreover, it is found to scale as $\Delta \propto (L_y/N)^2$, as seen in Fig.~\ref{fig:QHFgapEntSpec}(a). Details of how the spin wave gap is computed 
are provided in Appendix~\ref{sec:appF}.  

Within the DMRG approach, we further characterize the vacuum-dressed uniform QHF state and its lowest neutral excitation using the entanglement spectrum obtained from an orbital bipartition of the cylinder into two equal halves. For an electronic many-body state $|\psi_{\rm e}\rangle$, we write the Schmidt decomposition across this bipartition as
\[
|\psi_{\rm e}\rangle
=
\sum_i \sqrt{\lambda_i}\,
|i_{L/2}\rangle |i_{\overline{L/2}}\rangle ,
\]
where $\lambda_i$ are the eigenvalues  of its reduced density matrix $\rho_{L/2}$. 
Then the entanglement spectrum is obtained from $\xi_i=-\log \lambda_i$.  

The resulting entanglement spectra for the many-body ground state and the lowest excited state are shown in Fig.~\ref{fig:QHFgapEntSpec}(b).  For the ferromagnetic ground state, computed in the $S_z=0$ sector, the spectrum contains a single lowest entanglement level followed by a sequence of degenerate doublets.  This structure reflects the fact that the finite-size $S_z=0$ ground state is the zero magnetization member of the maximal $S^2$ state in a spin multiplet.  Under the orbital bipartition, Schmidt sectors with subsystem spin polarization $S_{L/2}^z=m$ can pair with sectors $S_{L/2}^z=-m$, producing the observed doublet structure, while the self-conjugate $m=0$ sector contributes an isolated, unpaired level. We provide a brief proof of this statement in Appendix~\ref{sec:entSpecQHFProof}, which also derives a closed-form expression for the Schmidt decomposition weights $\lambda_i = 
\binom{N/2}{N/4+i}^2/\binom{N}{N/2}$. This expression clearly shows $\lambda_i=\lambda_{-i}$, leading to a double degeneracy of the entanglement spectrum, except for $\lambda_0$, which is its own partner. From this, one can also analytically write the entanglement spectrum eigenvalues $\xi_i$ for the uniform QHF state, for example at $N=24$, and compare to the numerical results  found by DMRG [Fig.~\ref{fig:QHFgapEntSpec}(b)]. The results are found to match perfectly.

By contrast, the lowest excited state exhibits a twofold degeneracy throughout the entanglement spectrum, as illustrated by the red points of Fig.~\ref{fig:QHFgapEntSpec}(b). We note that this is a signature of a spin wave excitation which is odd under inversion symmetry with respect to the center of the cylinder. 
We provide a proof of this statement in Appendix~\ref{sec:SW-entSpec}. We indeed confirm that the excited state in Fig.~\ref{fig:QHFgapEntSpec}(b) is inversion-odd in our DMRG calculations. Due to the tensor product nature of the electron-photon hybrid wave function, this electronic signature is retained even when the 2DEG is subject to a linearly varying cavity field. 

\section {Properties of the compact core phases \label{sec:CompactCoreSec}} 

\subsection{Mean Field Theory \label{sec:MFTCompactCore}}

The DMRG results suggest that some compact core phases may be well-described by a mean-field theory.
This follows from two observations: 
(i) At all $g$, the ground state of the hybrid is a factorizable state between a photon mode and the electrons, $\vert \Psi_{\rm gs} \rangle = \vert \psi_{\rm e} \rangle \otimes \vert \psi_{\rm p} \rangle $. 
(ii) Once the uniform QHF state becomes unstable, the electronic part of the ground state becomes non-uniform, acquiring a wedding-cake structure, with doubly occupied sites -- the compact core -- near the node of the cavity electric field profile $u(x)$, singly occupied sites on its flanks, and empty sites beyond them. Furthermore, the analytical analysis of the spin wave instability of the QHF leads to the same conclusion since the instability occurs exactly at the wavevector corresponding to removing a spin-$\uparrow$ from the periphery and replacing it with a spin-$\downarrow$ at the center.

The structure of the Hamiltonian Eq.~\eqref{eq:DMRG_Ham} suggests that we choose the photon part of the variational ground state to be a shifted oscillator,
\begin{equation}
    |\psi_{\rm p}\rangle=e^{ip(\hat a+\hat a^\dagger)}|0_{\rm p}\rangle=e^{ip\hat a^\dagger}|0_{\rm p}\rangle, \label{eq:coherent_photon_dipolegauge}
\end{equation}
where $|0_p\rangle$ denotes a photon vacuum state, $a|0_p\rangle=0$.
In addition, the DMRG results suggest we choose the electronic part of the variational state to be non-uniform with a compact core. 
To realize this, we consider systems with an odd number of electrons, $N=2j_{max}+1$. As above, we will retain only guiding center states with $|X_j|\le 2\pi j_{max}\ell^2/L_y $, assuming that guiding center states outside this range have very high energy, and will thus never be populated. 

As in our DMRG study, we assume that $u(x)$ is antisymmetric, implying  $\varphi^{(1)}=0$. After setting $\varphi^{(0)}=0$, we have
\begin{equation}
\varphi_n=\varphi^{(2)} X_n^2.
\end{equation}
For a compact core structure, we choose to have $n_d$ doubly occupied guiding centers near the center of the sample. The guiding centers are assumed to be as close as possible to the node in $u(x)$. Thus, our electronic HF state $|\psi_{\rm e}^{\rm HF}\rangle \equiv |\psi_{\rm e}^{\rm HF}(n_d)\rangle$ has a definite number of electrons at each guiding center, and the number of doubly occupied sites $n_d$ characterizes the compact core size. Using this, we can easily evaluate the variational energy of our trial state, 
\begin{eqnarray}
\cal{E}_{\rm var}&=&\omega_0\left(p^2+\frac{1}{2}\right)+2(\varphi^{(2)}\zeta\ell-p)\omega_0\langle \psi_{\rm e}^{\rm HF}|{\Pi}|\psi_{\rm e}^{\rm HF}\rangle\nonumber\\
&+&\langle \psi_{\rm e}^{\rm HF}|H_{\rm int}|\psi_{\rm e}^{\rm HF}\rangle+\omega_0\left(\langle \psi_{\rm e}^{\rm HF}|\bar{\Pi}|\psi_{\rm e}^{\rm HF}\rangle\right)^2.
%&+&2\hbar\omega_0\varphi^{(2)}\langle \psi_{\rm e}^{\rm HF}|\bar{\Pi}|\psi_{\rm e}^{\rm HF}\rangle.
\end{eqnarray}
After some algebra one finds
\begin{equation}
    \langle \psi_{\rm e}^{\rm HF}(n_d)|\bar{\Pi}|\psi_{\rm e}^{\rm HF}(n_d)\rangle=\frac{\zeta}{\ell}\sum\limits_{\alpha,n} n_\alpha(X_n)\varphi_n \equiv \bar{\Pi}_{\rm av}(n_d),\notag
\end{equation}
where $n_\alpha(X_n)$ is the occupation of spin $\alpha$ at guiding center $X_n$. 

\begin{figure}[t]
 \centering
\includegraphics[width=1\columnwidth]{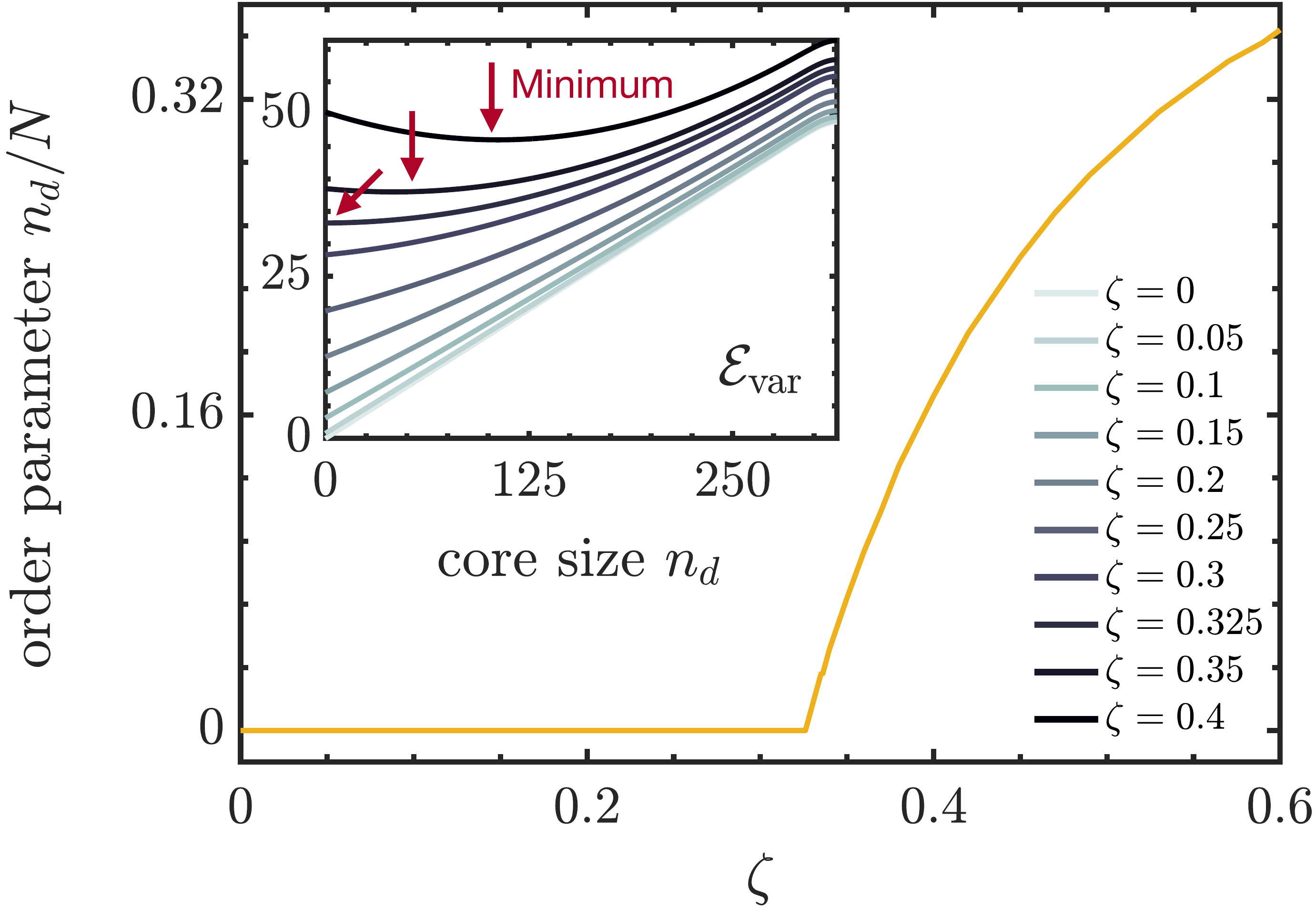}  
    \caption{The order parameter, which is the number of doubly occupied orbitals normalized by the total number of elctrons, $n_d/N$, with respect to the dimensionless light-matter coupling $\zeta$. The core size is found by minimizing $\varepsilon_{\rm var}$. Beyond $\zeta_c\approx0.326$, the optimal $n_d$ increases monotonically with $\zeta$. Inset: $\varepsilon_{\rm var}$, Eq.~\eqref{eq:var_wedding_cake}, as a function of the compact core size $n_d$ for various values of the dimensionless light-matter coupling $\zeta$ (in legend). The parameters are $\Delta X=\frac{2\pi\ell^2}{L_y}=0.1\ell$, $L_x=L_y$, $N=629$, and $v^{\rm ee}_0=\hbar\omega_0$. For small $\zeta$ the minimum lies at $n_d=0$, which corresponds to QHF phase. Beyond $\zeta_c\approx0.326$, the minimum of the variational energy occurs at $n_d \neq 0$.
    }   \label{fig:Fig4}
\end{figure}

The cases of $n_d$ odd and even must be considered separately. 
(i) $n_d \equiv 2j_d+1$. We choose the doubly occupied guiding centers to be $j\in[-j_d,j_d]$.  The singly occupied guiding centers are $j\in[-j_{\rm max}+j_d+1,-j_d-1]\bigcup [j_d+1,j_{\rm max}-j_d]$. The result is
\begin{eqnarray}
    \bar{\Pi}_{\rm av}(n_d)&=&\frac{\zeta}{\ell}\varphi^{(2)}\left(\frac{2\pi\ell}{L_y}\right)^2\bigg[\frac{N\left(N^2-1\right)}{12}\nonumber\\
    &-&\frac{1}{4}\left(N^2n_d-N(n_d^2+1)+n_d\right)\bigg]
\end{eqnarray}
(ii) $n_d=2j_d$.  Here we choose the double occupied guiding centers to be $j\in[-j_d+1,j_d]$, and the singly occupied guiding centers are $j\in[-j_{\rm max}+j_d,-j_d]\bigcup [j_d+1,j_{\rm max}-j_d]$. In this case, we obtain 
\begin{eqnarray}
    \bar{\Pi}_{\rm av}(n_d)&=&\frac{\zeta}{\ell}\varphi^{(2)}\left(\frac{2\pi\ell}{L_y}\right)^2\bigg[\frac{N\left(N^2-1\right)}{12}\nonumber\\
    &-&\frac{1}{4}\left(N^2n_d-Nn_d^2-n_d\right)\bigg]
\end{eqnarray}
Finally, we turn to the electron-electron interaction energy. As in the DMRG analysis we adopt a contact interaction. 
\begin{comment}, 
\begin{equation}
v_{\rm int}(\mathbf{r}-\mathbf{r}')= v^{\rm ee}_0\ell^2 \delta^2(\mathbf{r}-\mathbf{r}').
\end{equation}
\end{comment}
Assuming that $L_x,L_y\gg \ell$ and $n_d\ll N$, we obtain 
\begin{equation}
\langle \psi_{\rm e}^{\rm HF}(n_d)|H_{\rm int}|\psi_{\rm e}^{\rm HF}(n_d)\rangle\approx \frac{v^{\rm ee}_0}{2\pi} n_d.
\end{equation}
Putting the ingredients together, the variational energy of the state becomes
\begin{eqnarray}
{\cal{E}}_{\rm var}&=& \omega_0 \left[p-\bar{\Pi}_{\rm av}(n_d)\right]^2 \notag + \frac{\omega_0}{2}\notag \\
&+&2\zeta\ell\omega_0 \varphi^{(2)} \bar{\Pi}_{\rm av}(n_d)+\frac{v^{\rm ee}_0}{2\pi}n_d. \label{eq:var_wedding_cake}
\end{eqnarray}
Note that the coefficient in front of the linear term in $\bar{\Pi}_{\rm av}(n_d)$ is $2g=2\zeta\varphi^{(2)}\ell\omega_0$. The variational energy $\cal{E}_{\rm var}$ as a function of $n_d$ for different dimensionless light-matter couplings $\zeta$ are shown in the inset of Fig.~\ref{fig:Fig4}. It is evident that there will be a cascade of transitions from the uniform QHF (with $n_d=0$) as the lowest energy $n_d$ value takes on positive integer values. 

The value of the variational parameter $p$ that minimizes the energy is $p_{\rm opt}= \bar{\Pi}_{\rm av}(n_d)$, which determines the displacement of the {photon} mode oscillator.  With this substitution,
one has to minimize the the variational energy with respect to $n_d$.  This yields an approximate value for the ground state as a function of $\zeta$ and the other system parameters. 

In Fig.~\ref{fig:Fig4} we show how the compact core size normalized by the number of electrons, $n_d/N$,   changes 
as a function of $\zeta$. The functional form of the core size suggests that the breakdown of QHF phase into compact core phases occurs via a second order phase transition in the thermodynamic limit, where $n_d/N$ acts as an order parameter for the transition. Below, we derive an expression for the order parameter in terms of $g$. The critical $\zeta$ capturing the transition from $n_d=0\rightarrow n_d=1$, may be found analytically.  This is accomplished by examining the difference
\begin{eqnarray}
{\cal E}_{\rm var}(1)-{\cal E}_{\rm var}(0)&=&\frac{v^{\rm ee}_0}{2\pi}+2g\left[\bar{\Pi}_{\rm av}(1)-\bar{\Pi}_{\rm av}(0)\right]\nonumber\\
&=&\frac{v^{\rm ee}_0}{2\pi}-8\pi^2 j_{\rm max}^2 \frac{\ell^2}{L_y^2}\frac{g^2}{\omega_0}.
\end{eqnarray}
The transition occurs when the two energies are equal. Noting that $j_{\rm max}\approx N/2$ we obtain a condition for the critical coupling $g_c$,
\begin{equation}
g_c=\frac{1}{2\pi\ell }\sqrt{\frac{v^{\rm ee}_0 \omega_0}{\pi}} \frac{L_y}{N}.\notag 
\end{equation}
This is exactly the same prediction found in the SW analysis, Eq.~\eqref{eq:SW_instability_v1} for contact interactions. This corresponds to a dimensionless critical light-matter coupling expression $\zeta_c=\sqrt{\frac{v^{\rm ee}_0}{3\pi\omega_0}}$, and for a set of parameters, $L_x=L_y=20\pi\ell$, $N=629$, and $v^{\rm ee}_0=\omega_0$ this value is $\zeta_c\sim 0.326$, as visible in Fig.~\ref{fig:Fig4}.

In the thermodynamic limit ($L_x\gg\ell$ fixed, $N\gg n_d\gg1$, and $L_y\gg\ell$), we can approximate
\begin{equation}
\bar{\Pi}_{\rm av}(n_d)\approx \frac{g}{\omega_0}\left(\frac{2\pi N\ell}{L_y}\right)^2\bigg[\frac{N}{12}-\frac{n_d}{4}+\frac{n_d^2}{4N}\bigg]\label{eq:DexpressionTD}
\end{equation}
Now we can find the dependence of the ground state value of $n_d$ in the compact core phase close to the transition. Assuming $g=g_c+\delta g$, where $\delta g\ll g_c$, we obtain the energy functional to be
\begin{equation}
{\cal E}_{\rm var}=\frac{2g^2}{\omega_0}\left(\frac{2\pi N \ell}{L_y}\right)^2\bigg[\frac{N}{12}-\frac{n_d}{4}+\frac{n_d^2}{4N}\bigg]+\frac{v^{ee}_0}{2\pi}n_d
\end{equation}
Treating $n_d$ as a continuous variable and minimizing, and using Eq.~\eqref{eq:SW_instability_g}, we obtain the order parameter of the compact-core states to be 
\begin{equation}
\frac{n_d}{N}=\frac{g-g_c}{g_c}. \label{Eq:ndOrderparam}
\end{equation}
As mentioned earlier, the average number of photons in Coulomb gauge $\langle \tilde{\hat n}_{\rm ph}^C\rangle$ in the cavity depends on the coupling to the electronic system. Furthermore, when the 2DEG undergoes the phase transition, the dependence of $\langle \tilde{\hat n}_{\rm ph}^C\rangle$ on the coupling changes, Fig.~\ref{fig:Eenergy-EE-density}(f). Here by substituting $\bar{\Pi}_{\rm av}(n_d)$ in Eq.~\eqref{eq:physPhotonNo}, we express the physical photon number as
\begin{eqnarray}
\langle \tilde{\hat n}_{\rm ph}^C\rangle
&\sim &\frac{g^2}{\omega_0^2}\left(\frac{2\pi\ell N}{L_y}\right)^2\left[\frac{N}{6}-\frac{n_d}{2} + \mathcal{O}(1/N)\right],\label{eq:photonNoMFT}
\end{eqnarray}
valid at $N\gg n_d\gg 1$, and the terms of order $1/N$ in the square braces are dropped. Taking the thermodynamic limit while keeping $L_x$ fixed ($2\pi\ell N/L_y=L_x/\ell$)  shows that the number of photons is extensive in the number of electrons, and decreases as the number of doubly occupied sites increase. This is consistent with the suppressed photon number observed in DMRG as shown in Fig.~\ref{fig:Eenergy-EE-density}(f), once the electronic system transitions from uniform QHF to compact-core phase, where $n_d > 0$. It is also evident in Eq.~\eqref{eq:photonNoMFT} that the scaling of photon number in light-matter coupling $g$ changes at the transition due to the dependence of $n_d$ on $g$, Eq.~\eqref{Eq:ndOrderparam}, again echoing the scaling change in Fig.~\ref{fig:Eenergy-EE-density}(f). Finally we note that Eq.~\eqref{eq:photonNoMFT} clearly ties the normalized photon number $\langle \tilde{\hat n}_{\rm ph}^C\rangle/N$ to the order parameter, demonstrating how the fluctuations of the cavity field can be utilized to sense the phase transition from the uniform QHF state to the compact-core phases.

\subsection{The multiply degenerate states in the compact core phases \label{sec:CompactCoreQHFFlanks}}

All our DMRG results discussed so far were obtained in the largest $S^2$ spin sector, with magnetization $S_z=0$, for which there is a unique uniform QHF ground state.  We have also chosen our electron number $N$ to satisfy $\mod(N,4)=0$, e.g.,~$N=24$.  This last condition guarantees a unique ground state in the largest compact core phase, where there are only doubly occupied sites. For cases where $\mod(N,4)\neq 0$, the largest possible compact core will always be flanked by single spins, on one or both sides. For example, for $N=22$ the maximum size of the compact core is $10$, leaving individual spins on either side. Because these spins can point in either direction, such a state is essentially four-fold degenerate, since exchange interactions across the large core are negligible.   Note that only two of these four-fold degenerate states are allowed in $S_z=0$ sector: $\vert \uparrow_L \downarrow_R \rangle$ and $\vert \downarrow_L \uparrow_R \rangle$, where the $L,R$ subscripts denote which side of the compact core the spin is located. The other cases for which $\mod(N,4)\neq 0$, e.g.,~$N=21,23$, similarly lead to four-fold degenerate ground states, both of which harbor a single spin, in either direction, and on one or the other side of the compact core. Note that neither of these ground states belong to an $S_z=0$ sector.

\begin{figure}[t]
 \centering
\includegraphics[width=0.6\columnwidth]{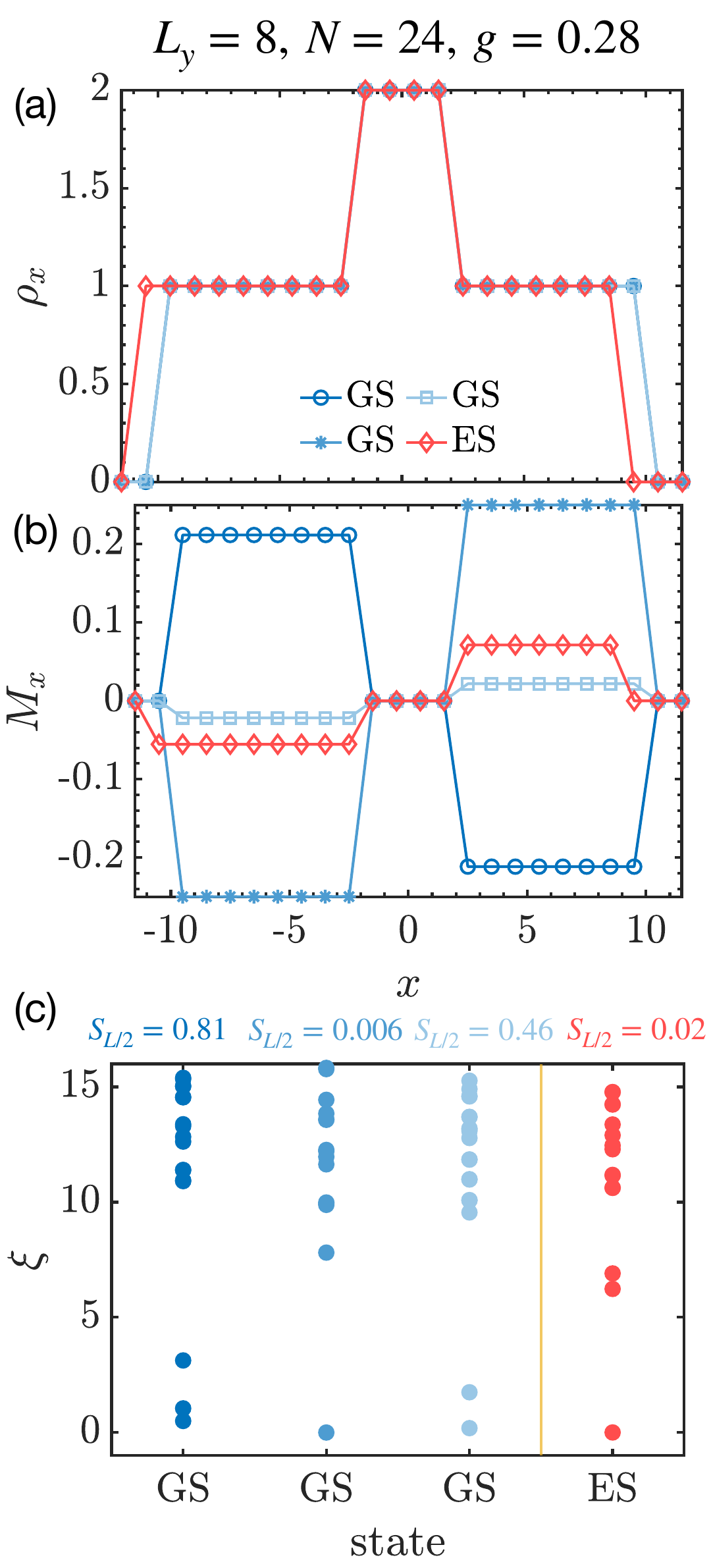}  
    \caption{The multiply-degenerate ground states and the first found excited state of the compact core phase (a,b,c) deep in the phase with $g/\omega_0=0.028$. (a) The electronic spatial profiles showing the wedding-cake structure  for all ground states (shades of blue) and the excited state (pink). (b) The magnetization profile and (c) the entanglement spectra for the same states. The bi-partite electronic entanglement entropy values are written on the top axis of (c) for each state.}   \label{fig:figMultiplyDeg}
\end{figure}

By examining the magnetization plateaus in the compact core phases that are not maximally compact, Fig.~\ref{fig:Eenergy-EE-density}(d), one understands that these ground states must be multiply-degenerate. Indeed, our DMRG routine can sometimes find multiple ground states. Fig.~\ref{fig:figMultiplyDeg}  shows the entanglement spectra, electron and magnetization densities for three different ground states of equal energy accessed by the DMRG, as well as a first excited state with a many-body gap of $0.19$ at $g/\omega_0=0.028$ for $L_y=8\ N=24$, deep in the compact core phase. The size of the compact core is $4$ for all states in (a), with various magnetization profiles apparent in (b). We confirm in our DMRG calculations that these flanks are strictly singly-occupied, and, importantly, have total spin quantum number $S=S_{f,\rm max} \equiv n_f/2$ when the flank contains $n_f$ electrons. Hence, the flanks are uniform QHF states with fixed average $S_z$ in each orbital, of equal and opposite values across the two flanks (the latter required for an $S_z=0$ state.)  For this reason, we expect the degeneracy of the ground state is considerably larger than three: for flanks each containing $n_f$ electrons [$n_f=8$ in Fig.~\ref{fig:figMultiplyDeg}], we expect a full degeneracy of $n_f+1$.  By the same token, the excited states should also be multiply degenerate. 

The entanglement spectra of the multiply degenerate ground states, when the entanglement cut is performed within the compact core, generically do not exhibit any degeneracies for states with sufficiently large compact cores. Note that the entanglement entropies for each of the ground states are also quite different, reflecting their differing entanglement spectra. 
Our DMRG procedure generically produces segments of uniform QHF on the flanks of the compact core, but these states do not appear to individually be eigenstates of $S_z$.  Since the {\it total} $S_z$ of the system is quantized at 0, such states are linear combinations of equal and opposite $S_z$ states in the flanks.  Because of this, and the fact that the entanglement cut performed is in the middle of the system within the core, the entanglement captured here is between the two uniform QHF flanks. 

To understand this in more detail, we consider a representative state in the $S_z=0$ sector, which from our DMRG results we expect to have the form
\begin{equation}
\vert \psi_e \rangle \simeq \sum_{m=-S_f}^{S_f} c_m \vert S_f, m \rangle_L \otimes \vert \textrm{Core} \rangle \otimes \vert S_f, -m \rangle_R,
\label{eq:flank_core_flank}
\end{equation}
where $\vert S_f, m \rangle_{L(R)} $ are left (right) states in the flanks, which we take to be uniform QHF states with spin quantum number $S_f$ and magnetization $m$.  Between these is a compact core state $|\textrm{Core}\rangle$.
Eq.~\eqref{eq:flank_core_flank} is already Schmidt-decomposed, resulting in $\xi_m=-\log |c_m|^2$. There is no general reason for the different $|c_m|^2$'s to be equal, and their values will in general depend on how the state was prepared. Hence the entanglement spectrum does not need to have a particular degeneracy structure.  If one begins with only the $S_z=0$ sector state in Eq. \eqref{eq:flank_core_flank}, one possible entangling operation between the two flanks is $\mathcal{S} \simeq S^-_L S^+_R +S^+_L S^-_R$, which transfers spin between the flanks without changing the global $S_z$. 
Similarly, repeated action of $\mathcal{S}$ on a state with fully polarized QHF flanks, $ \vert S_f,m_{\rm max} \rangle \otimes \vert \textrm{Core} \rangle \otimes \vert S_f,-m_{\rm max} \rangle$, where $m_{\rm max}$ is the largest magnetization consistent with $S_f$, can also produce a superposition over $m$ of the form in Eq. \eqref{eq:flank_core_flank}. We note that among the possible degenerate states, ones with a single $\pm m$ in the flanks are allowed, for which we expect minimal entanglement entropy. Indeed,
one such ground state appears in Fig. \ref{fig:figMultiplyDeg}, with essentially negligible entanglement entropy, $S_{L/2}=0.006$.  The presence of such states supports the MFT analysis of the compact core states discussed in the previous section.

\begin{figure}[t]
 \centering
\includegraphics[width=1\columnwidth]{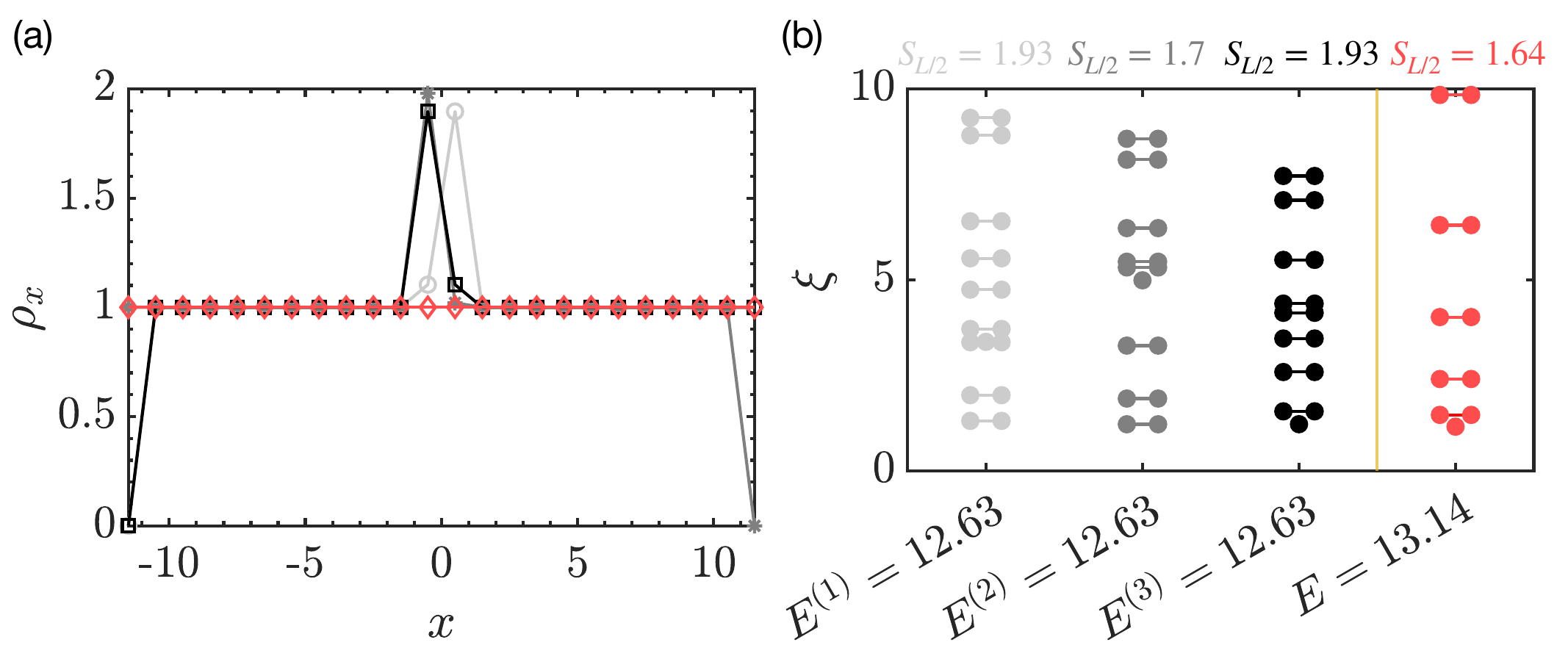}  
    \caption{(a) The electronic spatial profiles and (b) the corresponding entanglement spectra of the multiply degenerate low-lying excited states (light gray to black) and a Quantum Hall ferromagnet (QHF) excited state (red) near the first transition from QHF. The corresponding energies of the states (in the units of $v_0^{\rm ee}/2\pi$) are written at the bottom axis of (b) together with their bi-partite electronic entanglement entropies at the top. The system size and light-matter coupling are $L_y=12,\ N=24$ and light-matter coupling is $g/\omega_0=0.05$, respectively.}   \label{fig:triplyDegExcited}
\end{figure}

Multiple degeneracy is also present in the low-lying excited states. Fig.~\ref{fig:triplyDegExcited} shows one such example for a system of $L_y=12,\ N=24$ near the uniform QHF state breakdown, with light-matter coupling $g/\omega_0=0.046$. We determine the energies of a set of lowest-lying states, starting from the ground state: $E=12.12,\ 12.13,\ 12.37,\ 12.63,\ 13.14$. All these states feature a very narrow compact core, while the degenerate partners for some of them were also found. The spatial density profile and the entanglement spectra of three degenerate states at $E=12.63$ are plotted in light gray, dark gray, and black lines. The density profiles demonstrate how an electron is transferred from the edge to the center in panel (a), echoing the spin wave excitation on top of the uniform QHF. The magnetization density remains zero for all [not shown]. This behavior is accompanied by significantly different entanglement spectra, as shown in panel (b): all the states are doubly degenerate with a single non-degenerate Schmidt coefficient. The value of the non-degenerate Schmidt coefficient changes from one state to another. Since the cores form right at the center, the spectra capture the entanglement between the QHF flanks. Such an entanglement structure suggests that $c_{S_f}=c_{-S_f}$ in Eq.~\eqref{eq:flank_core_flank} with the unpaired $m=0$ quantum number representing the unpaired level, all together resulting in a zero total magnetization. We note that states of this type, where a narrow core emerges at the center with a paired entanglement spectrum, are also observed as ground states (Appendix~\ref{sec:appC2}). 

Remarkably, the next excitation at energy $E=13.14$ was found to have a uniform electron density and zero magnetization profile, together with a fully doubly-degenerate entanglement spectrum, strongly suggesting that it is a uniform QHF state. This is consistent with the fact that for finite size systems, all transitions become energy-level level-crossings. Indeed, as one sweeps over light-matter couplings beyond $g_c$
where the uniform QHF state is destabilized, a cascade of level-crossings occurs, characterized by changing quantized compact core sizes.  This implies a non-monotonic evolution of the many-body gap, which is exactly what is observed in the DMRG.

\subsection{Thin cylinder limit}

\begin{figure}[t]
 \centering
\includegraphics[width=1\columnwidth]{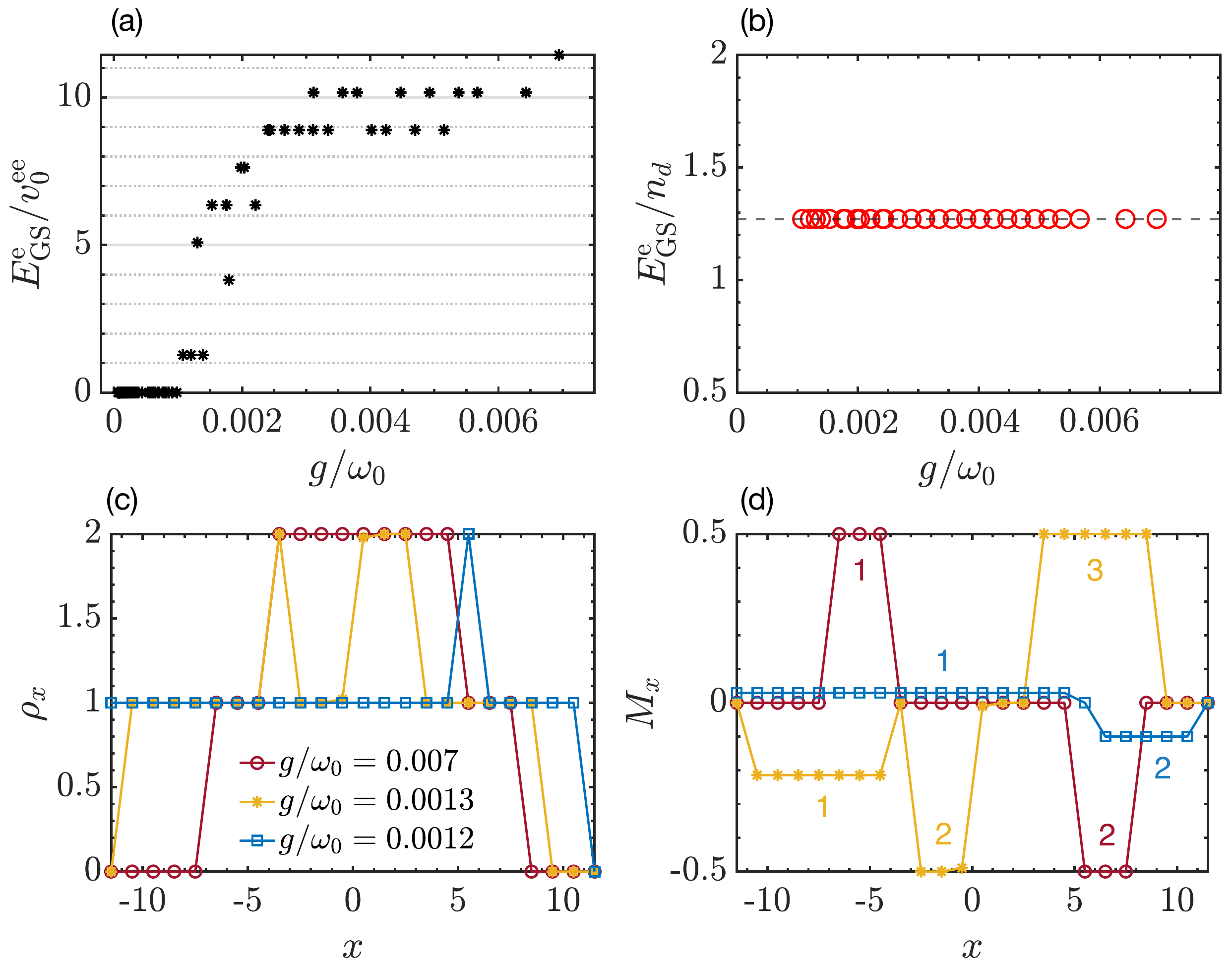}  
    \caption{(a) The electronic interaction energy $E^{\rm e}_{\rm GS}$, (b) the ratio of this energy to the core size, $E^{\rm e}_{\rm GS}/n_d$ with respect to light-matter coupling normalized by cavity frequency $g/\omega_0$,  
    (c) the electronic density profile and (d) the magnetization profile at different $g/\omega_0$, all in thin-cylinder limit, $L_y=2$ for $N=24$ electrons. $E^{\rm e}_{\rm GS}$ diverge from quantized values, which are marked with dashed lines in (a). The ratio $E^{\rm e}_{\rm GS}/n_d=1.27$ independent of $g$. At the breakdown of the uniform QHF, there may be multiple compact core regions in the interior the state, leading to multiple uniform QHF regions. An example appears in (d), showing a state with three such regions (yellow, numbered).}   \label{fig:thinCylinder}
\end{figure}

The DMRG calculations also reveal enhanced quantum fluctuations in the thin cylinder limit. To demonstrate this, in this section we exclusively focus on $L_y=2$ and the largest electron number, $N=24$. We first note that the physics qualitatively does not  change: after a critical light-matter coupling, a compact core nucleates, separating uniform QHF flanks. However, we observe quantitative changes at the breakdown of the uniform QHF state and when the system has a small-sized compact core, which we interpret as a result of enhanced quantum fluctuations.  
Fig.~\ref{fig:thinCylinder}(a) shows how the electronic interaction energies deviate from the quantized values, that we had found for sufficiently large $L_y \gg \ell$ (see Fig.~\ref{fig:Eenergy-EE-density}(a)). The integrally quantized values observed there are a result of our choice of energy unit, in which $v_0^{\rm ee}=2\pi$ in the DMRG, which results in $E_{\rm GS}^{\rm e}=n_d$, the number of orbitals in the core region. By contrast, in the thin cylinder limit, we observe $E^{\rm e}_{\rm GS} > n_d$, implying that there is an energy cost at the boundaries between the compact core and the uniform QHF-flanks. In fact, the ratio $E^e_{\rm GS}/n_d = 1.27$ was found for any light-matter coupling $g$, Fig.~\ref{fig:thinCylinder}(b). This likely reflects the larger $\Delta X$, the distance between guiding centers, associated with the small $L_y$, rather than a boundary energy.  Since interactions are short-range, any energy cost at the boundary should be instead independent of $n_d$, as $n_d \gg 1$, which is not what we observe in (b).

Importantly, the core that first emerges as $g$ increases from small values does not necessarily appear in the middle of the chain, as illustrated in Figs.~\ref{fig:thinCylinder}(c) and(d).  Moreover, as $n_d$ increases, doubly occupied states do not necessarily form the most compact configuration possible, and may host further uniform QHF regions beyond those in the flanks.
As light-matter coupling increases, the core compactifies in the middle of the cylinder, converging to the kinds of states we observe for $L_y \gg \ell$. These results suggest that the quantum fluctuations in the 2DEG are most pronounced at the breakdown of the uniform QHF, a qualitative behavior that likely reflects the enhanced quantum fluctuations associated with lower dimensionality, as the system become quasi-one-dimensional in the small $L_y$ limit.

\section{Conclusions and Outlook}
\label{sec:Conclusion}
In this work we reported the theoretical
discovery of a new phase of matter, induced by enhanced vacuum fluctuations in a 2DEG in a large magnetic field, at filling factor $\nu=1$, in situations where the cavity mode field profile is spatially non-uniform. These ``compact-core"  states replace the uniform state for strong enough coupling.  Their existence was demonstrated analytically in terms of an emerging instability in the spin wave excitation spectrum of the latter state with maximal magnetization, and more directly via a mean-field treatment of the compact-core states. The formalism led to an expression for the phase boundary that relates the critical light-matter coupling to other properties of the system: magnetic length, orbital number, circumference size, cavity frequency and electron-electron interaction energy. Extensive DMRG simulations for large circumferences $L_y \gg \ell$ 
confirm the results of the spin wave and mean-field analyses. 

A natural question arises as to the potential to realize compact core states in currently available 2DEG systems.
Assuming the 2DEG is GaAs with relative permitivity of $\epsilon_r\sim 13$~\cite{Hambleton_1961}, we can write the magnetic length as $\ell \sim 25.66\text{nm}/\sqrt{B\text{[T]}}$. This leads to an exchange Coulomb energy of $v^{\rm ee}_0\sim 5.45\text{meV}\sqrt{B\text{[T]}}$. For example, for $B=6.2$T magnetic field obtained from the carrier density of $n=1.5\times 10^{15} \text{m}^{-2}$~\cite{PhysRevLett.76.680}, we have $v^{\rm ee}_0 \sim 13.5$meV and $\ell\sim 10.3$nm. Then for a $\omega_0=1$THz cavity~\cite{Appugliese_2022,li2018vacuum}, we obtain a light-matter interaction scale of $\hbar g_c \sim 4.22\text{meV}\frac{10.3\text{nm}}{L_x}$, leading to $\hbar g_c\sim 8.7\mu$eV and $\hbar g_c\sim 0.435\mu$eV for a sample size of $L_x=5\mu$m and $L_x=100\mu$m, respectively. These are quite low light-matter interaction values~\cite{Uemoto:14}. Note that both cases lead to the dimensionless light-matter coupling of $\zeta_c \sim 0.59$. These parameters then yield the ratio $g_c/\omega_0 \sim 2\times 10^{-3}$  ($g_c/\omega_0\sim 10^{-4}$) for $L_x=5\mu$m ($L_x=100\mu$m), showing that the compact-core phase emerge at normalized effective couplings well below the conventional ultrastrong coupling threshold~\cite{li2018vacuum,Helmrich_2026}. 
A more challenging aspect of reaching this critical light-matter coupling is that it requires a rather small cavity mode volume, $\mathcal{V}\sim 6.67\times 10^{-4}\mu\text{m}^3$, when we assume that the cavity field extends across the entire 2DEG. This rather small effective mode volume can be increased by working at lower magnetic fields, or by screening the Coulomb interactions. For smaller 2DEG systems, the effective confinement length in the $z-$direction is also less demanding compared to larger samples. Interestingly, a transition between the uniform QHF and the compact core states
could be achieved within a single sample by changing the effective cavity volume, which in principle is possible by the use of a moving mirror~\cite{Graziotto_2026}.

Our study shows that coupling a QHF to a cavity mode with a non-uniform field profile can lead to states that would not be present without the cavity.  Many interesting questions remain to be investigated with respect to such systems.  Can other field realizations lead to further new quantum Hall states, both at integral $\nu$ and at fractional fillings?  What are the effects of higher Landau levels and disorder on the states we have found?  The compact-core states also offer interesting possibilities with respect to their low-lying excitations, particularly given their support of internal edges where gapless chiral charged excitations should be present.  What is the effect of the cavity on these modes?  How do these gapless excitations affect the phase transition between the compact core and uniform states?  Finally, how does the physics we have found in this work play out in more general multi-component quantum Hall systems \cite{Castro_Neto_2009,Goerbig_2011,Katsnelson_2020} beyond those in which spin plays a role,~e.g.,~those with valley or layer degrees of freedom?

Another set of questions revolve around the impact of the higher Landau levels ignored in our work.  Among these are how they affect the energetics and phase transitions of the states we have found.  More fundamentally, we found in the {\it dipole gauge} that our states are unentangled tensor products of photon and electron states. This separable nature of the hybrid photo-electronic states in the dipole gauge greatly helped us in achieving satisfactory convergence in DMRG, as well as motivated the analytical analyses performed in our work. However, the photon 
degrees of freedom 
in the dipole gauge are not physical. Indeed, in the Coulomb gauge, the ground state is not a product state between photons and electrons, because the unitary transformation of Eq.~\eqref{eq:dipole_gauge1}, that entangles the photonic subsystem with the electronic subsystem~\cite{Bacciconi_2025}. 
However, since the electron density remains invariant under this unitary transformation, we expect the compact-core phases and its transition unveiled here to remain.  
Under physical conditions where truncation to the lowest Landau level is not justified, we expect our states 
will host entanglement of the photons with the electrons even in dipole gauge. The impact of such quantum correlations 
on the quantum phase transitions found in this work offer another interesting line of inquiry. 

\section{Acknowledgments}
GM would like to acknowledge partial support from the US Department of Energy (grant no. DE-SC0024346), and to Indiana University, the Pennsylvania State University, and ICTS-TIFR Bangalore for their hospitality. HAF acknowledges the support of the NSF through Grant No. DMR-2531425. L.B. acknowledges financial support from the Ministerio de Ciencia e Innovacio ́n through the grant PID22024- 161156NB-I00. L.B. also acknowledges Severo Ochoa Centres of Excellence program through Grant CEX2024- 001445-S. CBD acknowledges F100 Initiative start-up funds by Indiana University.

\newpage
\onecolumngrid

\appendix

\section{\label{appSec:contact}Derivation of projected electronic contact  interactions }

Here we derive the electronic interactions in the LLL by computing
\begin{eqnarray}
    & &\frac{1}{2} \sum_{\br\br'} :\hat{c}^{\dg}(\br)\hat{c}(\br) V(\br-\br') \hat{c}^{\dg}(\br')\hat{c}(\br'): \notag \\
    &=& \sum_{\alpha \beta}\sum_{k_yp_yk_y'p_y'} :d^{\dg}_{\alpha,k_y}d_{\alpha,p_y}d^{\dg}_{\beta,k_y'}d_{\beta,p_y'}: \frac{1}{2} \left( \sum_{\br\br'} \phi^*_{\alpha,k_y}(\br)\phi_{\alpha,p_y}(\br) V(\br-\br') \phi^*_{\beta,k_y'}(\br')\phi_{\beta,p_y'}(\br') \right), 
\end{eqnarray}
taking note of the normal ordering. Writing the sums over $\br$ and $\br'$ as an integral over center-of-mass $(X,Y)$ and relative coordinates $(\tilde{x},\tilde{y})$, we find that the explicit form of the matrix element takes the form 
\begin{eqnarray}\label{Eqn:Vkm_suppl}
    V_{km}& =& \int d\br d\br' \phi^*_{\alpha,k_y}(\br)\phi_{\alpha,p_y}(\br)V(\br-\br')\phi^*_{\beta,k_y'}(\br')\phi_{\beta,p_y'}(\br') \notag\\
    &=& \frac{1}{\pi L_y^2 \ell^2}\int dY e^{iY(-k_y+p_y-k_y'+p_y')} \int dX d\tilde{y} d\tilde{x} e^{i\frac{\tilde{y}}{2}(-k_y+p_y+k_y'-p_y')} V(\tilde{x},\tilde{y}) \notag \\
    &\times & \exp\left( -\frac{2}{\ell^2}(X-\frac{\ell^2}{4}(k_y+p_y+k_y'+p_y'))^2 \right)  \exp\left( -\frac{1}{2\ell^2}(\tilde{x}+\frac{\ell^2}{2}(-k_y-p_y+k_y'+p_y'))^2 \right) \notag \\
    &\times & \exp\left( -\frac{\ell^2}{8}(k_y-p_y+k_y'-p_y')^2 \right) \times \exp\left( -\frac{\ell^2}{8}(k_y-p_y-k_y'+p_y')^2 \right) \notag,\\
    &=& \frac{1}{\sqrt{2\pi} L_y^2 \ell} L_y \delta_{k_y+k_y',p_y+p_y'} \exp\left( -\frac{\ell^2}{4}\left[(k_y-p_y)^2+(k_y'-p_y')^2 \right] \right) \notag \\
&\times & \int d\tilde{y} d\tilde{x} e^{i\frac{\tilde{y}}{2}(-k_y+p_y+k_y'-p_y')} V(\tilde{x},\tilde{y}) \exp\left( -\frac{1}{2\ell^2}(\tilde{x}+\frac{\ell^2}{2}(-k_y-p_y+k_y'+p_y'))^2 \right),
\end{eqnarray}
where we use the selection rule $p_y-k_y=k_y'-p_y' \rightarrow p_y+p_y'=k_y'+k_y$. Hence the final result reads
\begin{eqnarray}
  V_{km}&=& \frac{1}{\sqrt{2\pi} L_y \ell} \delta_{k_y'+k_y,p_y'+p_y} \exp\left( -\frac{\ell^2}{2}(k_y-p_y)^2 \right) \int d\tilde{y} d\tilde{x} e^{i\tilde{y}(p_y-k_y)} V(\tilde{x},\tilde{y}) \exp\left( -\frac{1}{2\ell^2}(\tilde{x}+\ell^2(-k_y+p_y'))^2 \right).
\end{eqnarray}
This is equivalent to Eq.~\eqref{Eqn:Vkm}. If we then choose the following coordinates:
\begin{eqnarray}
    k_y  &=& \frac{2\pi}{L_y}(n+k), \notag \\
    p_y  &=& \frac{2\pi}{L_y}n, \notag \\
    k_y' &=& \frac{2\pi}{L_y}(n+m), \notag \\
    p_y' &=& \frac{2\pi}{L_y}(n+m+k), \notag
\end{eqnarray}
one obtains the Eq.~\eqref{Eqn:H_halfQH}. When we use contact-like interactions, we obtain
\begin{eqnarray}
V_{lm}^{\delta} &= &\frac{v_0^{\rm ee}\ell}{\sqrt{2\pi} L_y }  \delta_{k_y'+k_y,p_y'+p_y} \exp\left( -\frac{\ell^2}{2}(k_y-p_y)^2 \right) \notag \\
&\times & \int d\tilde{y} d\tilde{x} e^{i\tilde{y}(p_y-k_y)} \delta(\tilde{x})\delta(\tilde{y}) \exp\left( -\frac{1}{2\ell^2}(\tilde{x}+\ell^2(-k_y+p_y'))^2 \right). \\
&=& \frac{v_0^{\rm ee}\ell}{\sqrt{2\pi} L_y} \delta_{k_y'+k_y,p_y'+p_y} \exp\left( -\frac{\ell^2}{2}(k_y-p_y)^2 \right)   \exp\left( -\frac{\ell^2}{2}(-k_y+p_y')^2 \right).
\end{eqnarray}
This is equivalent to Eq.~\eqref{Eqn:Vkm_contact}. 
 
\section{\label{sec:SW_Details} Details of the Spin Wave Hamiltonian Construction}
\subsection{
Expressions for  $\overline{ \Pi^2}$ and $\bar \Pi$}

We briefly explain the origin of Eq.~\eqref{eq:Pisq_main} in the main text.
We start with the unprojected operator in first quantization,
\begin{eqnarray}
\Pi^2 &=& \left(\frac{\zeta}{\ell}\right)^2
\sum_{i,j}\varphi(x_i)\varphi(x_j)\nonumber \\
&=&\left(\frac{\zeta}{\ell}\right)^2
\left\{
\sum_{i \ne j} \varphi(x_i)\varphi(x_j) + \sum_i \varphi(x_i)^2
\right\}.
\label{eq:Pisq}
\end{eqnarray}
This expression contains both an effective interaction term as well as an effective single particle potential.  We can now second quantize this expression, taking care to normal order operators appearing in what corresponds to the first term in the second line of Eq.~\eqref{eq:Pisq}, and project the expression into the LLL.  This yields
\begin{equation}
\overline{\Pi^2}
=
\left(\frac{\zeta}{\ell}\right)^2\sum_{n,n^\prime} \varphi_n\varphi_{n'}\sum_{\alpha,\alpha^\prime}d^\dag_{\alpha,n}d^\dag_{\alpha^\prime,n^\prime}d_{\alpha^\prime,n^\prime}d_{\alpha,n}+\left(\frac{\zeta}{\ell}\right)^2\sum_{n}\sum_{\alpha}\langle X_n |\varphi^2(x)| X_n \rangle d^\dag_{\alpha,n}d_{\alpha,n}.
\end{equation}
Reordering operators in the first term and taking into account their fermionic commutation relations, one finds
\begin{eqnarray}
\overline{\Pi^2}&=& \left[\left(\frac{\zeta}{\ell}\right)\sum_{n}\varphi_n\sum_{\alpha}d^\dag_{\alpha,n}d_{\alpha,n} \right]^2+\left(\frac{\zeta}{\ell}\right)^2\sum_{n}\sum_{\alpha}\left\{\langle X_n |\varphi^2(x)| X_n \rangle - \varphi_n^2\right\} d^\dag_{\alpha,n}d_{\alpha,n} \nonumber \\
&=&
\left(\bar\Pi\right)^2 +\left(\frac{\zeta}{\ell}\right)^2\sum_{n}\sum_{\alpha}
\delta\varphi^2_nd^\dag_{\alpha,n}d_{\alpha,n}
\end{eqnarray}
which appears as Eq.~\eqref{eq:Pisq_main} the main text.  The last term represents an effective single-particle potential in the Hamiltonian that dominates the behavior of the spin waves for parameters where the uniform ferromagnetic state is stable.  It is thus important to recognize that its appearance is a combined effect of normal-ordering, and thus the fermionic nature of the underlying electron system, and the projection into the LLL.

\subsection{$\pi_{\varphi}$ and $v_{\varphi}$ from Cavity Field: General Expressions
\label{sec:phis}
}
In our study, the most general spatial behavior for a cavity field we consider takes the form $u(x)=\varphi^{(1)}+2\varphi^{(2)}x$, ultimately leading to phase factors and, most importantly, an effective potential that plays an important role in the coupling between the electrons and the cavity mode.  In this subsection we provide the  forms for these quantities for this general $u(x)$.  We begin by recalling a few definitions.  In going from Coulomb to dipole gauge, we characterize $u(x)$ in terms of a potential-like quantity $\varphi(x)$, satisfying  $\frac{d\varphi}{dx}=u(x)$, which in this general case is expanded as $\varphi(x)=\varphi^{(0)}+\varphi^{(1)}x+\varphi^{(2)}x^2$. We then define
$\varphi_n \equiv \langle X_n | \varphi(x) | X_n \rangle$,
which takes the form
\begin{equation}
        \varphi_n = \varphi^{(0)}+\varphi^{(1)}X_n+\varphi^{(2)}\left(X_n^2+\frac{\ell^2}{2}\right).
        \label{eq:phi_n}
\end{equation}
For the spin wave Hamiltonian, it is convenient to choose $\varphi^{(0)}$ so that $\sum_n\varphi_n=0.$
After projecting into the LLL, we return the Hamiltonian to Coulomb gauge, using states $|X,Q_y\rangle$ that are eigenstates of $\bar{\Pi}$
(cf. Eqs. \ref{eq:single_spin_flips} and \ref{eq:pi_eval}) with eigenvalues $\pi_{\varphi}(X,Q_y)$.  The general form for these eigenvalues is
\begin{equation}
    \pi_\varphi(X,Q_y)=
    \left(\frac{\zeta}{\ell}\right)\left[\varphi^{(1)}Q_{y}\ell^2+\varphi^{(2)}\left( 2XQ_y\ell^2+Q_y^2\ell^4 \right)\right].
    \label{eq:pi_phi}
\end{equation}
As explained in the subsection just above, rewriting $\overline{\Pi^2}$ in terms of $\bar{\Pi}^2$ generates a commutator term involving $\delta\varphi_n^2 \equiv \langle X_n |\varphi^2(x) | X_n\rangle -\left(\varphi_n \right)^2$, which in this general case has the explicit form
\begin{equation}
        \delta\varphi_n^2 = 
    \frac{1}{2}\left(\varphi^{(1)}\ell\right)^2
+2\varphi^{(1)}\varphi^{(2)}X_n\ell^2
+
2\left(\varphi^{(2)}X_n\ell\right)^2
+{1 \over 2}\left(\varphi^{(2)}\ell^2\right)^2.
\label{eq:del_phi_n_sq}
\end{equation}
Finally, from Eq.~\eqref{eq:del_phi_n_sq} we can construct the effective single-particle potential
for a spin flip state $|X,Q_y\rangle$ with the definition
$v_{\varphi}(X_n,Q_{y,m})=\left(\frac{\zeta}{\ell} \right)^2\left[\delta\varphi_{n+m}^2-\delta\varphi_{n}^2\right]$, where $X_{n+m} \equiv X_n +Q_{y,m}\ell^2$ defines a  wavevector $Q_{y,m} \equiv 2\pi m/L_y$ that is consistent with the periodic boundary condition in the $y$-direction.
The effective potential
is then written as  
$V_{\varphi}|X,Q_y\rangle = \omega_0 v_\varphi(X,Q_y) |X,Q_y\rangle$, where in the most general case considered, the (unitless) eigenvalue of the potential has the form
\begin{equation}
    v_{\varphi}(X,Q_y) = \left(\zeta\varphi^{(2)}\right)^2 \left(4XQ_y\ell^2+2Q_y^2\ell^4\right)
+2\zeta^2\varphi^{(1)}\varphi^{(2)}Q_y\ell^2.
\label{eq:vphi_gen}
\end{equation}
Eqs. \ref{eq:pi_phi} and \ref{eq:vphi_gen} provide the information about the cavity field form $u(x)$ that enters the spin wave Hamiltonian, Eq.~\eqref{eq:HSF}.

\section{\label{sec:AppendixB} Lowest Landau Level Projections for DMRG}
\subsection{Spatially uniform field}

In this section we explicitly compute the projected operators for a spatially uniform field.  We begin by writing the relevant operators in second quantized form,
\begin{eqnarray}
 -i \frac{\zeta}{\ell} \omega_0  (\hat a-\hat a^{\dagger}) 
 \left(\sum_j x_j\right) 
 &=& -i \frac{\zeta}{\ell} \omega_0  (\hat a-\hat a^{\dagger}) \int d^2\br \ x \ \hat \rho(\br),\label{eq:C1}\\
 \omega_0 \left(\frac{\zeta}{\ell}\right)^2  
 \left(\sum_j  x_j \right)^2 
 &=&\ \omega_0 \left(\frac{\zeta}{\ell}\right)^2  \left[\int d^2\br  d^2 \br ' \ x \ x'  :\hat \rho (\br) \hat \rho (\br'): + \int d^2\br \,
x^2\hat \rho(\br). \right],\label{eq:C2} 
\end{eqnarray}
where $\hat \rho(\bm r)$ is the density operator in second quantized form.  Projecting into the LLL, these terms more explicitly read
\begin{eqnarray}
\int d^2 \br  \ x \ P_{\rm LLL}\hat \rho(\bm r) P_{\rm LLL}&=& \sum_{\alpha} \sum_{k_y p_y} d^{\dg}_{\alpha,k_y}d_{\alpha,p_y} \int d^2\br \ x \ \phi^*_{\alpha,k_y}(\br)\phi_{\alpha,p_y}(\br), \\
\int d^2\br  d^2 \br ' \ x \ x'  P_{\rm LLL}:\hat \rho (\br) \hat \rho (\br'):P_{\rm LLL} &=& \sum_{\alpha,\beta} \sum_{\substack{k_y\, p_y\\ k_y'\, p_y'}} \int d^2\br  d^2 \br' x\ x' \phi^*_{\alpha,k_y}(\br)\phi_{\alpha,p_y}(\br)\phi^*_{\beta,k_y'}(\br')\phi_{\beta,p_y'}(\br') \notag \\
&\times & : d^{\dg}_{\alpha,k_y}d_{\alpha,p_y}d^{\dg}_{\beta,k_y'}d_{\beta,p_y'} :.
\end{eqnarray}
The first of these is equivalent to $\bar{\Pi}$ in the previous appendix, computed for $\varphi(x) \equiv x$.  Let us now compute the integrals:
\begin{eqnarray}
  \int d^2\br \ x \ \phi^*_{\alpha,k_y}(\br)\phi_{\alpha,p_y}(\br)  &=& \frac{1}{\sqrt{\pi}L_y \ell} \int\limits_{0}^{L_y} dy \int\limits_{-\infty}^{\infty}dx \ x \ e^{ - iyk_y - \frac{1}{2\ell^2}(x - \ell^2k_y)^2 } e^{ iyp_y - \frac{1}{2\ell^2}(x - \ell^2p_y)^2 }, \notag \\ 
  &=&\frac{\delta_{k_y,p_y}}{\ell\sqrt{\pi}}\int\limits_{-\infty}^{\infty}dx\ x\ e^{-\frac{\left(x-k_y\ell^2\right)^2}{\ell^2}}=\frac{\delta_{k_y,p_y}}{\ell\sqrt{\pi}}\int\limits_{-\infty}^{\infty}dx\ \left(x+k_y\ell^2\right)\ e^{-\frac{x^2}{\ell^2}},\notag \\
  &=&\delta_{k_y,p_y} k_y \ell^2.
\end{eqnarray}
In the second equation we have used the fact that due to the periodic boundary conditions in the $y$-direction, all $y$-momenta are quantized as $2\pi n/L_y$, where $n$ is an integer. The result leads to the second quantized form
\begin{equation}
 P_{\rm LLL} \left(\sum_j x_j\right) P_{\rm LLL} 
 =\ell^2 \sum_{\alpha,k_y}   k_y \ d^{\dagger}_{\alpha,k_y}d_{\alpha,k_y}.
\end{equation}
Following our coordinate notation in the main text, we write $k_y=\frac{2\pi}{L_y}n$. Hence one finally obtains for Eq.~\eqref{eq:C1}
\begin{eqnarray}
  -i \frac{\zeta}{\ell} \omega_0  (\hat a-\hat a^{\dagger}) 
 \left(\sum_j x_j\right) \rightarrow-i \frac{\zeta}{\ell} \omega_0   \frac{2\pi \ell^2}{L_y}  (\hat a-\hat a^{\dagger}) \sum_{\alpha,n} n\ d^{\dg}_{\alpha,n}d_{\alpha,n}.
\end{eqnarray}

Now we consider the integral
\begin{equation}
    \int d^2\br  d^2 \br ' \ x \ x'  :\hat \rho (\br) \hat \rho (\br'):,
\end{equation}
arising from the first term on the RHS of Eq.~\eqref{eq:C2}. Clearly, apart from the normal ordering of the operators, the matrix element is simply the product of the matrix elements of $x$ and $x'$. Therefore
\begin{equation}
    \omega_0 \left(\frac{\zeta}{\ell}\right)^2\int d^2\br  d^2 \br ' \ x \ x'  :\hat \rho (\br) \hat \rho (\br'): =\omega_0 \left(\frac{\zeta}{\ell}\right)^2   \left(\frac{2\pi \ell^2}{L_y}\right)^2 \sum_{\alpha, \beta}\sum_{n,m}n m:d^{\dg}_{\alpha,n}d_{\alpha,n}d^{\dg}_{\beta,m}d_{\beta,m}:.
\end{equation}
Finally we compute the one-body  term which arises from the last term of the RHS of Eq.~\eqref{eq:C2},
\begin{eqnarray}
 \int d^2\br \ x^2 \ \phi^*_{\alpha,k_y}(\br)\phi_{\alpha,p_y}(\br) &=& \frac{1}{\sqrt{\pi}L_y \ell}\int dy dx \ x^2\ e^{ - iyk_y - \frac{1}{2\ell^2}(x - \ell^2k_y)^2 } e^{ iyp_y - \frac{1}{2\ell^2}(x - \ell^2p_y)^2 } , \notag \\
&=& \frac{1}{\sqrt{\pi}L_y \ell} \int dy dx \ x^2 e^{ - iyk_y+iyp_y} e^{ - \frac{1}{2\ell^2}(x - \ell^2k_y)^2 - \frac{1}{2\ell^2}(x - \ell^2p_y)^2 },\notag \\
&=&\frac{1}{\sqrt{\pi}L_y \ell} \int dy dx  \ x^2 e^{ - iyk_y+iyp_y} e^{ - \frac{1}{\ell^2}\left[ x - \frac{\ell^2}{2}(k_y+p_y)\right]^2}e^{ - \frac{\ell^2}{4}(k_y-p_y)^2}, \notag \\
&=&\ell^2 \delta_{k_y,p_y} e^{ - \frac{\ell^2}{4}(k_y-p_y)^2} \left[\frac{1}{2}+ \frac{\ell^2}{4}(k_y+p_y)^2\right],
\end{eqnarray}
leading to,
\begin{eqnarray}
\sum_{\alpha,k_y,p_y} \ell^2 \delta_{k_y,p_y} e^{ - \frac{\ell^2}{4}(k_y-p_y)^2} \left[\frac{1}{2}+ \frac{\ell^2}{4}(k_y+p_y)^2\right] \ d^{\dg}_{\alpha,k_y}d_{\alpha,p_y} &=& \ell^2 \sum_{\alpha,k_y}  \left[ \frac{1}{2} + \ell^2 k_y^2\right] \ d^{\dg}_{\alpha,k_y}d_{\alpha,k_y},\notag\\
&=& \ell^2 \sum_{\alpha,n}  \left[ \frac{1}{2} +\left( \frac{2\pi \ell}{L_y} \right)^2 n^2 \right] \ d^{\dg}_{\alpha,n}d_{\alpha,n}.\label{eq:oneBody}
\end{eqnarray}
Putting the pieces together, we arrive at
\begin{eqnarray}
 \omega_0 \left(\frac{\zeta}{\ell}\right)^2 P_{\rm LLL}\left(\sum_j  x_j \right)^2 P_{\rm LLL} &=& \omega_0 \left(\frac{\zeta}{\ell}\right)^2  \bigg[  \left(\frac{2\pi \ell^2}{L_y}\right)^2 \sum_{\alpha, \beta}\sum_{n,m}n m:d^{\dg}_{\alpha,n}d_{\alpha,n}d^{\dg}_{\beta,m}d_{\beta,m}: \notag \\
 &+& \ell^2 \sum_{\alpha,n}  \left[ \frac{1}{2} +\left( \frac{2\pi \ell}{L_y} \right)^2 n^2 \right] \ d^{\dg}_{\alpha,n}d_{\alpha,n} \bigg], \\
 &=& \omega_0 \left(\frac{\zeta}{\ell}\right)^2 \left(\frac{2\pi \ell^2}{L_y}\right)^2 \sum_{\alpha, \beta}\sum_{n,m}n m \ d^{\dg}_{\alpha,n}d_{\alpha,n}d^{\dg}_{\beta,m}d_{\beta,m} + \omega_0  \frac{ \zeta^2}{2} N.
\end{eqnarray}

\subsection{Linearly varying field \label{sec:AppBSSec2}}

We next compute the LLL-projected terms relevant to a linearly-varying cavity field, $u(x) = \sqrt{\mathcal{N}_c}x$, where $\mathcal{N}_c$ is a normalization constant with units of $1/\ell^2$. Let us fix the normalization constant. Recall $$\int d^2r\ u^2({\bf r}) = S,$$
where $S$ is the area in which the cavity mode is confined. Remembering that we defined the cavity mode volume as $\mathcal{V}=\chi L_x L_y$ with confinement length of $\chi$ in the $z-$direction, and $L_xL_y$ the area of the 2DEG. Setting $S=L_x L_y$ (i.e., cavity and 2DEG area the same), we arrive at a relation fixing $\mathcal{N}_c$,
\[
\mathcal{N}_c\int_{-L_x/2}^{L_x/2}\int_{-L_y/2}^{L_y/2} dx\ dy \ x^2 = L_x L_y, \qquad \to \sqrt{\mathcal{N}_c} = \frac{\sqrt{12}}{L_x}.
\]
This leads to 
\begin{eqnarray}
    u(x) = \sqrt{12} \frac{x}{L_x} = \frac{\sqrt{3}}{\pi} \frac{L_y}{N  \ell^2} x
    \equiv \sqrt{\mathcal{N}_c} x.
\label{eq:Ncdef}
\end{eqnarray}
The projected operators that we need to compute are:
\begin{eqnarray}
    -\frac{i}{2}\frac{\zeta}{\ell} \omega_0  (\hat a-\hat a^{\dagger})\mathcal{N}_c P_{\rm LLL} \left(\sum_j x_j^2\right) P_{\rm LLL} &=& -\frac{i}{2}\frac{\zeta}{\ell} \omega_0  (\hat a-\hat a^{\dagger}) \mathcal{N}_c \int d^2\br \ x^2 \hat P_{\rm LLL}\rho(\br) P_{\rm LLL},\\
 \omega_0 \left(\frac{\zeta}{\ell}\right)^2 \mathcal{N}_c^2 P_{\rm LLL}\left(\sum_j  \frac{x_j^2}{2} \right)^2 P_{\rm LLL}&=&\left(\frac{\zeta}{\ell}\right)^2 \frac{\omega_0}{4} \mathcal{N}_c^2 \Bigg[\int d^2\br  d^2 \br ' \ x^2 \ x'^2  P_{\rm LLL}:\hat \rho (\br) \hat \rho (\br'): P_{\rm LLL} \nonumber \\
 &+& \int d^2\br \,
x^4\hat P_{\rm LLL}\rho(\br) P_{\rm LLL}\Bigg], 
\label{eq:RHS_first}
\end{eqnarray}
The first integral was already computed above [cf. Eq.~\eqref{eq:oneBody}]. Hence,
\begin{eqnarray}
  -\frac{i}{2} \frac{\zeta}{\ell} \omega_0 \mathcal{N}_c   (\hat a-\hat a^{\dagger}) P_{\rm LLL} \left(\sum_j x_j^2\right) P_{\rm LLL} = -\frac{i}{2}\frac{\zeta}{\ell} \omega_0 \frac{\sqrt{3}}{\pi} \frac{L_y}{N } (\hat a-\hat a^{\dagger})   \sum_{\alpha,n}  \left[ \frac{1}{2} +\left( \frac{2\pi \ell}{L_y} \right)^2 n^2 \right] \ d^{\dg}_{\alpha,n}d_{\alpha,n}.
\end{eqnarray}
The first term on the right-hand side of Eq. \eqref{eq:RHS_first}
can also be most easily calculated by using our previous result, Eq.~\eqref{eq:oneBody}, leading to
\begin{eqnarray}
\int d^2\br  d^2 \br ' \ x^2 \ x'^2  P_{\rm LLL}:\hat \rho (\br) \hat \rho (\br'): P_{\rm LLL} &=& \ell^4 \sum_{\alpha,\beta}\sum_{n,m}  \left[ \frac{1}{2} +\left( \frac{2\pi \ell}{L_y} \right)^2 n^2 \right]\left[ \frac{1}{2} +\left( \frac{2\pi \ell}{L_y} \right)^2 m^2 \right] \nonumber \\ &\times& :d^{\dg}_{\alpha,n}d_{\alpha,n} d^{\dg}_{\beta,m}d_{\beta,m}: .
\end{eqnarray}
The remaining one-body term in Eq. \eqref{eq:RHS_first} involves the integral
\begin{eqnarray}
 \int d^2\br \ x^4 \ \phi^*_{\alpha,k_y}(\br)\phi_{\alpha,p_y}(\br) &=& \frac{1}{\sqrt{\pi}L_y \ell}\int dy dx \ x^4\ e^{ - iyk_y - \frac{1}{2\ell^2}(x - \ell^2k_y)^2 } e^{ iyp_y - \frac{1}{2\ell^2}(x - \ell^2p_y)^2 } , \notag \\
&=& \frac{1}{\sqrt{\pi}L_y \ell} \int dy dx \ x^4 e^{ - iyk_y+iyp_y} e^{ - \frac{1}{2\ell^2}(x - \ell^2k_y)^2 - \frac{1}{2\ell^2}(x - \ell^2p_y)^2 },\notag \\
&=&\frac{1}{\sqrt{\pi}L_y \ell} \int dy dx  \ x^4 e^{ - iyk_y+iyp_y} e^{ - \frac{1}{\ell^2}\left[ x - \frac{\ell^2}{2}(k_y+p_y)\right]^2}e^{ - \frac{\ell^2}{4}(k_y-p_y)^2}, \notag \\
&=&\frac{\ell^4}{16} \delta_{k_y,p_y} e^{ - \frac{\ell^2}{4}(k_y-p_y)^2} \left[ 12+\ell^2(k_y+p_y)^2 \left(12+\ell^2(k_y+p_y)^2 \right)\right],
\end{eqnarray}
leading to
\begin{eqnarray}
&&\sum_{\alpha,k_y,p_y} \frac{\ell^4}{16} \delta_{k_y,p_y} e^{ - \frac{\ell^2}{4}(k_y-p_y)^2} \left[ 12+\ell^2(k_y+p_y)^2 \left(12+\ell^2(k_y+p_y)^2 \right)\right] \ d^{\dg}_{\alpha,k_y}d_{\alpha,p_y} \notag \\
&=&\ell^4\sum_{\alpha,k_y}  \left[ \frac{3}{4}+3k_y^2\ell^2+k_y^4\ell^4\right] \ d^{\dg}_{\alpha,k_y}d_{\alpha,k_y} = \ell^4 \sum_{\alpha,n}  \left[ \frac{3}{4}+3 \left(\frac{2\pi \ell}{L_y}\right)^2 n^2+\left(\frac{2\pi \ell}{L_y}\right)^4n^4\right] \ d^{\dg}_{\alpha,n}d_{\alpha,n}.\label{eq:oneBody2}
\end{eqnarray}
Putting the various parts together, we obtain
\begin{eqnarray}
    \omega_0 \left(\frac{\zeta}{\ell}\right)^2 \mathcal{N}_c^2  P_{\rm LLL}\left(\sum_j  \frac{x_j^2}{2} \right)^2 P_{\rm LLL}&=& \frac{1}{4}\omega_0 \left(\frac{\zeta}{\ell}\right)^2 \frac{3}{\pi^2} \frac{L_y^2}{N^2  } \bigg\lbrace \sum_{\alpha,\beta}\sum_{n,m}  \left[ \frac{1}{2} +\left( \frac{2\pi \ell}{L_y} \right)^2 n^2 \right]\left[ \frac{1}{2} +\left( \frac{2\pi \ell}{L_y} \right)^2 m^2 \right] \\
    &\times &:d^{\dg}_{\alpha,n}d_{\alpha,n} d^{\dg}_{\beta,m}d_{\beta,m}: +  \sum_{\alpha,n}  \left[ \frac{3}{4}+3 \left(\frac{2\pi \ell}{L_y}\right)^2 n^2 +\left(\frac{2\pi \ell}{L_y}\right)^4 n^4\right] \ d^{\dg}_{\alpha,n}d_{\alpha,n}\bigg\rbrace\notag \\
    &=& \omega_0 \left(\frac{\zeta}{\ell}\right)^2 \frac{3}{4\pi^2} \frac{L_y^2}{N^2  } \bigg\lbrace \sum_{\alpha,\beta}\sum_{n,m} \left( \frac{1}{2} +\left( \frac{2\pi \ell}{L_y} \right)^2 n^2 \right) \left( \frac{1}{2} +\left( \frac{2\pi \ell}{L_y} \right)^2 m^2 \right) \notag \\
    &\times & d^{\dg}_{\alpha,n}d_{\alpha,n} d^{\dg}_{\beta,m}d_{\beta,m} + 2 \left( \frac{2\pi \ell}{L_y} \right)^2 \sum_{\alpha,n} n^2 d^{\dg}_{\alpha,n}d_{\alpha,n} + \frac{N}{2} \bigg\rbrace.\label{eq:C22}
\end{eqnarray}
Based on these results, we defined $(\hbar=1)$ our coupling constant for the linearly varying $u(x)$ as
\[
g = \frac{\zeta}{\ell} \frac{ \omega_0 \sqrt{3}}{2\pi}\frac{L_y}{N} = \frac{1}{2\pi}\frac{L_y}{N}\sqrt{\frac{ 3\omega_0}{2\epsilon_0\mathcal{V}}} = \frac{1}{2\pi}\frac{L_y}{N}\sqrt{\frac{ 3\omega_0}{2\epsilon_0\chi L_x L_y}} = \frac{1}{4\pi \ell}\frac{L_y}{N^{3/2}}\sqrt{\frac{ 3\omega_0}{ \pi \epsilon_0 \chi }}.
\]

\begin{figure}[t]
 \centering
\includegraphics[width=1\columnwidth]{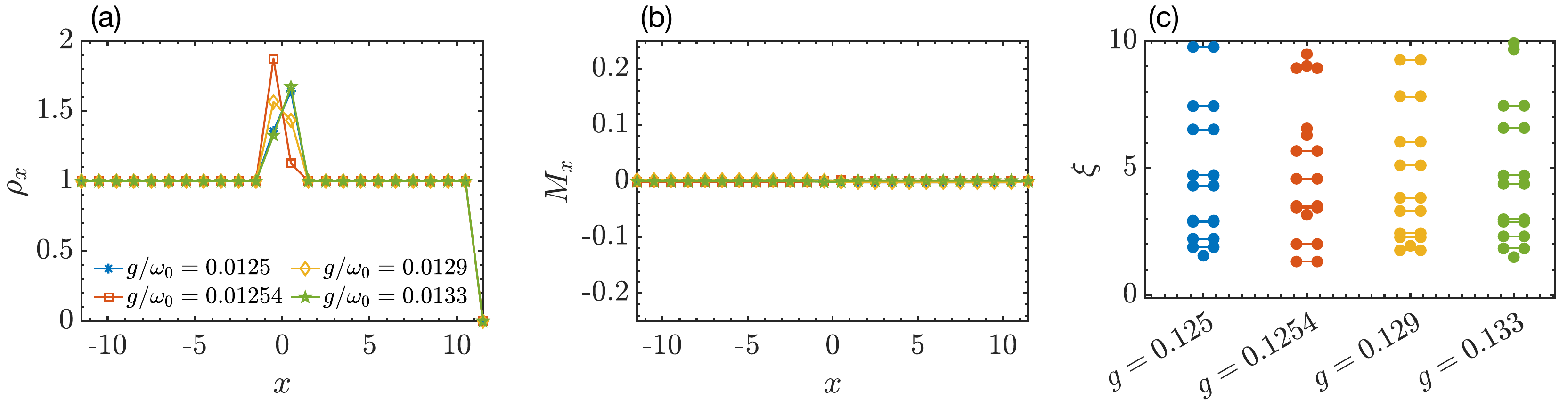}  
    \caption{(a) Electronic density, (b) magnetization profile and (c) entanglement spectra of narrow compact-core ground states at the breakdown of the uniform quantum Hall ferromagnet for $N=24,\ L_y=8$. One electron from the edge is transported to the center in all results.}   \label{fig:AppC2}
\end{figure}

Hence the cavity-induced interactions and potential have $g^2/\omega_0$ as their strengths. Then we define 
\[
g_0 = \frac{1}{4\pi \ell}\sqrt{\frac{ 3}{ \pi \epsilon_0 \chi }}, \qquad \to \qquad g = g_0 \frac{L_y \omega_0^{1/2}}{N^{3/2}}.
\]
Finally, we note that $g$ is defined differently for the spatially uniform cavity: $ g\equiv e A_0 \omega_0 = \frac{\zeta}{\ell} \omega_0$. 

\section{Density matrix renormalization group simulations \label{sec:DMRGApp}}

\subsection{Details of the DMRG simulations \label{sec:appC}}

We model the system with a finite-cylinder Hamiltonian using the density matrix
renormalization group (DMRG), representing the lowest Landau level
guiding center orbitals as a one-dimensional chain, as explained in the main text. The Hamiltonian is
encoded as a matrix product operator, while the many-body eigenstates
are represented as quantum-number conserving matrix product states (MPS).  We explicitly conserve the total electron number and the total
spin projection $S_z$, and the calculations presented here are performed
in the $S_z=0$ sector.  To reduce sensitivity to the initial
configuration, the optimization was initialized using a few different
MPS states. Each DMRG run was allowed to perform up to
$N_{\rm sw}=500$ sweeps, although the optimization could terminate earlier upon reaching
the prescribed energy error goal.  At the end of each run, the optimized
MPS was saved.  When the desired convergence was not obtained within
the first run, the calculation was continued using the saved MPS as the
initial state of a subsequent run.  In the calculations reported here,
we allowed up to a cumulative total of $N_{\rm sw}^{\rm tot}=1000$
sweeps for convergence. For the ground state, the maximum MPS bond dimension was increased
according to $\chi_{\max}=50,\;100,\;200,$
with $\chi_{\max}=200$ retained during the subsequent sweeps.  For the
low-lying excited states, we used $\chi_{\max}=100,\;200,\;400,
$
with $\chi_{\max}=400$ retained thereafter.  The singular-value
truncation cutoff was fixed to $\epsilon_{\rm SVD}=10^{-10}$.  To facilitate convergence during the initial sweeps, we employed the following
noise schedule $10^{-6},\;10^{-7},\;10^{-8},\;10^{-9},\;10^{-10},\;0,$
after which the noise was kept at zero.  The DMRG energy-error goal was
set to $\epsilon_E=10^{-8}$.

\begin{figure}[t]
 \centering
\includegraphics[width=1\columnwidth]{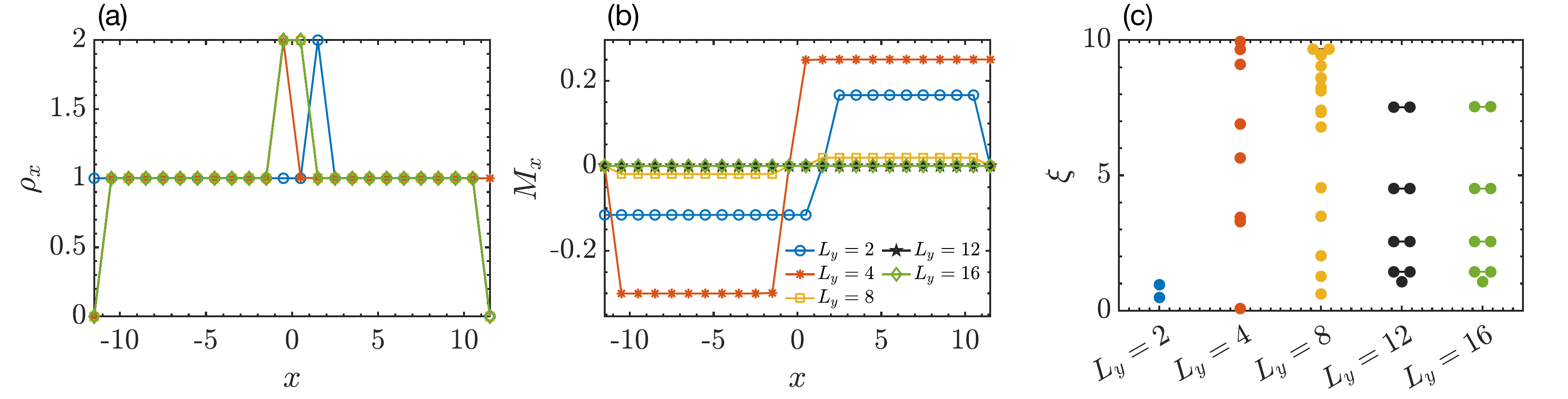}  
    \caption{(a) Electronic density, (b) magnetization profile and (c) entanglement spectra of the first found compact-core state after the transition from the uniform quantum Hall ferromagnet at $N=24$ electron number for all different cylinder circumferences $L_y=2, \cdots, 16$ (see legend in (b)).}   \label{fig:AppC2-2}
\end{figure}
We sequentially targeted the ground state and
the next few low-lying states. Excited states were obtained by imposing
orthogonality to all previously converged lower-energy MPSs, using the
penalty weight $W=20$. During the processing of data, we determine the lowest energy achieved by the DMRG applied to different initial MPS, and the smallest many-body gap found among those. This gap should be interpreted as an upper-bound due to the nature of variational algorithm. 

We perform some posterior convergence diagnostics, two of which are the energy variance
\[
\delta E^2
=
\langle\psi|\bar H^2|\psi\rangle
-
\langle\psi|\bar H|\psi\rangle^2
\]
and the mutual overlaps $\left|\langle\psi_i|\psi_j\rangle\right|$
between the independently targeted states. As a separate check we monitor the bare vacuum fluctuations, 
$\bigg\langle  \left(\tilde{\hat E}_C(x)/x\right)^2_1\bigg\rangle$ (see Appendix~\ref{sec:appD}) which should follow $\bigg\langle  \left(\tilde{\hat E}_C(x)/x\right)^2_1\bigg\rangle = 4g^2/\ell^4$
and lead to $0$ photon number, and hence to the zero-point energy of the cavity, for all light-matter couplings. This follows from the fact that the electronic states are eigenstates of polarizability~(Appendix~\ref{sec:appD}). We monitor all studied ground and excited states to ensure zero bare vacuum photon number. We observe that this quantity is very sensitive to convergence: when a result is not converged, it will exhibit nonzero bare vacuum photon numbers. 

We note that as light-matter coupling $g$ increases, the convergence becomes particularly slow, requiring the sweep numbers mentioned above. This slow convergence however was observed not to be caused by the entanglement build-up in the electronic subsystem: as $g$ increases the product fully-compact core states emerge with $S=0$. Rather it was found to be due to the slow evolution of the state in the photonic Hilbert space, due to possible local minima in the energy landscape.  

\subsection{Additional DMRG results \label{sec:appC2}}

In this section, we present further results emerged from our DMRG analysis. In Fig.~\ref{fig:AppC2} we demonstrate that narrow compact-core states can also appear as many-body ground states, similar to the excited states discussed in the main text. Such states are not fully compact even at the center of the system as seen in panel (a), showing the transfer of an edge electron to the center, albeit the electron is not localized at a site. For such states, we always find uniform zero magnetization [panel (b)].  Their entanglement spectra contain similar features to what is discussed in the main text, exhibiting a level of structure with some paired Schmidt coefficients, as illustrated in panel (c).

Fig.~\ref{fig:AppC2-2} depicts the density and magnetization profiles, along with the entanglement spectra, of the first compact core states found after the breakdown of the uniform QHF state. The results show how these features change with $L_y$, the cylinder circumference, at a fixed electron number $N=24$. Echoing the discussion in the main text, the nucleated core away from the center in $L_y \sim \ell$ converges to the center as $L_y$ increases. The entanglement spectra in panel (c) exhibit a significant change as $L_y$ increases, accompanying the conclusion stated in the main text: the electronic entanglement entropy tends to increase with increasing $L_y$ due to the guiding centers getting closer to one another. The difference in the entanglement spectra is consistent with the magnetization profiles shown in panel (b): The completely paired entanglement spectra with a single non-degenerate Schmidt coefficient for $L_y=12,16$ suggests an equal contribution from $\pm m \neq0$ with an unpaired $m=0$ contribution, leading to uniform and zero magnetization profile in an $S_z=0$ symmetry sector.

\section{Dressed cavity operators \label{sec:appD}}

Here we express the dressed cavity operators, which are computed in the DMRG. For this we define a quantity closely related to $\Pi$, evaluated for specific case of the linear field profile $u(x)$ discussed in Appendix \ref{sec:AppBSSec2}.  This is a generalized polarizability $\hat{\mathcal{P}}$, defined via
\begin{eqnarray}
\frac{\zeta}{\ell}\omega_0\mathcal N_c
\left(
\sum_j \frac{x_j^2}{2}
\right)
=
g\,\hat{\mathcal P}=\omega_0 \Pi, 
\end{eqnarray}
where $\mathcal N_c$ is defined in Eq. \eqref{eq:Ncdef}.
Note that the projection to the LLL has not been carried out yet. After projection to the LLL, we will denote the quantity by $\hat {\bar {\mathcal P}}$
\begin{eqnarray}
\hat{\bar{\mathcal P}}
=
\sum_{\alpha,n}
\left[
\frac{1}{2}
+
\left(\frac{2\pi\ell}{L_y}\right)^2 n^2
\right]
d^\dagger_{\alpha,n}d_{\alpha,n}.
\end{eqnarray}

\begin{figure}[t]
 \centering
\includegraphics[width=0.75\columnwidth]{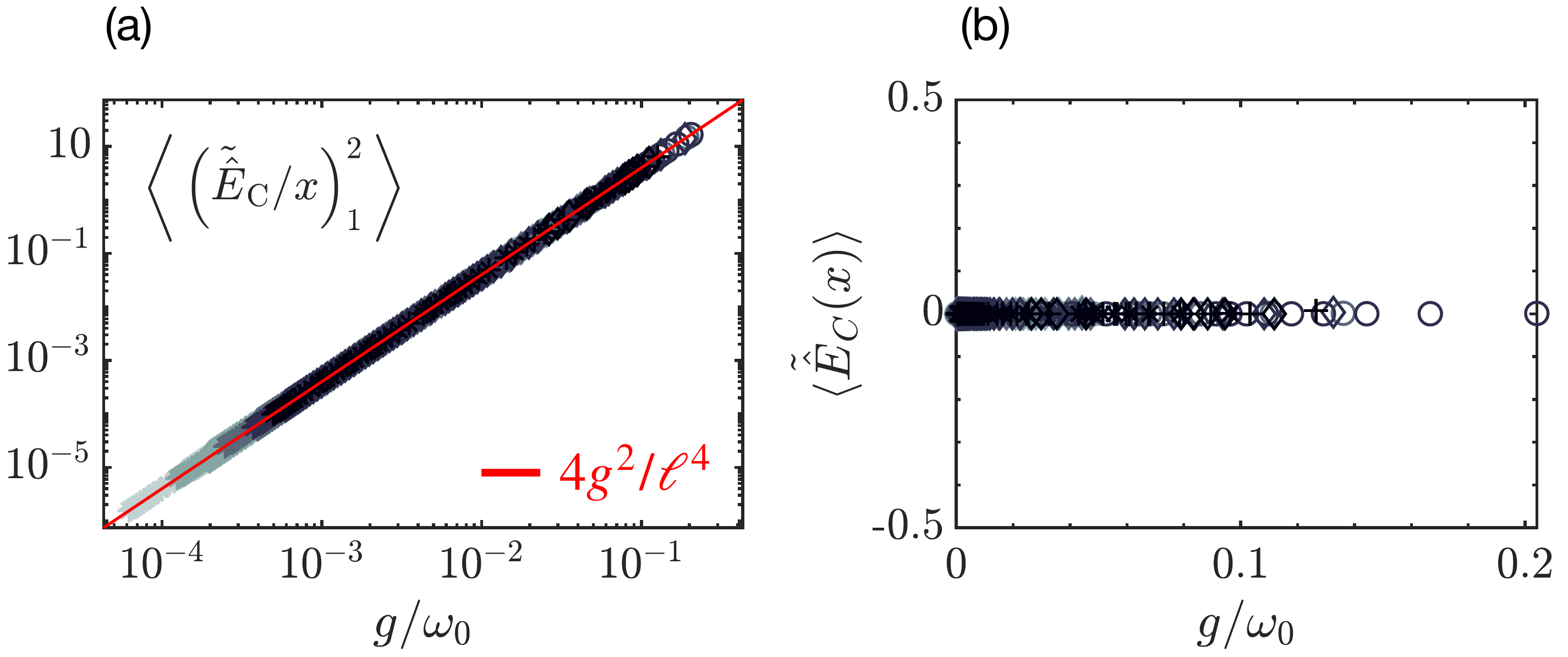}  
    \caption{(a) The vacuum zero point quantum fluctuations, that is the first term of Eq.~\eqref{eq:E13} $\left(\tilde{\hat E}_C(x)\right)_1^2$ for all DMRG ground states (light gray to black data for $L_y=2,4,8,12,16$ and $N=12,16,20,24$--- note we have 5 $L_y$ and 4 $N$ values)
    collapse on the same curve, $4g^2/\ell^4$, confirming the step from Eq.~\eqref{eq:E13} to~Eq.~\eqref{eq:E14}. (b) The cavity displacement is zero for the same data, independent of light-matter coupling.}   \label{fig:AppD}
\end{figure}

In this subsection, since we will be dealing with multiple gauges, we will use the following notation. Any operator in the Coulomb (original) gauge will have a subscript $C$. For example, the cavity mode destruction operator in the Coulomb gauge is $\hat{a}_C$. Recall that we used the unitary transformation of Eqs.~\eqref{eq:dipole_gauge1}-\eqref{eq:dipole_gauge2} to transform the Hamiltonian to dipole gauge.  An operator ${\cal O}_C$ in the Coulomb gauge transformed into the dipole gauge will be represented as ${\tilde{\cal O}}_C$; explicitly   
\begin{equation}
{\tilde{\cal O}}_C=U {\cal O}_C U^{\dagger}.
\end{equation}
We will frequently take the expectation value of various operators in the ground state. Clearly, the ground state depends on the gauge. Explicitly,
\begin{equation}
|\Psi_C\rangle=U^\dagger |\Psi_D\rangle\Rightarrow \langle\Psi_C|{\cal O}_C|\Psi_C\rangle=\langle\Psi_D|\tilde{\cal O}_C|\Psi_D\rangle.
\end{equation}
Using the notation defined above we obtain
\begin{eqnarray}
\tilde{\hat a}_C^\dagger=U{\hat a}_C^\dagger U^\dagger=\hat a_D^\dagger
+
i\frac{g}{\omega_0}\hat{\mathcal P},
\qquad
\tilde{\hat a}_C=U{\hat a}_C U^\dagger=
{\hat a}_D
-
i\frac{g}{\omega_0}\hat{\mathcal P},\label{eq:E5}
\end{eqnarray}
where $\hat{a}_D$ is a destruction operator for a dipole gauge photon.
It is important to note that since the unitary transformation depends only on position, it commutes with $\hat{\mathcal P}$. Thus, $\hat{\mathcal P}$ is the same in either gauge and does not need a suffix. It is also important to note that there is no projection to the LLL in Eq.~\eqref{eq:E5}. Recalling that $eA_0=\zeta/\ell$, the displacement field in the Coulomb gauge is 
\begin{eqnarray}
\hat{D}_C(x)
=
i\,u(x)\,\frac{\zeta}{e\ell}\omega_0\epsilon_0
\left[
 {\hat a}_C-{\hat a}_C^\dagger
\right],
\end{eqnarray}
After the unitary transformation we obtain 
\begin{eqnarray}
\tilde{\hat D}_C(x)
=i u(x)\omega_0\frac{\zeta}{e\ell}\left[{\tilde{\hat a}}_C-{\tilde{\hat a}}_C^\dagger\right]=
\frac{u(x)}{e}
\left[
\hat{\cal D}_D
+
2\epsilon_0 \frac{\zeta}{\ell} g\,\hat{\mathcal P}
\right],\label{eq:E6}
\end{eqnarray}
where we define $\hat{\cal D}_D
=
i\,\frac{\zeta}{\ell}\omega_0\epsilon_0
\left[
 {\hat a}_D-{\hat a}_D^\dagger
\right]$.
The corresponding physical 
electric field (in dipole gauge) is
\begin{eqnarray}
\tilde{\hat E}_C(x)
=
\frac{1}{\epsilon_0}
\tilde{\hat D}_C(x)
=
\frac{u(x)}{e\epsilon_0}
\left[
\hat{\cal D}_D
+
2\epsilon_0 \frac{\zeta}{\ell} g\,\hat{\mathcal P}
\right]=\frac{\zeta u(x)}{e\ell}\left[i\omega_0\left({\hat a}_D-{\hat a}_D^\dagger\right)+2g{\hat {\cal P}}\right]. \label{eq:E8}
\end{eqnarray}
In terms of our definition of $g$, we obtain 
\begin{eqnarray}
\frac{\tilde{\hat E}_C(x)}{x} = \frac{2g}{e\ell^2} \left[i \left(\hat a_D - \hat a_D^{\dagger} \right) + \frac{2g}{\omega_0}\mathcal{\hat  P}\right].\label{eq:E9}
\end{eqnarray}
Once again, we emphasize that no LLL projection appears in the above equation. We find the cavity field displacement $\langle \tilde{\hat E}_C(x)\rangle$ to be zero for all ground and excited states. For example, Fig.~\ref{fig:AppD}(b) shows $\langle \tilde{\hat E}_C(x)\rangle$ for the ground states at different $N$ and $L_y$ as computed in our DMRG. Recalling that the low energy states we consider are restricted to the LLL, this immediately implies that
\begin{eqnarray}
    \langle \mathcal{\hat  P} \rangle= \langle \hat{\bar{\mathcal P}}\rangle=-i\frac{\omega_0}{2g} \bigg \langle \left(\hat a_D - \hat a_D^{\dagger} \right)\bigg \rangle.  
\end{eqnarray}
The photonic part of the ground state wave function in dipole gauge is a coherent state as will be shown in the next section. Additionally, since the electric field in the dipole gauge is also zero, i.e.,~$\langle \mathcal{\hat D}_D \rangle=0$, the real part of $\langle \hat a_D \rangle$ has to vanish:
\[
\langle \hat a_D \rangle = ip, \qquad p \in \mathbb{R},
\]
consistent with Eq.~\eqref{eq:coherent_photon_dipolegauge}.
Then,
\begin{equation}
\langle \mathcal{\hat  P}\rangle =   \frac{\omega_0}{g}p =    \frac{\omega_0}{g}\sqrt{\langle \hat n^D_{\rm ph}\rangle}.\label{eq:Pandn_D}
\end{equation}
where $\hat n^D_{\rm ph} \equiv \hat{a}_D^\dagger \hat{a}_D$.
Hence the average polarizability of the electronic system can be immediately determined with the coherent state amplitude, or the photon number, of the photonic state in the dipole gauge. 

Now we compute the second moment of the dressed electric field $\left(\tilde{\hat E}_C(x)\right)^2$ and the physical photon number $\tilde{\hat n}_{\rm ph}^C$. One has to be extra careful for this calculation, because there is no LLL projection in Eq.~\eqref{eq:E9}. On the other hand, the electronic part of the wave function is restricted to the LLL. Thus, we need the following identity, which is a consequence of Eq.~\eqref{eq:C22}. 
\begin{eqnarray}
 P_{\rm LLL}{\hat {\mathcal P}}^2 P_{\rm LLL}=\overline{{{\hat{\mathcal P}}}^2}={\hat{\bar{\mathcal P}}}^2+2{\hat{\bar{\mathcal P}}}-\frac{{\hat N}_e}{2},\label{eq:E11}
\end{eqnarray}
where $\hat N_{\rm e}$ is the number operator for the total electron number. 
The second moment of the dressed electric field is
\begin{eqnarray}
\left(\tilde{\hat E}_C(x)\right)^2
&=&
\frac{u^2(x)}{e^2\epsilon_0^2}
\left[
 \hat{\cal D}_D^2
+
4\epsilon_0 \frac{\zeta}{\ell}g
\hat{\cal D}_D\hat{\bar{\mathcal P}}
+
4\epsilon_0^2\left(\frac{\zeta}{\ell}\right)^2g^2
\left(\hat{\bar{\mathcal P}}^2 + 2 \hat{\bar{\mathcal P}} - \frac{\hat N_{\rm e}}{2}\right)
\right],\label{eq:E12}\\
\left(\frac{\tilde{\hat E}_C(x)}{x}\right)^2 &=& \frac{4g^2}{e^2\ell^4} \left[-\left(\hat a_D +\hat a_D^{\dagger}\right)^2 +i \frac{4g}{\omega_0} \left(\hat a_D - \hat a_D^{\dagger}\right) {\hat{\bar{\mathcal P}}} + \frac{4g^2}{\omega_0^2}{\hat{\bar{\mathcal P}}}^2 + \frac{8g^2}{\omega_0^2}{\hat{\bar{\mathcal P}}} - \frac{2g^2}{\omega_0^2} \hat N_{\rm e} \right], \notag \\
&=& \frac{4g^2}{e^2\ell^4} \left[\bigg\lbrace i \left(\hat a_D - \hat a_D^{\dagger} \right) + \frac{2g}{\omega_0}{\hat {\bar{\mathcal  P}}} \bigg\rbrace^2 + \frac{8g^2}{\omega_0^2}{\hat{\bar{\mathcal P}}} - \frac{2g^2}{\omega_0^2} \hat N_{\rm e} \right]\label{eq:E13}
\end{eqnarray}
Because $\langle \hat{\bar{\mathcal P}}\rangle$ is known, and the electronic ground state is found to be, within our DMRG simulations, an eigenstate of $\hat{\bar{\mathcal P}}$ 
leading to the variance $\langle \Delta {\hat{\bar{\mathcal P}}} \rangle = 0$, this expression reduces to
\begin{eqnarray}
\bigg\langle  \left(\frac{\tilde{\hat E}_C(x)}{x}\right)^2\bigg\rangle &=& \frac{4g^2}{e^2\ell^4} \left[1 + \frac{8g}{\omega_0}\sqrt{\langle \hat n^D_{\rm ph}\rangle} - \frac{2g^2}{\omega_0^2} N \right].\label{eq:E14}
\end{eqnarray}
The first term of this expression gives rise to an energy equivalent to the vacuum zero point energy of the isolated cavity mode. Now we proceed to verify this statement. Before computing the integrated $E$-field contribution of the first term, let us show explicitly the simplification of the first term in Eq.~\eqref{eq:E13}:
\begin{equation}
\langle\Psi_D|\bigg\lbrace i \left(\hat a_D - \hat a_D^{\dagger} \right) + \frac{2g}{\omega_0}{\hat {\bar{\mathcal  P}}} \bigg\rbrace^2|\Psi_D\rangle.
\end{equation}
Rewrite the operator inside the expectation value as 
\begin{eqnarray}
\bigg\lbrace i \left(\hat a_D - \hat a_D^{\dagger} \right) + \frac{2g}{\omega_0}{\hat {\bar{\mathcal  P}}} \bigg\rbrace^2 &=& -\bigg\lbrace \left(\hat a_D - i\frac{g}{\omega_0}{\hat{\bar{\mathcal P}}}\right) -\left(a_D^{\dagger}  + i\frac{g}{\omega_0}{\hat {\bar{\mathcal  P}}}\right) \bigg\rbrace^2, \notag\\
&=&-\left(\hat a_D - i\frac{g}{\omega_0}{\hat{\bar{\mathcal P}}}\right)^2-\left(a_D^{\dagger}  + i\frac{g}{\omega_0}{\hat {\bar{\mathcal  P}}}\right)^2+\left(a_D^{\dagger}  + i\frac{g}{\omega_0}{\hat {\bar{\mathcal  P}}}\right)\left(a_D  - i\frac{g}{\omega_0}{\hat {\bar{\mathcal  P}}}\right), \notag \\
&+&\left(a_D  - i\frac{g}{\omega_0}{\hat {\bar{\mathcal  P}}}\right)\left(a_D^{\dagger}  + i\frac{g}{\omega_0}{\hat {\bar{\mathcal  P}}}\right).\label{eq:E17}
\end{eqnarray}
The ground state expectation value of the first three terms of Eq.~\eqref{eq:E17} is zero due to the fact that $|\Psi_D\rangle$ is a coherent state both within MFT, Eq.~\eqref{eq:coherent_photon_dipolegauge}, and also in DMRG calculations, as will be shown in Appendix~\ref{sec:AppE}. Moreover, for both the DMRG states and the mean-field ground states, we find that the electron number operator for each $X_n$ is a good quantum number.   This implies 
\begin{equation}
    \left({\hat a}_D-i\frac{g}{\omega_0}{\hat {\bar {\mathcal P}}}\right)|\Psi_D\rangle=0,\quad
   \langle\Psi_D|\left({\hat a}_D^\dagger+i\frac{g}{\omega_0}{\hat {\bar{\mathcal P}}}\right)=0.\label{eq:E18}
\end{equation}
We normal order the last term of Eq.~\eqref{eq:E17} and take the ground state expectation value to finally obtain 
\begin{equation}
\langle\Psi_D|\bigg\lbrace i \left(\hat a_D - \hat a_D^{\dagger} \right) + \frac{2g}{\omega_0}{\hat {\bar{\mathcal  P}}} \bigg\rbrace^2|\Psi_D\rangle=1,
\end{equation}
Now let us integrate this first term of Eq.~\eqref{eq:E13} over the sample to obtain
\begin{equation}
\int\limits_{-\frac{L_x}{2}}^{\frac{L_x}{2}}dx \int\limits_{-\frac{L_y}{2}}^{\frac{L_y}{2}} dy \frac{1}{2}\epsilon_0\frac{\zeta^2(u(x))^2\omega_0^2}{e^2\ell^2}\langle\Psi_D|\left\{i \left(\hat a_D - \hat a_D^{\dagger} \right) - \frac{2g}{\omega_0}{\hat {\bar{\mathcal  P}}}\right\}^2|\Psi_D\rangle=\frac{1}{2}\epsilon_0\frac{\zeta^2\omega_0^2}{e^2\ell^2}\int\limits_{-\frac{L_x}{2}}^{\frac{L_x}{2}}dx \int\limits_{-\frac{L_y}{2}}^{\frac{L_y}{2}} dy (u(x))^2
\end{equation}
Now, we recall that
\begin{equation}
    A_0=\frac{\zeta}{e\ell}=\frac{1}{\sqrt{2\epsilon_0\omega_0{\mathcal V}}}=\frac{1}{\sqrt{2\epsilon_0\omega_0L_xL_y\chi}}, 
\end{equation}
where $\chi$ is the effective length of the cavity field in the $z$ direction, and that
\begin{equation}
    \int\limits_{-\frac{L_x}{2}}^{\frac{L_x}{2}}dx \int\limits_{-\frac{L_y}{2}}^{\frac{L_y}{2}} dy \left(u(x)\right)^2=L_xL_y.
\end{equation}
We finally obtain
\begin{equation}
    \int\limits_{-\frac{L_x}{2}}^{\frac{L_x}{2}}dx \int\limits_{-\frac{L_y}{2}}^{\frac{L_y}{2}} dy \frac{1}{2}\epsilon_0\langle\Psi_D|\left(\tilde{\hat E}_C(x)\right)_1^2|\Psi_D\rangle=\frac{\omega_0}{4\chi}
\end{equation}
where the subscript 1 refers to the first term of Eq.~\eqref{eq:E13}.
This still has to be integrated over $z$ before one obtains the electric contribution to the zero-point energy, which accounts for the factor of $\chi$ in the denominator. Remembering that the electric and magnetic contributions to the zero point energy have to be identical, we obtain the usual cavity zero-point energy of $\omega_0/2$ due to the first term of Eq.~\eqref{eq:E13}. We indeed confirmed this for the states determined by DMRG.

Hence, the integrated result over the remaining terms in the second moment of electric field are purely induced by coupling to the electronic system. Because $\langle \tilde{\hat E}_C(x) \rangle$=0, the vacuum field fluctuations are controlled by the total electron number and the photon number in the dipole gauge. 
Correspondingly, we calculate the physical photon number
\begin{eqnarray}
\tilde{\hat n}_{\rm ph}^C
&=&
\left(
\hat a_D^\dagger
-
i\frac{g}{\omega_0}\hat{\mathcal P}
\right)
\left(
\hat a_D
+
i\frac{g}{\omega_0}\hat{\mathcal P}
\right)
=
\hat a_D^\dagger\hat a_D
-
\frac{g}{\omega_0}
i\left(\hat a_D-\hat a_D^\dagger\right)
\hat { {\mathcal P}}
+
\frac{g^2}{\omega_0^2}
\hat {{\mathcal P}}^2.\label{eq:D10}
\end{eqnarray}
Note that there is no projection on the LLL in the left side of the above expression. Taking the ground state expectation value, recalling that the electronic part of the ground state is restricted to the LLL, and using the identity Eq.~\eqref{eq:E11}, we find
\begin{eqnarray}
\langle\Psi_D|\tilde{\hat n}_{\rm ph}^C|\Psi_D\rangle
&=&
\langle\Psi_D|\hat a_D^\dagger\hat a_D
-
\frac{g}{\omega_0}
i\left(\hat a_D-\hat a_D^\dagger\right)
\hat {\bar {\mathcal P}}
+
\frac{g^2}{\omega_0^2}
\hat {\bar {\mathcal P}}^2 + \frac{2g^2}{\omega_0^2}\hat {\bar {\mathcal P}} - \frac{g^2}{2\omega_0^2} \hat N_{\rm e}|\Psi_D\rangle.
\end{eqnarray}
Now we use Eq.~\eqref{eq:E18} to finally obtain the average physical photon number in the Coulomb gauge
\begin{eqnarray}
\langle \tilde{\hat n}_{\rm ph}^C\rangle
&=&\frac{2g^2}{\omega_0^2}\langle{\hat{\bar{\mathcal P}}}\rangle-\frac{g^2}{2\omega_0^2}N\\
&=&\frac{2g}{\omega_0}\sqrt{\langle \hat n^D_{\rm ph}\rangle} - \frac{g^2}{2\omega_0^2} N, \label{eq:D11}
\end{eqnarray}
where we have used Eq.~\eqref{eq:Pandn_D} in the second line. 

\section{The cavity state in the dipole gauge \label{sec:AppE}}

\begin{figure}[t]
 \centering
\includegraphics[width=0.8\columnwidth]{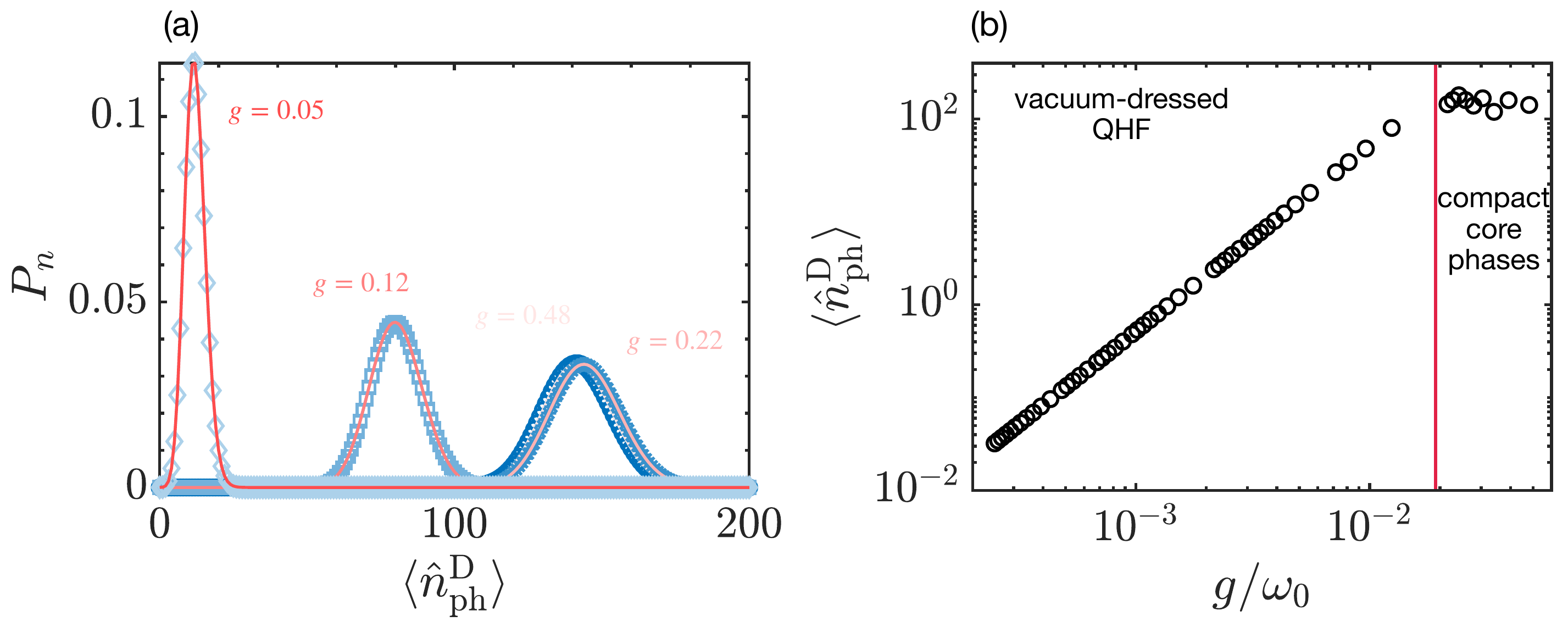}  
    \caption{(a) The dipole gauge photon state distribution for four different light-matter coupling values normalized by cavity frequency, $g/\omega_0=0.005,0.01,0.02,0.05$, each can be fitted with a Poisson distribution $e^{-\lambda}\lambda^n/n!$, where $\lambda = \langle \hat n^{\rm D}_{\rm ph}\rangle$ is the mean photon number in the dipole gauge, which is plotted in (b) with respect to light-matter coupling. $ \langle \hat n^{\rm D}_{\rm ph} \rangle$ shows distinct behaviors for many-body states in QHF and compact core phases.}   \label{fig:AppE}
\end{figure}

In dipole gauge, the system ground state of the cavity mode is that of a displaced oscillator, i.e., a coherent state. For any non-vanishing light-matter coupling between the 2DEG and the cavity mode, the photonic distribution exhibits a Poisson distribution, as seen in Fig.~\ref{fig:AppE}(a) where $\lambda = \langle \hat n^{\rm D}_{\rm ph}\rangle = \langle \hat a_D^\dagger \hat a_D \rangle$  is the mean photon number in the dipole gauge. We emphasize that this is not the physical photon number, which was computed in Eq.~\eqref{eq:D10}. 
It is important to monitor this fictitious mean photon number and the spread of the photonic distribution of this coherent state to allow for a sufficiently large Hilbert space for the photons, i.e.,~setting the correct photonic Hilbert space truncation in the dipole gauge. We find that $\langle \hat n^{\rm D}_{\rm ph}\rangle \propto g^2$, when the matter is in the uniform QHF phase, whereas it is nearly constant around a non-vanishing value once the matter transitions to the compact core phases, as illustrated in Fig.~\ref{fig:AppE}(b). This sensitivity to the breakdown of uniform QHF  in the electronic state is also the reason why the physical photon number can probe the electronic transition, which is discussed in the main text.

\section{Many-Body Gap in Spin Wave Spectrum \label{sec:appF}}

\begin{figure}[t]
 \centering
\includegraphics[width=0.45\columnwidth]{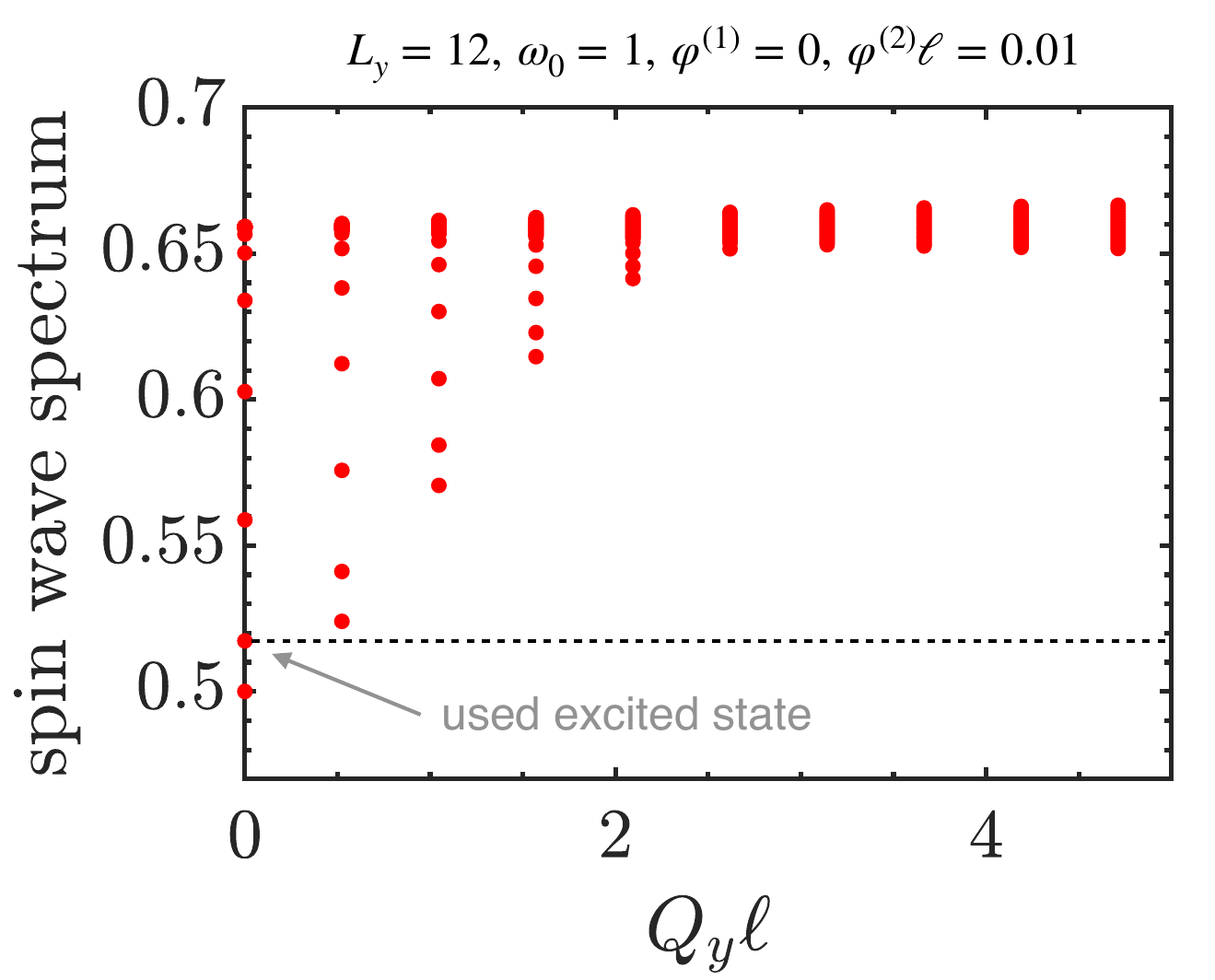}  
    \caption{The spin wave spectrum for $L_y=12$ and $N=25$ with cavity frequency $\omega_0=1$ in an antisymmetric cavity field with strength $\varphi^{(2)}\ell =0.01$. We find that the lowest energy excited state is at $Q_y\ell = 0$, as shown with a gray arrow.  We find this to be the momentum of the lowest energy excited state when $N$ and $L_y$ are small.}   \label{fig:AppF}
\end{figure}

Because of the relatively small system sizes involved in our comparison between the DMRG and spin wave computations, some care must be taken in extracting  the many-body energy gap in the latter case. For large enough systems one expects the lowest lying excited state to reside at the lowest non-vanishing $Q_y$ consistent with periodic boundary conditions, representing the lowest energy spin wave state as $L_y$ becomes large.  However, for smaller systems where $L_y$ is not so large, the quantization of the modes can bring the first excited $Q_y=0$ mode below the energy of any mode with non-vanishing $Q_y$.  This is illustrated in Fig. \ref{fig:AppF}.  

For our comparison of the many-body gap as computed in the DMRG and the spin wave analysis, we find this to be the case for all the system sizes considered.  For this reason the gap computed from our spin wave analysis uses the energy of the first excited state in the $Q_y=0$ sector.  A simplification in this situation allows an improved estimate of the many-body gap from the spin wave approach: examination of Eqs. \eqref{eq:pi_eval} and \eqref{eq:vphi} show that $\pi_{\varphi}(X,Q_y=0)$ and $v_{\varphi}(X, Q_y=0)$  vanish identically when $Q_y=0$.  This means that the photon degree of freedom is completely decoupled from spin wave excitations for this wave vector even in Coulomb gauge.  We can exploit this to approximately include edge effects in the spin wave spectrum, in particular by adopting von Neumann (i.e., vanishing derivative) boundary conditions at the system edges $X-\pm L_x/2$.  In this case our spin wave excitations will have the form
$$
|m,Q_y=0\rangle \propto
\sum_{X_n=-L_x/2}^{L_x/2} \cos\left(\frac{\pi m X_n}{L_x}\right)|X_n,Q_y=0\rangle,
$$
with $m \ge 0$ an integer.  The lowest excited state in this case occurs for $m=1$, and we use this value for the results presented in Fig. \ref{fig:QHFgapEntSpec}.  

We note that in the limit of large $N$ and $L_y$, differences between the periodic boundary conditions used for our other calculations and the von Neumann boundary conditions become less important, as the spin wave spectrum becomes dense.  The use of periodic boundary conditions for small system sizes in the spin wave calculations, however, leads to significant discrepancy from the gap values obtained using DMRG.

\section{Derivation of entanglement spectrum of the QHF many-body ground state \label{sec:entSpecQHFProof}}

For even $N$, the finite-size $S_z=0$ ferromagnetic state is the maximally symmetric Dicke state
\[
|\psi_{\rm FM}\rangle
=
\binom{N}{N/2}^{-1/2}
\sum_{C:\, |C|=N/2} |C\rangle ,
\]
where $C$ denotes a choice of the $N/2$ orbitals carrying $\downarrow$ spin.  We cut the cylinder into two equal halves, each containing $N/2$ orbitals.  The Schmidt states of each half are labeled by the maximal subsystem spin $S_{L/2}=\frac{N}{4}$, and its $z$-component $m$. The Schmidt decomposition is
\[
|\psi_{\rm FM}\rangle
=
\sum_{m=-N/4}^{N/4}
\sqrt{\lambda_m}\,
\left|S_{L/2},m\right\rangle
\left|S_{\overline{L/2}},-m\right\rangle ,
\]
with
\[
\lambda_m
=
\frac{
\binom{N/2}{N/4+m}
\binom{N/2}{N/4-m}
}{
\binom{N}{N/2}
}
=
\frac{
\binom{N/2}{N/4+m}^2
}{
\binom{N}{N/2}
}.
\]
The corresponding entanglement spectrum eigenvalues follow straightforwardly, using $\xi_m=-\log \lambda_m$. Since $\lambda_m=\lambda_{-m}$, the entanglement spectrum is symmetric under $m\leftrightarrow -m$. Thus all nonzero-$m$ Schmidt sectors occur in degenerate pairs, while the self-conjugate $m=0$ sector, when present, gives an unpaired entanglement level. This explains the characteristic structure of the $S_z=0$ quantum Hall ferromagnet entanglement spectrum: an isolated lowest level associated with the $m=0$ sector, followed by doublets associated with opposite subsystem spin polarizations. 

\section{Derivation of entanglement spectrum of a spin wave many-body state \label{sec:SW-entSpec}}

We now present a simple symmetry argument for the twofold degeneracy of 
the entanglement spectrum of a spin wave excitation.  Consider an even
number of electrons (and hence orbitals) $N$ 
at filling factor $\nu=1$, and restrict to the singly occupied
ferromagnetic Hilbert space. The fully polarized quantum Hall
ferromagnet was already defined in the main text as
\[
|\Omega\rangle
=
\prod_n d_{\uparrow,n}^{\dagger}|0\rangle .
\]
We define the local and uniform spin-lowering operators by
\[
s_n^-
=
d_{\downarrow,n}^{\dagger}d_{\uparrow,n},
\qquad
S^-=\sum_n s_n^- ,
\]
and introduce a nonuniform spin-flip operator
\[
S_f^-
=
\sum_n f_n s_n^- .
\]
We require the spin wave envelope to be odd under inversion of the
cylinder $f_{-n}=-f_n$, hence $\sum_n f_n=0$ holds. An example for such an envelope is $f_n \propto \sin(Q_x X_n)$ where we have $X_{-n}=-X_n$. 
The state $S_f^-|\Omega\rangle$ has $S_z=N/2-1$, hence the $S_z=0$ member of this spin wave multiplet is
\[
|\Psi_{\rm sw}^{(-)}\rangle
\propto
\left(S^-\right)^{N/2-1}
S_f^-|\Omega\rangle ,
\]
up to a normalization constant. 

Now let $\mathcal I$ denote inversion about the center of the
cylinder.  Up to a fixed phase convention, its action on
the electronic operators is
\begin{equation}
\mathcal I
d_{\sigma,n}^{\dagger}
\mathcal I^{-1}
=
d_{\sigma,-n}^{\dagger}.
\label{eq:inversion}
\end{equation}
We choose the overall phase of $\mathcal I$ such that $\mathcal I|\Omega\rangle=|\Omega\rangle$. With this choice the uniform QHF state is even under inversion, which we indeed confirm with our DMRG simulations. Note that the uniform lowering operator is also inversion even,
\[
\mathcal I S^-\mathcal I^{-1}=S^-,
\]
whereas the nonuniform spin-flip operator is inversion odd:
\[
\begin{aligned}
\mathcal I S_f^-\mathcal I^{-1} =
\sum_n f_n s_{-n}^- =
\sum_n f_{-n}s_n^- = -S_f^- .
\end{aligned}
\]
It follows then this spin wave state is odd under the inversion that exchanges the
two equal halves of the orbital chain, 
\[
\mathcal I|\Psi_{\rm sw}^{(-)}\rangle
=
-|\Psi_{\rm sw}^{(-)}\rangle. 
\]
We now bipartition the cylinder into two equal halves about
$X=0$ and expand the state as
\[
|\Psi_{\rm sw}^{(-)}\rangle
=
\sum_{\alpha,\beta}
\Gamma_{\alpha\beta}
|\alpha_{L/2}\rangle
|\beta_{\overline{L/2}}\rangle .
\]
Here $\Gamma$ is the Schmidt coefficient matrix of the bipartite wavefunction.
We choose matched bases for the two halves such that a basis state for the $L/2$-half labeled as $\alpha_{L/2}$ is the inversion image of the basis state for the other half labeled as $\alpha_{\overline{L/2}}$. 
Consequently, inversion exchanges the two basis labels:
\[
\mathcal I
\left(
|\alpha_{L/2}\rangle
|\beta_{\overline{L/2}}\rangle
\right)
=
|\beta_{L/2}\rangle
|\alpha_{\overline{L/2}}\rangle .
\]
Because the spin wave remains within the Hilbert space of states with strictly singly occupied
orbitals, which we have confirmed with DMRG,  inversion of the states can be formally implemented using the action of the $\mathcal{I}$ operator in Eq. \eqref{eq:inversion}.  This operation inverts the orbital order of the $d_{\sigma,n}^{\dagger}$ operators of the basis states in which the spin wave state is expanded. Reordering them to their original form introduces an overall sign that is the {\it same} for every basis state, yielding a possible overall change in sign that can be contained in the Schmidt coefficients $\Gamma_{\alpha\beta}$, leading to no impact on the entanglement spectrum eigenvalues. Thus, acting on the bipartite expansion of a state with the inversion operator yields
\[
\begin{aligned}
\mathcal I|\Psi_{\rm sw}^{(-)}\rangle
&=
\sum_{\alpha,\beta}
\Gamma_{\alpha\beta}
|\beta_{L/2}\rangle
|\alpha_{\overline{L/2}}\rangle =
\sum_{\alpha,\beta}
\Gamma_{\beta\alpha}
|\alpha_{L/2}\rangle
|\beta_{\overline{L/2}}\rangle .
\end{aligned}
\]
On the other hand, inversion oddness requires
\[
\mathcal I|\Psi_{\rm sw}^{(-)}\rangle
=
-\sum_{\alpha,\beta}
\Gamma_{\alpha\beta}
|\alpha_{L/2}\rangle
|\beta_{\overline{L/2}}\rangle .
\]
Comparing the coefficients of the basis states yields
\[
\Gamma_{\beta\alpha}
=
-\Gamma_{\alpha\beta},
\]
or equivalently in matrix notation $\Gamma^T=-\Gamma$. The bipartite coefficient matrix is therefore antisymmetric. The eigenvalues of the reduced density matrix $\rho_{L/2}
=
\operatorname{Tr}_{\overline{L/2}}
\left[
|\Psi_{\rm sw}^{(-)}\rangle
\langle\Psi_{\rm sw}^{(-)}|
\right]$ are the squares of the singular values of $\Gamma$, since $\rho_{L/2}=\Gamma\Gamma^{\dagger}$. 
Any complex antisymmetric matrix admits a decomposition of the form
\[
\Gamma
=
U
\left[
\bigoplus_{j=1}^{r}
\begin{pmatrix}
0&s_j\\
-s_j&0
\end{pmatrix}
\oplus 0
\right]
U^T ,
\]
where $s_j>0$.  The optional zero block represents the null space of
$\Gamma$.  It follows that
\[
\rho_{L/2}
=
U
\left[
\bigoplus_{j=1}^{r}
s_j^2
\begin{pmatrix}
1&0\\
0&1
\end{pmatrix}
\oplus 0
\right]
U^{\dagger}.
\]
Therefore every nonzero eigenvalue of $\rho_{L/2}$ occurs twice $\lambda_{2j-1}=\lambda_{2j}=s_j^2 $. 
The eigenvalues in the null space do not appear in the
finite entanglement spectrum. All together this leads to $\xi_{2j-1}=\xi_{2j}$. 
Thus the finite entanglement spectrum of an inversion-odd spin wave
state is exactly twofold degenerate.

This result does not predict the numerical values of the eigenvalues.  Rather, it establishes their pairwise degeneracy from
the inversion parity of the state. Because we find that the DMRG excited state whose entanglement spectrum is doubly degenerate throughout, is indeed inversion-odd, (i) our proof explains why the entanglement spectrum in Fig.~\ref{fig:QHFgapEntSpec}(b) for the excited state must be doubly degenerate, and (ii) the assumed spin wave state above is a viable many-body excited state on the QHF in $S_z=0$ sector.

\bibliographystyle{apsrev4-1}
%\bibliography{Bibliography}

\begin{thebibliography}{71}
\bibitem [{com()}]{com_phi1size}%
  \BibitemOpen
  \href@noop {} {}\bibinfo {note} {Note that we are assuming $\varphi^{(1)}$ is not so large that the value of $Q_y\ell^2$ is outside the system.}\BibitemShut {Stop}%
\end{thebibliography}


\end{document}